\documentclass[twocolumn,prd,superscriptaddress,nofootinbib,preprintnumbers]{revtex4-1}

\usepackage{mathtools,slashed}
\usepackage{booktabs,multirow}
\usepackage{graphicx}
\usepackage{amsmath,bm,amssymb,amsfonts,dsfont,xspace}
\usepackage{textgreek}
\usepackage{derivative}
\usepackage[usenames,dvipsnames]{xcolor}
\usepackage[normalem]{ulem}
\usepackage{url}
\usepackage{footmisc}
\usepackage{multirow}
\usepackage[colorlinks  = true,
            linkcolor   = NavyBlue,
            urlcolor    = NavyBlue,
            citecolor   = NavyBlue,
            anchorcolor = NavyBlue, unicode]{hyperref}

\newcommand{\lsim}{\mathrel{\mathop{\kern 0pt \rlap
  {\raise.2ex\hbox{$<$}}}
  \lower.9ex\hbox{\kern-.190em $\approx$}}}
\newcommand{\gsim}{\mathrel{\mathop{\kern 0pt \rlap
  {\raise.2ex\hbox{$>$}}}
  \lower.9ex\hbox{\kern-.190em $\approx$}}}

\newcommand{\vMol}{v}
\newcommand{\sigmav}{$\langle\sigma \vMol\rangle$ }
  
\renewcommand{\vec}[1]{\boldsymbol{#1}}

\newcommand{\maddm}{{\sc MadDM}\xspace}

\newcommand{\micromegas}{{\sc micrOMEGAs}\xspace}

\newcommand{\drake}{{\sc DRAKE}\xspace}

\usepackage[normalem]{ulem} 

\begin{document}


\title{Dark matter Simplified models in the Resonance Region}

\author{Mattia Di Mauro}\email{dimauro.mattia@gmail.com}
\affiliation{Istituto Nazionale di Fisica Nucleare, Sezione di Torino, Via P. Giuria 1, 10125 Torino, Italy}

\author{Bohan Xie}\email{zhd1xbh@gmail.com}
\affiliation{School of Physical Science and Technology, Southwest Jiaotong University, Chengdu 610031, PRC}

\begin{abstract}
The particle-physics nature of dark matter (DM) remains one of the central open questions in modern physics.
A widely used framework to investigate DM properties is provided by simplified models (\texttt{DMSimps}), which extend the Standard Model with a DM particle and a mediator that connects the visible and dark sectors.
Much of the \texttt{DMSimps} parameter space is already constrained by direct and indirect detection, collider searches, and the measured DM relic abundance.
We show, however, that the \emph{resonant} regime ($m_{\rm DM}\simeq m_{\rm med}/2$) remains viable under current bounds and will be stringently tested by forthcoming experiments.
Using a full Boltzmann treatment that allows for departures from kinetic equilibrium near resonance, we demonstrate that this regime can reproduce the observed relic density with couplings compatible with direct-detection limits. 
We also show that models with $s$-wave–dominated annihilation can explain the \textit{Fermi}-LAT Galactic Center Excess with couplings consistent with relic-density and direct-detection constraints.
Finally, we propose two minimal constructions that naturally realize $m_{\rm med}\!\approx\!2m_{\rm DM}$, making the resonant scenario generic rather than fine-tuned.
\end{abstract}

\maketitle

\section{Introduction}
\label{sec:intro}

The origin of dark matter (DM) is one of the most puzzling problems in physics. So far, only the gravitational effects of DM have been observed, while no such particles have been detected in laboratory experiments~\cite{Bertone:2016nfn,Bertone:2010zza,Cirelli:2024ssz}.
If DM is composed of particles, its existence implies physics beyond the Standard Model (BSM), since no particles in the Standard Model (SM) can fully account for it.

A viable DM candidate should satisfy several properties~\cite{Cirelli:2024ssz}.
It should be stable (or have a lifetime longer than the age of the Universe), electrically neutral, and weakly interacting with ordinary matter.
Moreover, the underlying BSM should contain a mechanism that leads from the early Universe to the currently observed DM abundance.
DM should primarily consist of non-relativistic particles to be consistent with structure formation.
Its self-annihilation rate must be small enough to agree with observations of the Bullet Cluster.
Finally, the DM properties and couplings must comply with constraints from laboratory searches.

Weakly Interacting Massive Particles (WIMPs) are among the best-motivated DM candidates that can fulfill all of the aforementioned conditions.
Several BSM theories, such as Supersymmetry, predict new particles with properties suitable to act as WIMPs.
Furthermore, WIMPs naturally undergo thermal freeze-out, which can account for the observed relic abundance.
In particular, WIMPs with masses of the order of electroweak gauge bosons ($\sim$\,GeV--TeV) and interaction cross sections comparable to electroweak ones would be produced through freeze-out with a relic density matching the observed value ($\Omega_{\rm{DM}} h^2 \simeq 0.12$).
This phenomenon is often referred to as the ``WIMP miracle.''

DM particle interactions are typically searched for through four strategies.
Measurements of anisotropies in the Cosmic Microwave Background (CMB) provide one of the most precise estimates of the DM density, determined by Planck to be $\Omega_{\rm DM} h^2 = 0.120$ with a $1\%$ uncertainty~\cite{Aghanim:2018eyx}.
Direct detection experiments aim to observe the scattering of cosmic DM particles off nuclei or electrons in detectors. These detectors, often located underground to shield them from cosmic rays (CRs), use ton-scale noble gas targets (e.g., xenon and argon)~\cite{Schumann:2019eaa}.
Collider experiments, such as those at LEP and the LHC, attempt to produce DM particles in high-energy collisions of SM particles (e.g., $e^{\pm}$ or $pp$); they look for signatures such as missing transverse energy in event reconstructions~\cite{Boveia:2018yeb}.
Indirect detection experiments, which may be ground-based or satellite-based, attempt to disentangle potential DM-induced CRs from those produced by known astrophysical sources, focusing on rare CRs such as $\gamma$ rays, neutrinos, and antimatter (antiprotons, positrons, and antinuclei)~\cite{Gaskins:2016cha}.

Several independent studies have reported an excess of $\gamma$-ray emission detected by the {\sc Fermi} Large Area Telescope ({\sc Fermi-LAT}) toward the Galactic Center (GC) (see, e.g., Refs.~\cite{Goodenough:2009gk,Hooper:2010mq,Boyarsky:2010dr,Hooper:2011ti,Abazajian:2012pn,Gordon:2013vta,Abazajian:2014fta,Daylan:2014rsa,Calore:2014nla,Calore:2014xka,TheFermi-LAT:2015kwa,TheFermi-LAT:2017vmf,DiMauro:2019frs,DiMauro:2021raz,Cholis:2021rpp}).
This feature, known as the Galactic Center Excess (GCE), has been observed under various background models—including point sources, extended sources, interstellar emission (IEM), {\sc Fermi} bubbles, and isotropic components—and with diverse data selections and analysis techniques.
Refs.~\cite{DiMauro:2021raz,Cholis:2021rpp} (hereafter Di~Mauro+21 and Cholis+22, respectively) have recently provided among the most accurate characterizations of the spectrum and morphology of the GCE using a template-fitting approach.
Their analyses confirm that the GCE is approximately spherically symmetric; the spectral energy distribution peaks at a few~GeV and extends up to $\sim 50$~GeV, with normalization variations of up to $60\%$ across different IEMs and analysis choices.
The GCE flux is consistent with DM annihilating into hadronic channels with a DM mass of about $30$--$60$~GeV and a cross section close to the thermal one, which provides the right relic density.
Finally, the spatial morphology follows a generalized Navarro--Frenk--White (NFW) profile with slope $\gamma=1.2$--$1.3$.
Therefore, the characteristics of the GCE are consistent with a DM origin.

DM models are often studied using three theoretical frameworks.
The first and simplest is Effective Field Theory (EFT), where only the DM particle is added to the SM.
These theories are valid as long as the energy scale of the processes is much lower than the mediator mass (which could correspond to a new particle).
In this case, the exact particle content of the BSM theory is irrelevant, as only the DM particle is accessible to current experiments.
EFTs have been widely used in indirect \cite{DiMauro:2015jxa,Fermi-LAT:2016afa,Gaskins:2016cha} and direct detection studies, as well as in collider searches during Run~1 and~2~\cite{ABERCROMBIE2020100371,ATLAS:2024kpy}.

When the process energy becomes comparable to the mediator mass, the microphysics of the BSM model—particularly the mediator properties—must be considered explicitly.
In such scenarios, the mediator is included in the model alongside the DM particle. The typical free parameters in this case are the masses of the DM and mediator particles, and the couplings between the mediator and the DM and SM particles.
If the mediator is an SM particle, such as in Higgs or $Z$ portal models~\cite{Cline:2013gha,Arcadi:2014lta,Beniwal:2015sdl,Arcadi:2019lka,Arcadi:2024ukq,DiMauro:2023tho}, the number of new parameters can be reduced to two.

When the mediator is a new particle, the models are referred to as Simplified Models (or {\tt DMSimps}), which typically involve four parameters: the DM and mediator masses, and two couplings.
{\tt DMSimps} provide a standard framework to study WIMPs via direct detection, indirect detection, collider experiments and cosmological constraints~\cite{ABDALLAH20158,Arina:2018zcq,Chang:2022jgo,Chang:2023cki,Arcadi:2024ukq}.
These models allow for various spin assignments for the DM and mediator particles, and different types of interactions (e.g., scalar, pseudoscalar, vector, axial-vector) in a systematic and standardized way.
{\tt DMSimps} are popular due to their minimal structure and represent the simplest extension of EFTs by introducing a mediator.

Finally, UV-complete theories may also be considered to study DM properties. However, these models often involve many parameters, making their exploration significantly more challenging.

{\tt DMSimps} have been tested against direct detection, indirect detection, collider searches, and cosmological constraints. Recently, Ref.~\cite{Arcadi:2024ukq} provided a comprehensive overview of these models considering different choices for the DM and mediator spins.
The authors show that most of the parameter space is excluded mainly due to very strong constraints from direct detection.
This is true for most of the parameter space except for the {\it resonance region}, in which the DM mass $m_{\rm DM}$ is about half the mediator mass $m_{\rm med}$.

In this paper, we perform a dedicated analysis of {\tt DMSimps} in the resonance region, which appears to be the only parameter space that can plausibly survive current constraints~\cite{Arcadi:2024ukq}.
In particular, when solving the Boltzmann equation, we calculate the relic density while taking into account the possibility of kinetic decoupling occurring simultaneously with, or prior to, chemical decoupling when $m_{\rm DM} \approx m_{\rm med}/2$.
Moreover, we include the most recent constraints from both indirect, using the most updated combined analysis of {\it Fermi}-LAT $\gamma$-ray data from Milky Way dwarf spheroidal galaxies (dSphs) \cite{McDaniel:2023bju}, and direct detection from {\sc XENONnT} and {\sc LZ} \cite{XENON:2023cxc,XENON:2024znc,LZ:2024zvo}.
We show that many {\tt DMSimps} scenarios remain viable even as future direct detection experiments approach the neutrino floor.
Therefore, although the DM and mediator masses require some tuning, {\tt DMSimps} can still be considered viable WIMP models.
Since the GCE is arguably the most striking hint of DM particle interactions, we also investigate its possible interpretation within a few DM simplified models.
In particular, in case future direct, indirect and collider experiments will detect WIMP DM consistent with a {\tt DMSimps} \emph{resonant} scenario, we would know for free also the mediator mass that is tied to the DM mass as $m_{\rm med}\simeq 2\,m_{\rm DM}$.

The paper is organized as follows: in Sec.~\ref{sec:model} we report the details of the models with the Lagrangians and the theoretical expressions for annihilation and nuclear cross sections.
In Sec.~\ref{sec:relicdensity} we present the evaluation of the relic density that takes into account the full solution of the Boltzmann equation.
In Sec.~\ref{sec:results} we display the combined results for each of the models presented in the paper.
In Sec.~\ref{sec:natural_resonance} we show two UV complete BSMs with a mechanism that would naturally force the model to sit in the resonant region. Finally, in Sec.~\ref{sec:conclusions} we draw our conclusions.

%
%
\section{Dark Matter Simplified Models}
\label{sec:model}

{\tt DMSimps} rely on a small number of foundational assumptions:

\begin{itemize}
\item The only new states accessible to current experiments are the DM particle and a mediator that connects the DM to the SM particles. The mediator is assumed to be the sole portal between the dark and visible sectors, coupling to both SM and DM particles. The stability of the DM candidate is enforced by an exact $\mathcal{Z}_2$ symmetry. Under these assumptions, the model typically introduces four new parameters: the masses of the DM ($m_{\rm DM}$) and mediator ($m_{\rm med}$), and the couplings of the mediator to DM ($\lambda$) and to SM particles ($\beta$). In the rest of the paper we assume the two coupling parameters take the same value and label it $g_{\rm{DM}}=\lambda=\beta$. Both the DM and the mediator can have spin-0, spin-$\tfrac{1}{2}$, spin-1, or spin-2.

\item The new interactions must respect all exact and approximate accidental global symmetries of the SM. This ensures, for example, that baryon- and lepton-number conservation remain intact. In addition, we assume Minimal Flavour Violation, which aligns any new flavour structures with the SM Yukawa matrices and thereby suppresses potential flavour-changing and CP-violating effects. Consequently, for Yukawa-like interactions (as in the scalar–mediator case), the mediator–fermion couplings are proportional to the SM fermion masses, \(\beta \, y_f = \beta\, m_f/(v_h\sqrt{2})\), where $v_h$ is the vacuum expectation value (VEV) of the SM Higgs boson.

\item Collider searches are generally insensitive to the DM spin, as missing-transverse-energy analyses often rely on cut-and-count strategies; differences in kinematic distributions induced by the DM spin usually have only a marginal impact (though Majorana fermions can forbid processes allowed for Dirac fermions). For this reason, DM is often taken to be a Dirac fermion in collider studies. By contrast, the DM spin can play a more prominent role in direct and indirect detection. In this work we explore scalar, fermionic, and vector DM candidates.
\end{itemize}

In the next sections we present the model Lagrangians and the direct- and indirect-detection cross sections. We consider {\tt DMSimps} with scalar and vector boson mediators, and DM particles as scalar, Dirac-fermion, and vector-boson.

\begin{table*}[t]
\centering
\setlength{\tabcolsep}{6pt}
\renewcommand{\arraystretch}{1.2}
\begin{tabular}{|c|c|c|c|c|}
\hline
\multirow{2}{*}{Mediator} & \multirow{2}{*}{DM spin} & \multicolumn{2}{c|}{Direct detection (DD)} & \multirow{2}{*}{ID $\langle\sigma v\rangle_0$ (s-channel)} \\
\cline{3-4}
 &  & SI & SD &  \\
\hline
\multirow{3}{*}{$S$ (scalar)} 
  & $S$ (scalar $\chi$)  & OK & NO & s-wave, helicity suppressed $\propto m_f^2$ \\
\cline{2-5}
  & $D$ (Dirac $\psi$)   & OK & NO & $p$-wave \\
\cline{2-5}
  & $V$ (vector $V$)     & OK & NO & s-wave, helicity suppressed $\propto m_f^2$ \\
\hline
\multirow{1}{*}{$P$ (pseudoscalar)} 
  & $D$ (Dirac $\psi$)   & loop (small)$^{\ast}$ & SD ($\propto q^4$)$^{\ast}$ & s-wave \\
\hline
\multirow{4}{*}{$Z'$ (vector)} 
  & $S$ (scalar $\chi$)  & OK & NO & $p$-wave \\
\cline{2-5}
  & $D$ (Dirac $\psi$), pure vector & OK & NO & s-wave \\
\cline{2-5}
  & $D$ (Dirac $\psi$), pure axial  & NO & OK & helicity suppressed $\propto m_f^2$ \\
\cline{2-5}
  & $V$ (vector $V$)     & OK & NO & typically s-wave$^{\S}$ \\
\hline
\end{tabular}
\caption{Summary for simplified DM models with scalar ($S$), pseudoscalar ($P$), and vector ($Z'$) mediators and scalar/Dirac/vector DM. 
The columns SI and SD refer to the spin-independent and dependent nuclear cross sections.
The ID column refers to $s$-channel annihilation to SM fermions in the $v\!\to\!0$ limit, which is relevant for indirect detection.
$t$-channel or mediator-pair final states (e.g.\ $\chi\chi\!\to\! SS$) are not shown.
\newline
$^{\ast}$~For a pseudoscalar mediator, tree-level scattering is SD and momentum-suppressed ($\propto q^4$); SI scattering arises only at loop level and is typically small.\\
$^{\S}$~For vector DM with a vector mediator, the precise $v$-dependence is model dependent; an $s$-wave piece is common in minimal setups.}
\label{tab:simp_summary_grid_ps_ax_vec}
\end{table*}

We summarize in Tab.~\ref{tab:simp_summary_grid_ps_ax_vec} whether the nuclear direct detection cross section is spin independent (SI) or spin dependent (SD) and we report the leading factor (s or p-wave or helicity suppressed) for the annihilation cross section that is relevant for indirect searches.

\subsection{Spin-0 mediator}
\label{sec:spin0}

\subsubsection{Scalar couplings}

We consider a scalar mediator $S$ with CP-even couplings to DM (pure scalar, no pseudoscalar component). 
Below we list the interaction Lagrangians added to the SM for spin-0 $\chi$, spin-$1/2$ $\psi$, and spin-1 $V^\mu$ DM, respectively:
\begin{eqnarray}
\label{eq:mediatorL}
\mathcal{L}^{\chi}_{\rm int} \!&=&\! \xi\,\mu_\chi \lambda_\chi\, \chi^2 S + \xi\,\lambda_\chi^2\,\chi^2 S^2 + \sum_f \frac{\beta}{\sqrt{2}} \frac{m_f}{v_h}\, \bar{f} f\, S \,,\\
\mathcal{L}^{\psi}_{\rm int} \!&=&\! \xi\,\lambda_\psi\, \bar{\psi}\psi\, S + \sum_f \frac{\beta}{\sqrt{2}} \frac{m_f}{v_h}\, \bar{f} f\, S \,, \nonumber\\
\mathcal{L}^{V}_{\rm int} \!&=&\! \mu_V \lambda_V\, V^\mu V_\mu S + \frac{1}{2}\lambda_V^2\, V^\mu V_\mu\, S S + \sum_f\frac{\beta}{\sqrt{2}} \frac{m_f}{v_h}\, \bar{f} f\, S \,.\nonumber
\end{eqnarray}
Here $\xi=1/2$ for a real scalar and a Majorana fermion, while $\xi=1$ for a complex scalar and a Dirac fermion\footnote{This definition of $\xi$ differs from that in \cite{Arcadi:2017kky,Arcadi:2024ukq} but matches, e.g., \cite{Arina:2018zcq} and references therein.}. 
The mediator–fermion interaction can be written in Yukawa form by defining $y_f \equiv (\beta/\sqrt{2})\, m_f/v_h$.
Each model also includes the kinetic terms for $S$ and for the DM field, and may contain a cubic self-interaction for the mediator\footnote{The cubic term $g_S S^3$ is generally not included in {\tt DMSimps}. It can become relevant for $g_S\!>\!1$ and when $m_\chi>m_{\rm med}$ in secluded DM scenarios~\cite{Pospelov:2007mp}.}. 
The parameters $\mu_\chi$ and $\mu_V$ have mass dimension; throughout we set $\mu_\chi=m_\chi$ and $\mu_V=m_V$ for scalar and vector DM, respectively\footnote{This choice differs from \cite{Arcadi:2017kky,Arcadi:2024ukq}, where $\mu_\chi$ is set equal to the mediator mass.}.

Given the simplicity of Eq.~\eqref{eq:mediatorL}, we can provide compact analytic expressions for the cross sections relevant to indirect and direct detection and to relic-density calculations, highlighting key features of these models.

\begin{figure}
\centering
\includegraphics[width=0.99\linewidth]{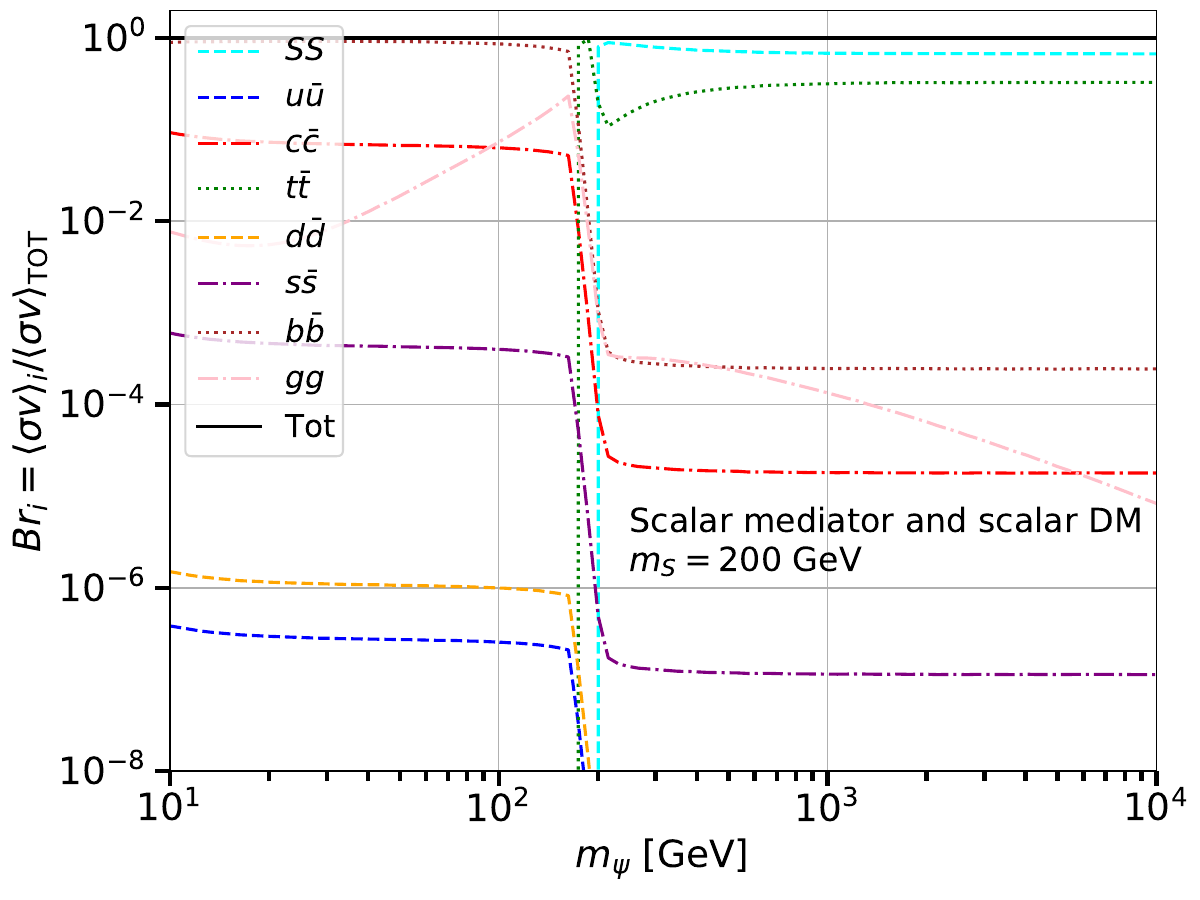}
\includegraphics[width=0.99\linewidth]{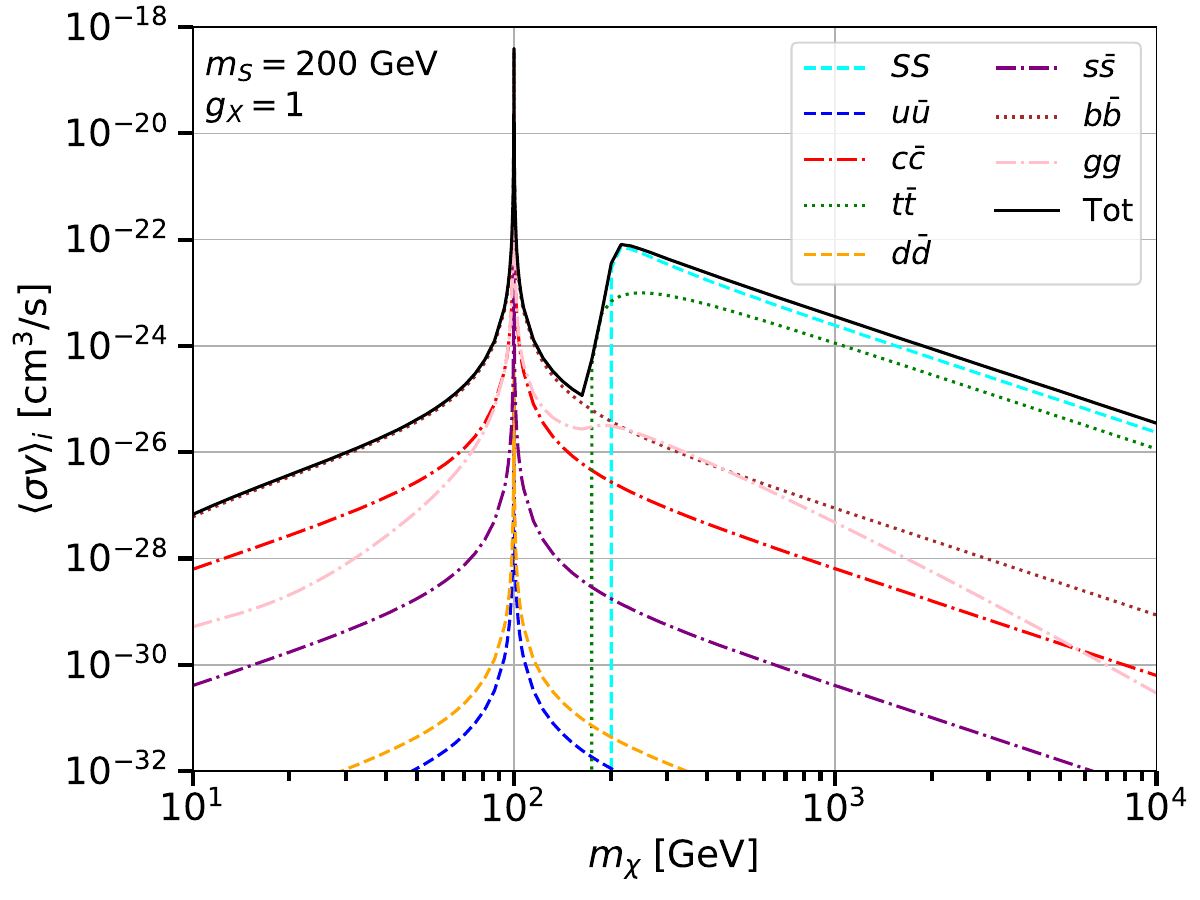}
\caption{Branching ratios (top) and velocity-averaged annihilation cross sections (bottom) as functions of the DM mass for a scalar mediator and scalar DM. We fix $m_{S}=200~\mathrm{GeV}$ and $g_X=\lambda=\beta=1$. 
The hierarchy of channels follows the SM fermion masses (Yukawa couplings); for $m_\chi\lesssim 170~\mathrm{GeV}$ the $b\bar b$ channel dominates, while above threshold the $t\bar t$ channel becomes important. 
The $SS$ final state opens for $m_\chi>m_{S}$. 
Near $m_{\rm S}\simeq 2m_\chi$ the $s$-channel resonance enhances the total cross section, allowing smaller couplings to achieve the correct relic abundance.}
\label{fig:brsigmav}
\end{figure}

The most relevant annihilation process for both relic density and indirect detection is DM annihilation into SM fermion pairs. 
Assuming $\xi=1$, the corresponding velocity-averaged cross sections for the three DM spins are approximately
\begin{eqnarray}
\label{eq:annihscalar}
& \langle\sigma v\rangle_{\bar f f} & \!\approx\! \frac{n_c^f}{4\pi}\, \lambda_\chi^2 \beta^2 \frac{m_f^2}{v_h^2}\, 
\frac{m_\chi^2 \left(1-\frac{m_f^2}{m_\chi^2}\right)^{3/2}}{\left(4 m_\chi^2-m_S^2\right)^2+m_S^2 \Gamma_S^2} \,,\\
& \langle\sigma v\rangle_{\bar f f} & \!\approx\! \frac{n_c^f\, v^2}{8\pi}\, \lambda_\psi^2 \beta^2 \frac{m_f^2}{v_h^2}\, 
\frac{m_\psi^2 \left(1-\frac{m_f^2}{m_\psi^2}\right)^{3/2}}{\left(4 m_\psi^2-m_S^2\right)^2+m_S^2 \Gamma_S^2} \,, \nonumber\\ 
& \langle\sigma v\rangle_{\bar f f} & \!\approx\! \frac{n_c^f}{3\pi}\, \lambda_V^2 \beta^2 \frac{m_f^2}{v_h^2}\, 
\frac{m_V^2 \left(1-\frac{m_f^2}{m_V^2}\right)^{3/2}}{\left(4 m_V^2-m_S^2\right)^2+m_S^2 \Gamma_S^2} \,, \nonumber
\end{eqnarray}
where $n_c^f$ is the color factor ($n_c^f=3$ for quarks and $n_c^f=1$ otherwise) and $\Gamma_S$ is the mediator decay width.
All previously reported annihilation cross sections are helicity suppressed ($\langle \sigma v\rangle \propto m_f^2$).
In Fig.~\ref{fig:brsigmav} we show the branching ratio for each annihilation channel $Br_i$ evaluated as the ratio between annihilation cross section divided for the total one: $Br_i=\langle \sigma v \rangle_i/\langle \sigma v \rangle_{\rm{TOT}}$.
We note that $Br_i$ follows the hierarchy of the Yukawa couplings and the SM fermion masses. In particular, for DM masses below the top mass, the annihilation into $b\bar{b}$ is the most relevant, because this quark has the highest mass among the kinematically accessible SM fermions. For DM masses above the top mass, instead, the annihilation channel into $t\bar{t}$ becomes one of the dominant.

If kinematically accessible ($m_{\rm DM} \ge m_S$), annihilation into mediator pairs ($SS$) should also be considered. 
This channel can affect the estimate of the relic density when both DM and $S$ remain in equilibrium with the thermal bath, even for relatively small couplings. 
In fact, from Fig.~\ref{fig:brsigmav} that when $m_{\rm{DM}}>m_s$ the annihilation into $SS$ pairs becomes the largest.
Moreover, the subsequent $S\!\to\! f\bar f$ decays produce four-body final states relevant for indirect detection:
\begin{equation}
\label{eq:4b}
{\rm DM}\,{\rm DM} \;\to\; S\,S \;\to\; 4 f \, .
\end{equation}
In the resonance region $m_{\rm DM}\!\approx\! m_{\rm med}/2$, which is the main region of the parameter space considered in this paper, the DM annihilation into on shell $SS$ pairs is kinematically forbidden for indirect detection, but we include the $4f$ processes of Eq.~\eqref{eq:4b} in our calculations for the relic density calculations.

While all three cross sections above share similar parametric dependencies, scalar and vector DM annihilations are s-wave dominated (velocity independent), whereas fermionic DM is p-wave dominated (velocity dependent). 
Thus, for the same $(m_{\rm DM},m_{\rm med})$ values, fermionic DM typically requires larger couplings to achieve the thermal target cross section.
Because p-wave annihilation is suppressed at late times, parameter regions yielding the correct relic abundance for fermionic DM are generally not testable with indirect detection constraints, unlike the scalar and vector cases. 
Typical DM velocities in astrophysical systems lie between $10^{-5}\,c$ and $10^{-3}\,c$, so $\langle \sigma v \rangle \propto v^2$ suffers a strong suppression today. This is not an issue for relic density, since before freeze-out DM was semi-relativistic ($v/c\sim 1$).

All $s$-channel annihilation cross sections include the Breit-Wigner structure.
\[
\langle \sigma v \rangle \;\propto\; \frac{1}{\left(4m_{\rm DM}^2 - m_{\rm med}^2\right)^2 + \Gamma_S^2 m_S^2}\,,
\]
so in the resonance region $m_{\rm med}\!\simeq\! 2 m_{\rm DM}$ the annihilation rate is strongly enhanced. This is visible in the bottom panel of Fig.~\ref{fig:brsigmav} where $\langle \sigma v\rangle$ increases by orders of magnitude when approaching the resonance region.
Consequently, much smaller couplings $\lambda$ and $\beta$ are sufficient to reproduce the relic abundance, making the resonant regime more easily compatible with direct-detection and collider constraints.

For direct detection, starting from Eq.~\eqref{eq:mediatorL} and working in the non-relativistic (NR) limit with $|q|\ll m_S$, one can integrate out $S$ using its equation of motion. 
Defining the DM currents
\[
\mathcal{O}_\chi \equiv \chi^\dagger\chi,\qquad 
\mathcal{O}_\psi \equiv \bar\psi\psi,\qquad
\mathcal{O}_V \equiv V_\mu V^\mu,
\]
and the short-hands
\[
C_\chi \equiv \xi\,\mu_\chi \lambda_\chi,\qquad 
C_\psi \equiv \xi\,\lambda_\psi,\qquad
C_V \equiv \mu_V \lambda_V,
\]
we obtain, to leading order in $q^2/m_S^2$, the following effective lagrangians
\begin{equation}
\mathcal{L}^{i}_{\rm eff} \!=\! 
\frac{C_i}{m_S^{2}}\,\mathcal{O}_i \!\left( \sum_{f=u,d,s,c,b,t} y_f\, \bar f f \right)
+ \mathcal{O}\!\left(\frac{q^2}{m_S^2}\right) , \, i=\chi,\psi,V.
\label{eq:Leff_quark}
\end{equation}

This yields the SI DM–proton scattering cross sections
\begin{equation}\label{scalarscalar}
\begin{aligned}
\sigma_{\chi p}^{\mathrm{SI}} &= \frac{\mu_{\chi p}^2}{4\pi}\, \frac{\lambda_\chi^2 \beta^2}{m_S^4}\, \frac{m_p^2}{v_h^2}\, \left[ f_p \frac{Z}{A} + f_n \!\left(1 - \frac{Z}{A} \right) \right]^2, \\
\sigma_{\psi p}^{\mathrm{SI}} &= \frac{\mu_{\psi p}^2}{\pi}\, \frac{\lambda_\psi^2 \beta^2}{m_S^4}\, \frac{m_p^2}{v_h^2}\, \left[ f_p \frac{Z}{A} + f_n \!\left(1 - \frac{Z}{A} \right) \right]^2, \\
\sigma_{V p}^{\mathrm{SI}} &= \frac{\mu_{V p}^2}{4\pi}\, \frac{\lambda_V^2 \beta^2}{m_S^4}\, \frac{m_p^2}{v_h^2}\, \left[ f_p \frac{Z}{A} + f_n \!\left(1 - \frac{Z}{A} \right) \right]^2,
\end{aligned}
\end{equation}
where $A$ and $Z$ are the target’s mass and proton numbers, and $\mu_{{\rm DM}p}=m_{\rm DM} m_p/(m_{\rm DM}+m_p)$ is the reduced DM–proton mass (with ${\rm DM}=\chi,\psi,V$). 
The coefficients $f_p$ and $f_n$ are the effective scalar couplings to protons and neutrons; for the CP-even scalar case with Yukawa-like couplings they are nearly equal, justifying the isospin-symmetric approximation often used in experimental analyses.

For light quarks $q=u,d,s$ we define
\begin{equation}
\langle N|\,m_q \bar q q\,|N\rangle \equiv m_N\, f_q^N,
\qquad N=p,n,
\label{eq:def_fqN}
\end{equation}
and
\begin{equation}
f_{TG}^N \equiv 1-\sum_{q=u,d,s} f_q^N,
\label{eq:def_fTGN}
\end{equation}
while heavy flavors $Q=c,b,t$ are integrated out via the SVZ relation
\begin{equation}
m_Q \bar Q Q = -\,\frac{\alpha_s}{12\pi}\,G_{\mu\nu}^a G^{a\,\mu\nu},
\label{eq:SVZ}
\end{equation}
which implies
\begin{equation}
\Big\langle N \Big| \frac{\alpha_s}{\pi} G_{\mu\nu}^a G^{a\,\mu\nu} \Big| N \Big\rangle
= -\,\frac{8}{9}\, m_N\, f_{TG}^N.
\label{eq:G2me_correct}
\end{equation}
The resulting scalar coupling of $S$ to a nucleon is
\begin{equation}
g_{SNN} =
m_N\!\left[\sum_{q=u,d,s} c_q\, f_q^N
+ \frac{2}{27}\, f_{TG}^N \sum_{Q=c,b,t} c_Q \right],
\label{eq:gSNN_master}
\end{equation}
and for Yukawa-like $c_q=c$ (with no hard gluon term) one finds
\begin{equation}
g_{SNN} = c\, m_N\!\left[\sum_{q=u,d,s} f_q^N + \frac{2}{9} f_{TG}^N\right]
\equiv c\, m_N\, f_N,
\label{eq:fN_simple}
\end{equation}
so that
\begin{equation}
f_N = \sum_{q=u,d,s} f_q^N + \frac{2}{9} f_{TG}^N.
\label{eq:fN_simple_val}
\end{equation}

We evaluate direct-detection cross sections using {\tt micrOMEGAs}~\cite{Belanger:2006is,Belanger:2013oya,Belanger:2018ccd,Alguero:2023zol}, which employs
\begin{eqnarray}
f_u^p&=&0.0153,\, f_u^n=0.011,\, 
f_d^p=0.0191,\, \\
f_d^n&=&0.0273,\, 
f_s^p=f_s^n=0.0447,
\end{eqnarray}
implying $f_{TG}^p\simeq f_{TG}^n\simeq 0.918$ and, with Eq.~\eqref{eq:fN_simple_val}, $f_p\simeq f_n$.

The {\sc LZ} and {\sc XENONnT} experiments currently set the strongest limits on the SI DM–nucleon cross section~\cite{LZ:2022ufs,XENON:2023cxc}. 
For $m_{\rm DM}\sim 10$--$1000~\mathrm{GeV}$ these bounds translate (within our setup) to $\sqrt{\lambda\,\beta}\lesssim (5\times 10^{-2})$--$10^{-1}$. 
As shown below, such small couplings typically produce away from the resonance $\langle\sigma v\rangle$ values much smaller than the thermal cross section and would thus overclose the Universe ($\Omega_{\rm{DM}}h^2>0.12$). 
Instead, the near-resonant regime $m_{\rm med}\simeq 2\,m_{\rm DM}$ provides a way to reconcile direct detection and relic density: the enhanced thermally averaged cross section allows the correct $\Omega_{\rm{DM}} h^2$ with much smaller couplings, remaining compatible with direct detection upper limits.

\subsubsection{Pseudoscalar couplings}

We now consider a simplified model featuring a CP-odd (pseudoscalar) spin-0 mediator $A$. 
Assuming CP conservation, this setup is consistent only with fermionic DM $\psi$. 
The relevant interaction Lagrangian is
\begin{equation}
\label{eq:pseudoDM}
\mathcal{L}^{\psi}_{\rm int} \;=\; i\,\lambda\, \bar{\psi}\gamma_5 \psi\, A 
\;+\; i \sum_f \frac{\beta}{\sqrt{2}} \frac{m_f}{v_h}\, \bar{f}\gamma_5 f\, A\, .
\end{equation}

The parity of the mediator has important phenomenological consequences. 
For annihilation into SM fermions one finds an s-wave (velocity–independent) rate that is helicity–suppressed by $m_f^2$:
\begin{equation}
\label{eq:ps-ff}
\langle\sigma v\rangle_{\bar f f} \;\approx\; 
\frac{n_c^f}{2\pi}\, \lambda^2 \beta^2\, \frac{m_f^2}{v_h^2}\;
\frac{m_\psi^2\,\sqrt{1-\tfrac{m_f^2}{m_\psi^2}}}{\bigl(4m_\psi^2-m_A^2\bigr)^2 + m_A^2 \Gamma_A^2}\, .
\end{equation}
In contrast, annihilation into a pair of mediators, $\psi\bar\psi\to A A$, is p-wave suppressed, 
$\langle\sigma v\rangle_{A A}\propto v^2$. 
In our analysis we focus on the resonant regime $2m_\psi \simeq m_A$, for which the $\psi \bar{\psi }\to A A$ channel is either kinematically inaccessible or subdominant.

The most distinctive feature of pseudoscalar mediator appears in direct detection. 
At tree level the interaction proceeds via the effetive operator $(i\bar{\psi}\gamma_5\psi)(i\bar q\gamma_5 q)$, which does not map onto the usual SI/SD operators. 
In the NR basis it contributes to the direct-detection operator $\mathcal{O}_6^{\rm NR}$~\cite{Fitzpatrick:2012ix}, yielding a nuclear recoil rate suppressed by the fourth power of the momentum transfer~\cite{Arina:2014yna,Dolan:2014ska}:
\begin{equation}
\label{eq:q4suppr}
\frac{d\sigma_T}{dE_R} =
\frac{\lambda^{2}\beta^{2}}{128\pi}
\frac{q^{4}}{m_{A}^{4}}
\frac{m_T^{2}}{m_\psi m_N v_E^{2}}
\frac{1}{E_R}
\sum_{N,N'=p,n} 
g_{N}\,g_{N'}\,
F^{NN'}_{\Sigma''}(q^{2})\,,
\end{equation}
with
\begin{equation}
\label{eq:gN-ps}
g_{N} \;\equiv\; 
\sum_{q=u,d,s}
\frac{m_{N}}{v_h}\,
\Biggl(1-\frac{\bar m}{m_q}\Biggr)\,
\Delta_{q}^{N}\,,
\bar m \equiv \Bigl(\tfrac{1}{m_u}+\tfrac{1}{m_d}+\tfrac{1}{m_s}\Bigr)^{-1}.
\end{equation}
Here $m_T$ is the nuclear mass, $m_N$ the nucleon mass, $E_R$ the recoil energy, $q^2\simeq 2m_T E_R$ in the NR limit, 
$\Delta_q^N$ encode the quark spin content of the nucleon, and $F_{\Sigma''}^{NN'}$ are the corresponding nuclear response functions~\cite{Fitzpatrick:2012ix}. 
Because typical elastic scatters have $|\vec q|\sim 10$–$100~\mathrm{MeV}$, the tree–level rate is extremely small.

However, in pseudoscalar models a coherent SI interaction is generated at loop level and, crucially, the loop–induced amplitude does not carry the same $q^4$ suppression. 
The resulting nuclear coherence can compensate the loop penalty, placing sizeable regions of parameter space within reach of current and near–future experiments. 
We follow the calculation in Ref.~\cite{Abe:2018emu} (see also~\cite{Ertas:2019dew}) for the loop–induced SI cross section.

\begin{figure}
\centering
\includegraphics[width=0.99\linewidth]{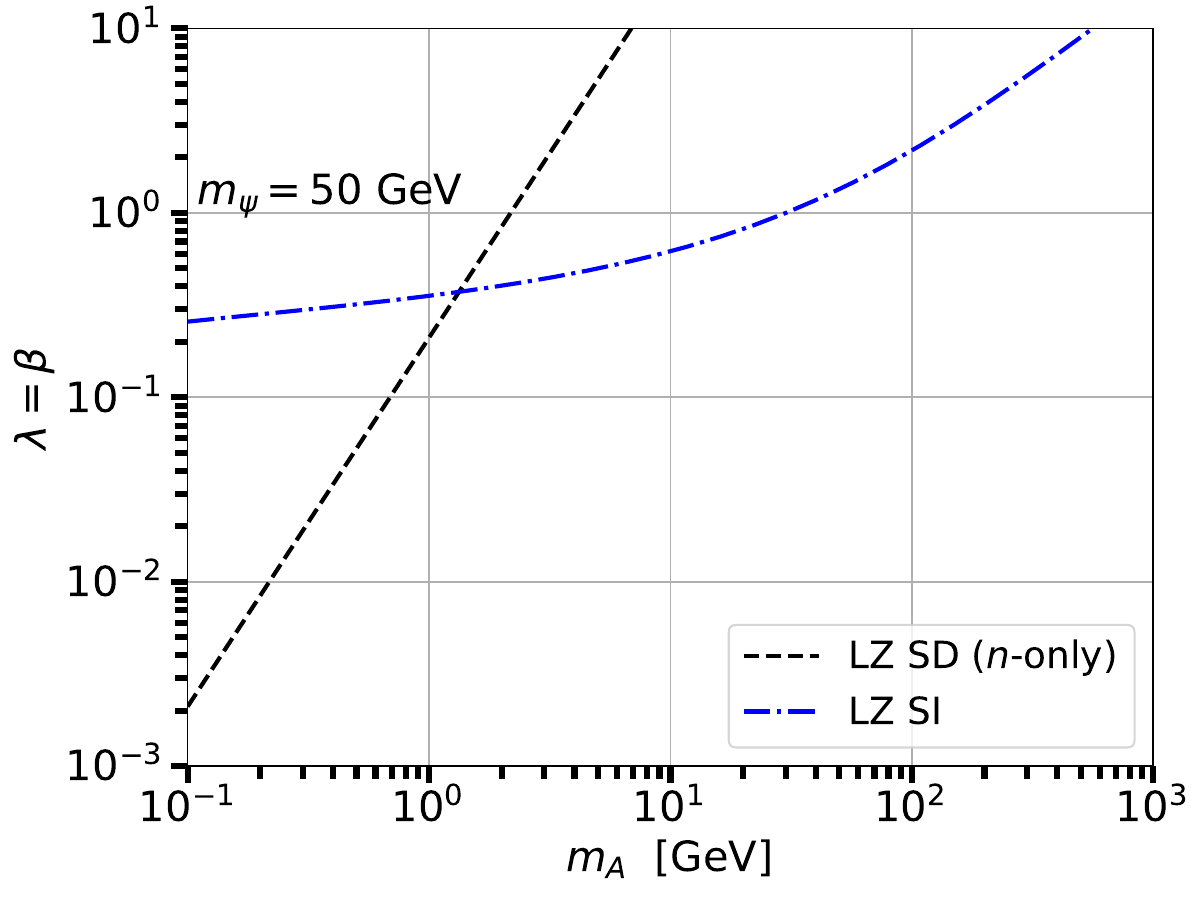}
\caption{Upper limits on the couplings $\lambda=\beta$ for the pseudoscalar mediator model of Eq.~\ref{eq:pseudoDM}, obtained by comparing the tree–level (momentum–suppressed) SD and the one–loop SI nuclear cross sections with the LZ data~\cite{LZ:2024zvo}. 
We fix $m_\psi=50~\mathrm{GeV}$ and show limits as a function of $m_A$.}
\label{fig:pseudolimits}
\end{figure}

Fig.~\ref{fig:pseudolimits} shows the upper limits obtained for $\lambda=\beta$ when the DM mass is fixed to 50 GeV.
For light mediators, $m_A\!\lesssim\!1~\mathrm{GeV}$, the tree–level term dominates and yields strong limits (e.g.\ $\lambda\!\lesssim\!2.5 \times 10^{-3}$ at $m_A=0.1~\mathrm{GeV}$). 
For $m_A\!\gtrsim\!1~\mathrm{GeV}$ the loop–induced SI contribution becomes dominant, leading to bounds of order unity for mediator masses of a few tens of GeV. 
As we will show in the results section, direct–detection constraints for pseudoscalar mediator remain much weaker than in scenarios where the nuclear scattering arises at tree level, is SI, and lacks momentum suppression.

\subsection{Spin-1 Mediator}
\label{sec:spin1}

We now turn to the case of a spin-1 mediator $Z'_{\mu}$. We consider two simplified models: one with a complex scalar DM field $\chi$, and one with a Dirac or Majorana fermion DM $\psi$. The interaction Lagrangians are
\begin{eqnarray}
\mathcal{L}_{\chi} &=& 
i\,g_{\chi}\,\bigl(\chi^* \partial_{\mu}\chi - \chi\,\partial_{\mu}\chi^*\bigr) Z'^{\mu} 
+ g_{\chi}^{2}\,|\chi|^{2}\,Z'_{\mu}Z'^{\mu} \nonumber\\
&&+\, g_{\chi}\,\bar{f}\,\gamma^{\mu}\bigl(V_{f}-A_{f}\gamma_{5}\bigr) f\,Z'_{\mu},
\label{eq:Lchi} \\[2mm]
\mathcal{L}_{\psi} &=&
g_{\psi}\,\xi\,\bar{\psi}\,\gamma^{\mu}\bigl(V_{\psi} - A_{\psi}\gamma_{5}\bigr) \psi\,Z'_{\mu}  \nonumber\\
&&+\, g_{\psi}\,\bar{f}\,\gamma^{\mu}\bigl(V_{f}-A_{f}\gamma_{5}\bigr) f\,Z'_{\mu},
\label{eq:Lpsi}
\end{eqnarray}
where $f$ denotes SM fermions, and $V$, $A$ are the vector and axial couplings. 
We take a universal general coupling $g_{\chi}$ or $g_{\psi}$ to both the DM and SM currents, with the SM flavor structure encoded in $V_f$, $A_f$.

\subsubsection{Scalar Dark Matter}

The DM annihilation channels $\chi\chi^* \to \bar f f$ and $\chi\chi^* \to Z'Z'$ behave as follows. 
The leading contribution to $\chi\chi^*\to \bar f f$ is $p$-wave:
\begin{align}
\langle\sigma v\rangle_{f f}
&\simeq \frac{g_{\chi}^{4}\, n_c^f\, m_{\chi}^{2}\, v^{2}}{8\pi}\,
\frac{\bigl(|V_{f}|^{2}+|A_{f}|^{2}\bigr)\left(1-\tfrac{m_f^2}{m_\chi^2}\right)^{1/2}}
{(4m_{\chi}^{2}-m_{Z'}^{2})^{2} + m_{Z'}^{2}\Gamma_{Z'}^{2}} \,,
\label{eq:annff}
\end{align}
In contrast, $\chi\chi^*\to Z'Z'$ proceeds in $s$-wave. 
Therefore, indirect-detection limits are generally weak for the $\bar f f$ channel.

For direct detection, and focusing first on purely vector couplings to quarks ($A_{f}=0$), the NR reduction yields a SI operator. 
The per-nucleon cross section is
\begin{eqnarray}
\label{eq:scalarSI}
\sigma_{\chi N}^{\rm SI}
&=& \frac{\mu_{\chi p}^{2}}{\pi}\,\frac{g_{\chi}^{4}}{m_{Z'}^{4}}
\;\frac{\bigl[Z\,f_{p}+(A-Z)\,f_{n}\bigr]^{2}}{A^{2}}, \\
f_{p}&=&2V_{u}+V_{d},\,
f_{n}=V_{u}+2V_{d}, \nonumber
\end{eqnarray}
with $\mu_{\chi p}$ the reduced mass. Since $f_{p}\neq f_{n}$ in general, the nuclear target composition must be accounted for when comparing to experimental limits. 
Without a significant isospin cancellation between $f_p$ and $f_n$, the \textsc{LZ} and \textsc{XENONnT} bounds~\cite{LZ:2022ufs,XENON:2023cxc} currently rule out most of the parameter space away from the resonance \cite{Arcadi:2024ukq}.

If instead only axial couplings to quarks are present ($V_{f}=0$), the leading operator in direct detection is velocity-suppressed and SD, corresponding to $O_{7}^{\rm NR}$ in the NR basis~\cite{Fitzpatrick:2012ix}:
\begin{equation}
\sigma^{\rm SD}_{\chi N}(v) \;=\; 
\frac{4\,\mu_{N}^{2}}{\pi} 
\left( \frac{g_\chi\, C_A^{(N)}}{m_{Z'}^{2}} \right)^{2} 
v^{2},
\end{equation}
where $C_A^{(p)} = A_{u}\,\Delta u^{(p)} + A_{d}\,\Delta d^{(p)} + A_{s}\,\Delta s^{(p)}$ (and similarly for $n$). 
The nucleon spin fractions $\Delta q^{(N)}$ are defined by
\begin{equation}
\langle N \,|\, \bar q \gamma^\mu \gamma^5 q \,|\, N \rangle 
= 2\, S^\mu \, \Delta q^{(N)} ,
\label{eq:DeltaS}
\end{equation}
with $S^\mu$ the nucleon spin four-vector.

\subsubsection{Fermionic Dark Matter}

Assuming vector couplings only ($A_{\psi,f}=0$), the annihilation cross section into SM fermions is $s$-wave:
\begin{align}
\langle\sigma v\rangle_{f f}
& \approx \frac{g_{\psi}^{4}\, V_{\psi}^{2} V_{f}^{2}\, n_{c}^{f}}{2\pi}\,
\frac{m_{\psi}^{2}\, \sqrt{1-\tfrac{m_f^2}{m_\psi^2}} \left( 1+ \tfrac{m_f^2}{2 m_\psi^2} \right)}
{(4m_{\psi}^{2}-m_{Z'}^{2})^{2} + m_{Z'}^{2}\Gamma_{Z'}^{2}} \,,
\label{eq:ffbar}
\end{align}
and the SI direct-detection cross section takes the same form as in Eq.~\eqref{eq:scalarSI} with $g_{\chi}\!\to\! g_{\psi}$.

For mixed axial–vector couplings, or for purely axial couplings, the annihilation into SM fermions acquires a-wave ($\propto v^2$) and helicity ($\propto m_f^2$) suppression
\begin{equation}
\langle\sigma v\rangle_{f f}
\approx \frac{g_{\psi}^{4}\, A_{\psi}^{2} A_{f}^{2}\, n_{c}^{f}}{4\pi}\,
\frac{ m_f^2 \,\sqrt{1-\tfrac{m_f^2}{m_\psi^2}}}
{(4m_{\psi}^{2}-m_{Z'}^{2})^{2} + m_{Z'}^{2}\Gamma_{Z'}^{2}} \,.
\label{eq:ffbaraxial}
\end{equation}
For purely axial couplings of both DM and quarks, the spin‐dependent DM–proton cross section is SD
\begin{equation}
\sigma_{\psi p}^{\rm SD}
= \frac{3\,\mu_{\psi p}^{2}}{\pi}\,\frac{g_{\psi}^{4}}{m_{Z'}^{4}}
\,\bigl|A_{\psi}^{Z'}\bigr|^{2}
\Bigl[A_{u}^{Z'}\,\Delta_{u}^{p}
+A_{d}^{Z'}\,(\Delta_{d}^{p}+\Delta_{s}^{p})\Bigr]^{2},
\label{eq:fermionSD}
\end{equation}
where the $\Delta_{q}^{N}$ are defined in Eq.~\eqref{eq:DeltaS}.

\medskip

In what follows we explore the $(m_{Z'},\,m_{\rm DM})$ plane, varying $g_{\chi}$ or $g_{\psi}$ and switching on/off $V_f$ and $A_f$ to delineate the phenomenologically distinct scenarios, with special attention to the resonant regime $m_{Z'}\simeq 2 m_{\rm DM}$.

\section{Relic density}
\label{sec:relicdensity}

The average DM density is measured by \textsc{Planck} with a $1\%$ uncertainty to be $\Omega_{\rm DM} h^2 = 0.120$~\cite{Planck:2018vyg}.
A first–principles prediction of the relic abundance requires solving the Boltzmann equation for the DM phase–space density $f_\chi(\vec{p})$. In an expanding Friedmann–Robertson–Lema\^itre–Walker Universe the equation reads~\cite{Kolb:1990vq,Edsjo:1997bg}
\begin{equation}
E\!\left(\partial_t - H\,\vec{p}\!\cdot\!\nabla_{\vec{p}}\right) f_{\chi}
= \mathcal{C}\!\left[f_{\chi}\right]\,,
\label{eq:RD1}
\end{equation}
where $E$ and $\vec{p}$ are the DM energy and momentum, $H$ is the Hubble rate, and $\mathcal{C}$ is the collision operator. In general, $\mathcal{C}$ contains the elastic–scattering term ($\mathcal{C}_{\rm el}$), which controls kinetic equilibrium, and the annihilation term ($\mathcal{C}_{\rm ann}$), which controls chemical equilibrium (see, e.g.,~\cite{Binder:2017rgn}).

When kinetic equilibrium is maintained the elastic scatterings with the SM bath are fast enough to continually (re)thermalize DM momenta, so that the SM and DM sector have the same temperature $T_\chi=T$ and the DM phase space distribution can be written in the Maxwell–Boltzmann form $f_\chi(p,t)\propto e^{-E/T}$. 
Instead, chemical equilibrium regulates number–changing reactions and keeps the DM abundance at its thermal value, $n_\chi\simeq n_{\chi,\rm eq}(T)$. 
Freeze–out occurs when the annihilation rate $\Gamma_{\rm ann}\equiv n_\chi\langle\sigma v\rangle$ drops below the Hubble rate $H$, while kinetic decoupling occurs when the elastic–scattering rate $\Gamma_{\rm el}$ falls below $H$.

In the canonical WIMP freeze–out treatment, several simplifying assumptions are made. The most important is that DM remains in kinetic equilibrium with the SM bath throughout chemical decoupling. Then the DM distribution tracks the thermal one, $f_\chi \propto f_{\chi,{\rm eq}}$, and Eq.~\eqref{eq:RD1} reduces to the Zel’dovich–Okun–Pikelner–Lee–Weinberg equation for the number density $n_\chi$,
\begin{equation}
\frac{{\rm d} n_{\chi}}{{\rm d} t}+3 H n_{\chi}
= -\left\langle\sigma \vMol\right\rangle_T \left(n_{\chi}^2-n_{\chi, \mathrm{eq}}^2\right),
\label{eq:RD2}
\end{equation}
with $n_{\chi}=g_\chi \!\int\! {\rm d}^3 p /(2 \pi)^3\, f_{\chi}(p)$ and the thermal average (for non–relativistic DM following Maxwell–Boltzmann statistics) given by~\cite{GONDOLO1991145}
\begin{equation}
\left\langle\sigma \vMol\right\rangle_T
=\int_{4 m_{\chi}^2}^{\infty} \!\!{\rm d} s\;
\frac{s \sqrt{s-4 m_{\chi}^2}\, K_1(\sqrt{s}/ T)\, \sigma \vMol}
{16 \, T \, m_{\chi}^4 \, K_2^2\!\left(m_{\chi}/ T\right)}\,,
\end{equation}
where $K_i$ are modified Bessel functions.

Typically, elastic scattering rates exceed annihilation rates by orders of magnitude, because DM can scatter off the abundant light SM species while the DM number density is already Boltzmann suppressed near freeze–out. Hence the kinetic–equilibrium assumption is often justified. However, this reasoning can fail near an $s$–channel resonance, $m_{\rm DM}\simeq m_{\rm med}/2$. In that case the annihilation rate is resonantly enhanced while elastic scattering is not; furthermore, in models where mediator–SM couplings scale with SM fermion masses, interactions with light species (which dominate the bath) are Yukawa–suppressed. Kinetic equilibrium may then break down during chemical decoupling, invalidating the reduction to Eq.~\eqref{eq:RD2}~\cite{Binder:2017rgn}. A full solution of Eq.~\eqref{eq:RD1}, including $\mathcal{C}_{\rm el}$, is required. Ref.~\cite{Binder:2017rgn} showed that, close to resonance, the relic abundance obtained from the full treatment can differ by up to an order of magnitude from the standard approximation.

\begin{figure*}[t]
\includegraphics[width=0.49\linewidth]{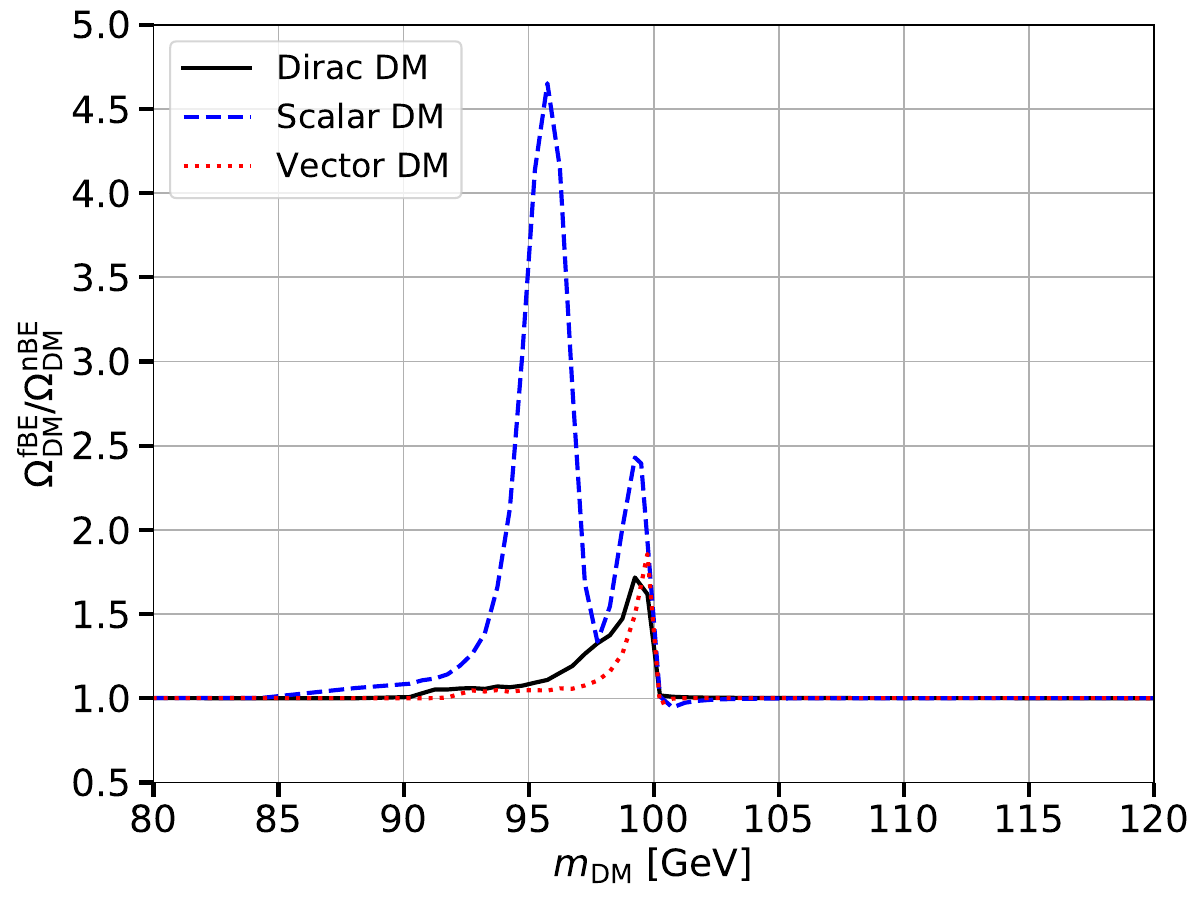}
\includegraphics[width=0.49\linewidth]{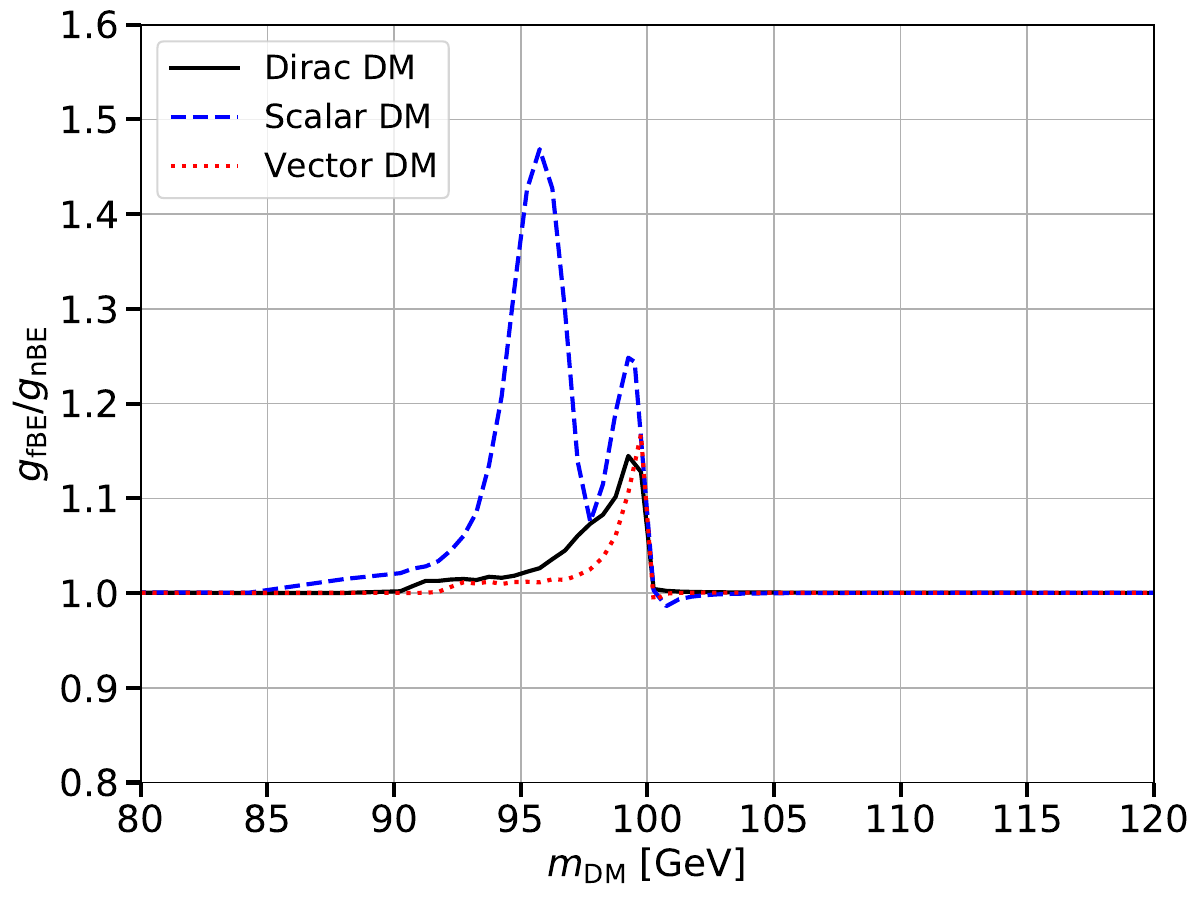}
\caption{Left: ratio $\Omega_{\rm DM}^{\tt fBE}/\Omega_{\rm DM}^{\tt nBE}$ obtained by solving the Boltzmann equation with \drake Eq.~\eqref{eq:RD1} under the {\tt nBE} (kinetic equilibrium assumed) and {\tt fBE} (full) prescriptions, for a scalar mediator and three DM spins. Right: ratio of the couplings $g=\lambda=\beta$ that reproduce $\Omega_{\rm DM} h^2$ for {\tt fBE} over {\tt nBE}. The mediator mass is fixed to $200~\mathrm{GeV}$.}
\label{fig:RD1}
\end{figure*}

In this work we compute the relic abundance $\Omega_{\rm DM} h^2$ in both setups:
\begin{itemize}
\item {\tt fBE}: full solution of Eq.~\eqref{eq:RD1};
\item {\tt nBE}: solution of Eq.~\eqref{eq:RD2} assuming kinetic equilibrium during freeze–out.
\end{itemize}
We use \drake~\cite{Binder:2021bmg}, which implements both prescriptions~\cite{Binder:2017rgn}. For comparison, we also compute $\Omega_{\rm DM} h^2$ with \micromegas~\cite{Belanger:2006is,Belanger:2013oya,Barducci:2016pcb,Belanger:2018ccd} under {\tt nBE}, finding agreement with \drake in the same setup.

Figure~\ref{fig:RD1} illustrates the impact of kinetic decoupling for a scalar mediator with $m_{\rm med}=200~\mathrm{GeV}$ and scalar, Dirac, or vector DM. 
In the left panel we show that the largest deviation occurs for scalar DM, where $\Omega_{\rm DM}^{\tt fBE}$ can exceed $\Omega_{\rm DM}^{\tt nBE}$ by up to a factor $\sim 4.5$; for Dirac and vector DM the enhancement peaks around a factor $\sim 2$. In all cases the effect is most pronounced on the \emph{left} side of the resonance, $m_{\rm DM}\lesssim m_{\rm med}/2$. 
In the right panel we show the ratio of the couplings $g=\lambda=\beta$ required to reproduce $\Omega_{\rm DM} h^2=0.120$ in the two treatments. Differences up to a factor $\sim 1.5$ appear for $m_{\rm DM}\in(90,100)~\mathrm{GeV}$. 
This demonstrates the importance of taking into account the solution of the full Boltzmann equation when the DM is close to the resonance region.


%
%

\section{Results}
\label{sec:results}

In this section we present results for several choices of DM and mediator spins. 
Each simplified model is specified by four parameters: the DM mass $m_{\rm DM}$, the mediator mass $m_{\rm med}$, the DM–mediator and the mediator–SM fermion couplings. For clarity, we fix two of these at a time and scan over the remaining ones.

Unless stated otherwise, we assume flavour-universal SM–fermion couplings and take the DM–mediator and mediator–fermion couplings to be equal:
\begin{itemize}
\item \textbf{Scalar mediator:} the DM–mediator coupling is $\lambda$ and the mediator–fermion coupling is $\beta$. We set $\lambda=\beta\equiv g_{\rm DM}$.
\item \textbf{Vector mediator:} we assume flavour-universal vector or axial couplings, $V_f$ or $A_f$, and set $g_\chi V_f=g_{\rm DM}$ (or $g_\chi A_f =g_{\rm DM}$) and similarly for the Dirac DM case.
\end{itemize}
With this choice, the models effectively depend on three parameters, $\{m_{\rm DM},\,m_{\rm med},\,g_{\rm DM}\}$. We explore this 3D space in two complementary ways: (i) fix $g_{\rm DM}$ and vary $(m_{\rm DM},m_{\rm med})$; (ii) fix $m_{\rm med}$ and scan over $(m_{\rm DM},g_{\rm DM})$.

Our analysis includes the following constraints:
\begin{itemize}
\item \textbf{Cosmology:} the relic abundance should match $\Omega_{\rm DM} h^2 \simeq 0.120$ as reported by \cite{Planck:2018vyg}.
\item \textbf{Direct detection:} nuclear cross sections should be consistent with the upper limits on SI and SD DM–nucleus scattering from LZ~\cite{LZ:2024zvo} and projected sensitivities from DARWIN, which reaches the neutrino floor,~\cite{DARWIN:2016hyl}.
\item \textbf{Indirect detection:} annihilation cross sections must comply with the upper limits on $\langle\sigma v\rangle$ from the combined $\gamma$-ray analysis of dSphs~\cite{McDaniel:2023bju}.
\item \textbf{GCE:} we report in Sec.~\ref{sec:GCE} the result of a fit of the {\tt DMSimp} parameters to the GCE flux data taken from Di~Mauro+21 and Cholis+22 \cite{DiMauro:2021raz,Cholis:2021rpp}.
\end{itemize}

In this work we do not perform a dedicated collider analysis. This choice is well justified for models with a dominant SI scattering rate: current direct–detection bounds are substantially stronger than collider limits for $m_\chi \gtrsim \text{few GeV}$ (see, e.g., \cite{ATLAS:2024kpy}).
For a \emph{pseudoscalar} mediator, tree–level scattering is SD and momentum suppressed ($\propto q^4$) and the SI rate arises only at loop level.
In this case collider searches can be competitive and often leading the model constraints for $m_\chi$ below the resonant region. Closer to the $s$–channel pole, however, the translation of collider null results into limits on $\langle\sigma v\rangle$ becomes weak, especially under the ``minimal width'' assumption with benchmark couplings $g_q=g_\chi \sim1$ used in Ref.~\cite{ATLAS:2024kpy}.
Because those limits are presented for fixed mediator choices and profiling assumptions that are not aligned with our parameter scan, they are not directly applicable here. In any case, for this model the indirect–detection constraints we consider are typically stronger near resonance.
For an \emph{axial–vector} mediator with Dirac DM, Ref.~\cite{ATLAS:2024kpy} reports collider upper limits that are stronger than SD direct detection bounds over wide mass ranges. Yet the published results are provided at specific mediator masses and in a format that precludes a faithful recast across our full parameter space, so we refrain from using them quantitatively in this analysis.

For each {\tt DMSimp} model we generate the associated {\sc UFO}~\cite{Degrande:2011ua} and {\sc CalcHEP}~\cite{Belyaev:2012qa} files with {\sc FeynRules}~\cite{Alloul:2013bka}, and pass them to \maddm~\cite{Backovic:2013dpa,Ambrogi:2018jqj,Arina:2021gfn} to compute the prompt $\gamma$-ray yields used in the GCE fits and to derive coupling limits from the combined dSph analysis in \textit{Fermi}-LAT data~\cite{McDaniel:2023bju}. 
We use \micromegas~\cite{Belanger:2006is,Belanger:2013oya,Belanger:2018ccd,Alguero:2023zol} to evaluate the relic density (when appropriate) and direct-detection signals. 
Close to the resonance, $m_{\rm DM}\simeq m_{\rm med}/2$, kinetic decoupling can occur during or even before chemical decoupling (Sec.~\ref{sec:relicdensity}). 
Therefore, in this regime we compute the relic abundance with the {\tt fBE} method implemented in \drake~\cite{Binder:2017rgn,Binder:2021bmg}.\footnote{We use \micromegas\ {\tt 6.0.4} and \maddm\ {\tt 3.2}.}

\subsection{Scalar mediator}
\label{sec:scalarmediator}

\subsubsection{Dirac dark matter: Scalar interactions}
\label{sec:scalardirac}

\begin{figure*}[t]
\includegraphics[width=0.49\linewidth]{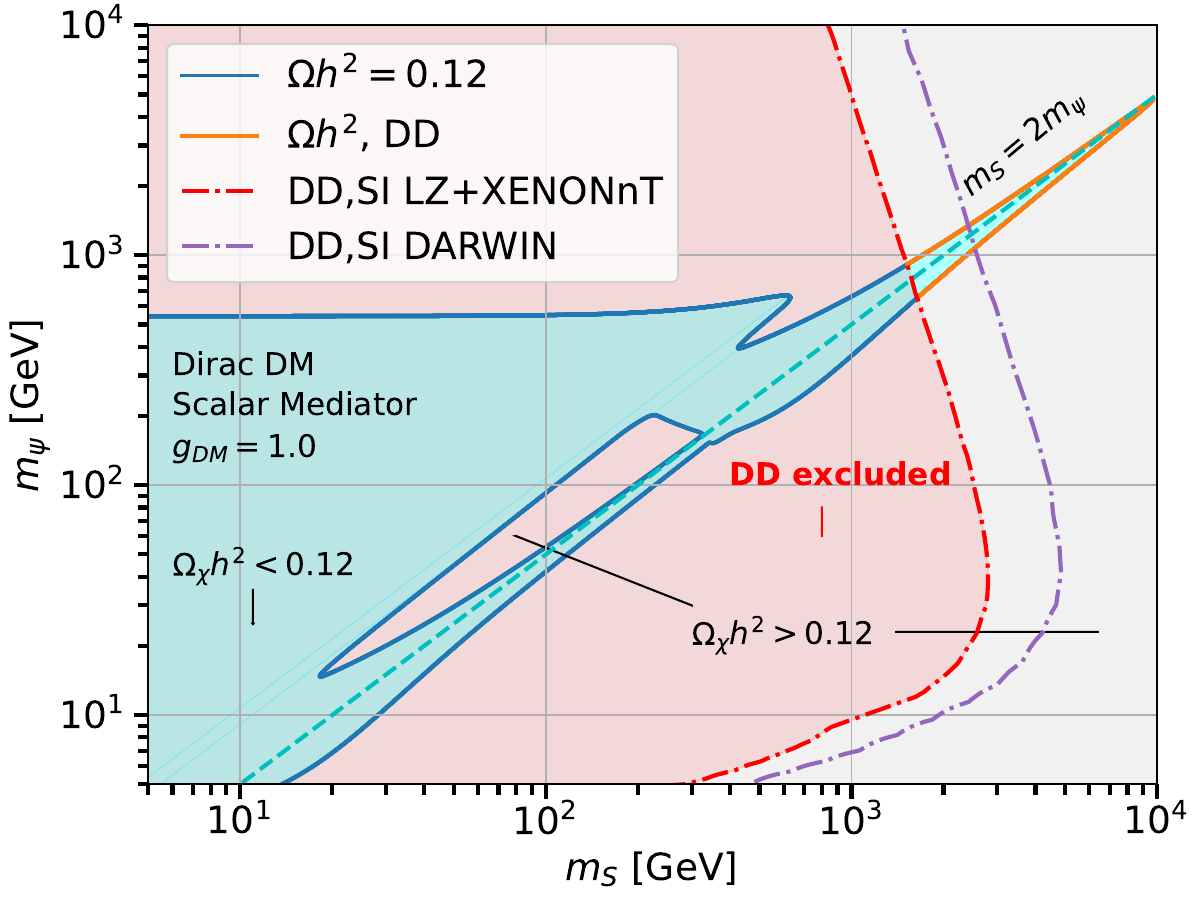}
\includegraphics[width=0.49\linewidth]{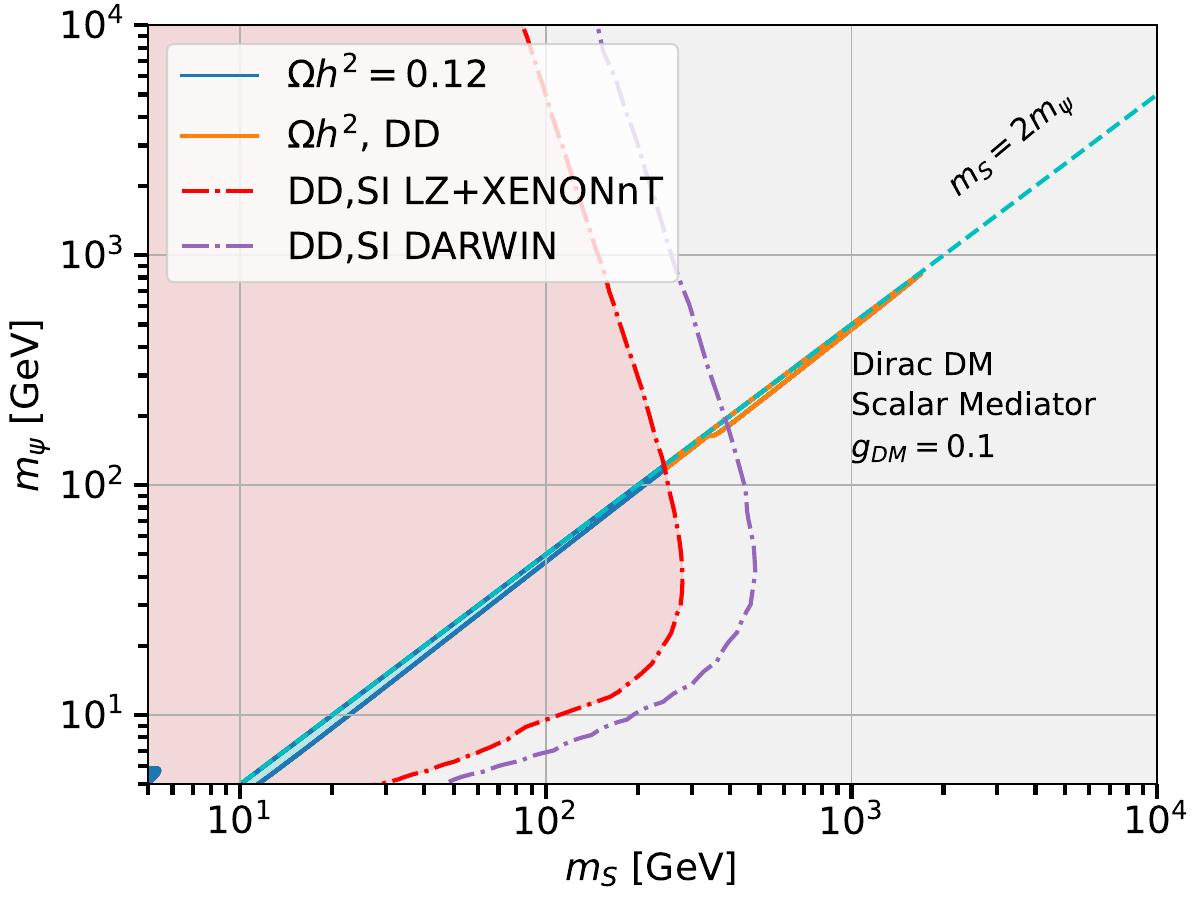}
\includegraphics[width=0.49\linewidth]{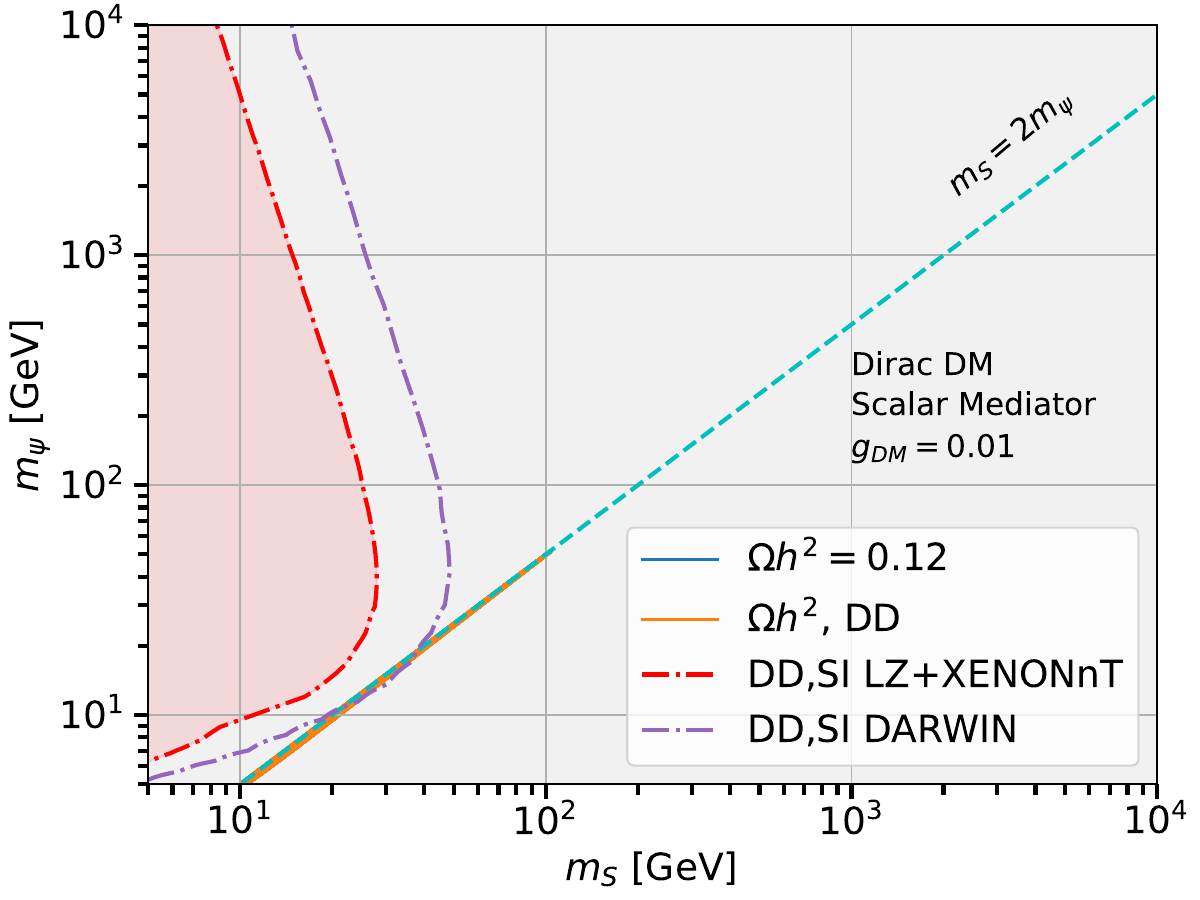}
\includegraphics[width=0.49\linewidth]{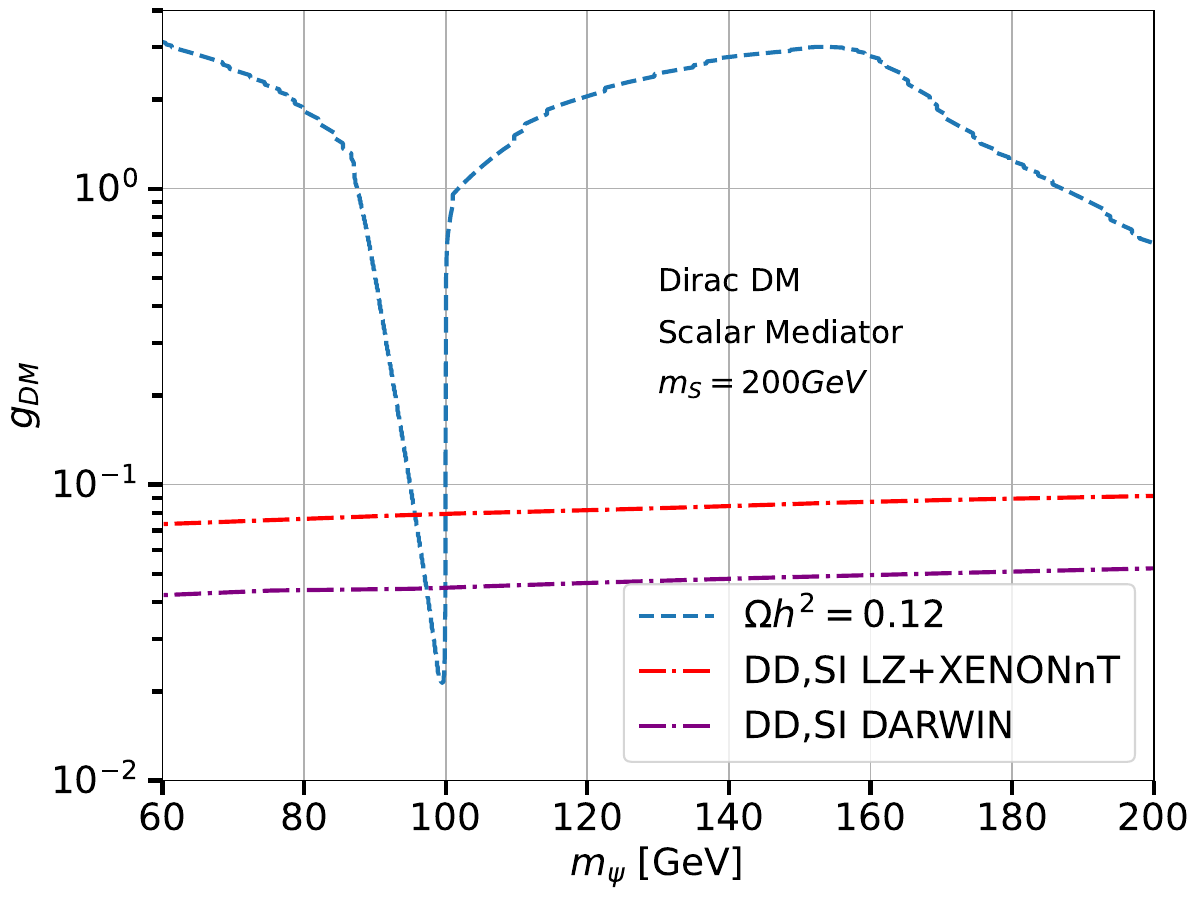}
\caption{Upper-left panel: Constraints in the $(m_{\psi},m_{S})$ plane for a simplified model with Dirac DM interacting via an s-channel scalar mediator.
In this case, $\lambda=\beta=g_{\rm{DM}}=1.0$.
The blue colored curve corresponds to parameter space that provides the correct DM relic density. The gray region denotes the region where DM is over-abundant $\Omega_{\rm{DM}}h^2>0.12$, while the cyan region is for under-abundant DM $\Omega_{\rm{DM}}h^2<0.12$. 
The red (purple) colored curves corresponds to the current (projected) exclusion limits from {\sc LZ} ({\sc DARWIN}) on $\sigma^{SI}_{\psi p}$, while we mark the exclusion region due to LZ constraints with a light red band. 
The orange curve represents the viable parameter space that satisfies relic density and LZ direct detection constraints. 
Upper-right and lower-left panels: Same as the upper-left panel but $g_{DM}=0.1$ and $g_{DM}=0.01$.
Lower-right panel: Constraints in the $(g_{DM},m_{\psi})$ plane. 
The mediator mass is fixed to 200 GeV, which implies a resonant DM mass around $m_{\psi}\sim 100$ GeV.
We show the direct detection upper limits based on LZ data \cite{LZ:2022ufs} (red dot-dashed line) and projections to DARWIN \cite{DARWIN:2016hyl} (purple dot-dashed line). The blue dot-dashed line corresponds to the correct DM relic density.}
\label{fig:SmedDDM}
\end{figure*}

\begin{figure*}[t]
\includegraphics[width=0.49\linewidth]{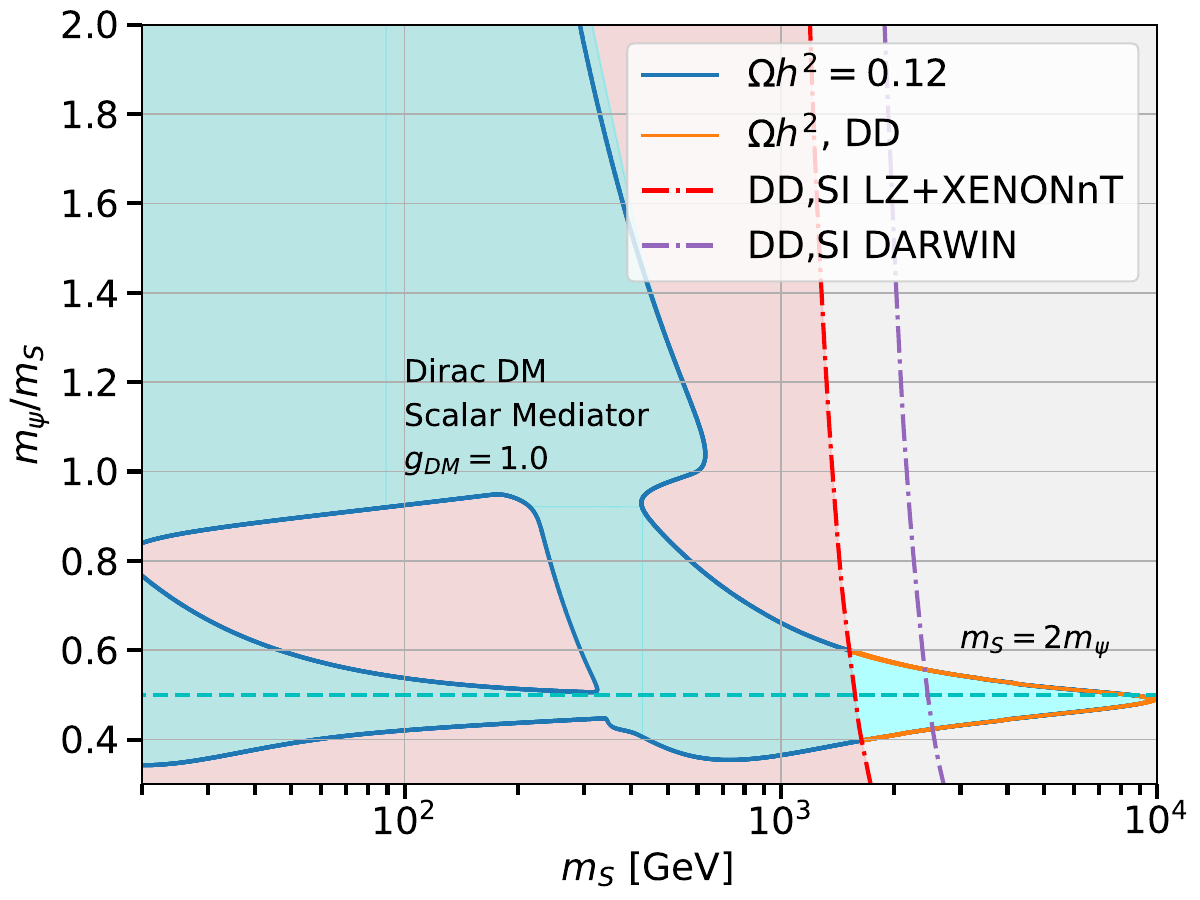}
\includegraphics[width=0.49\linewidth]{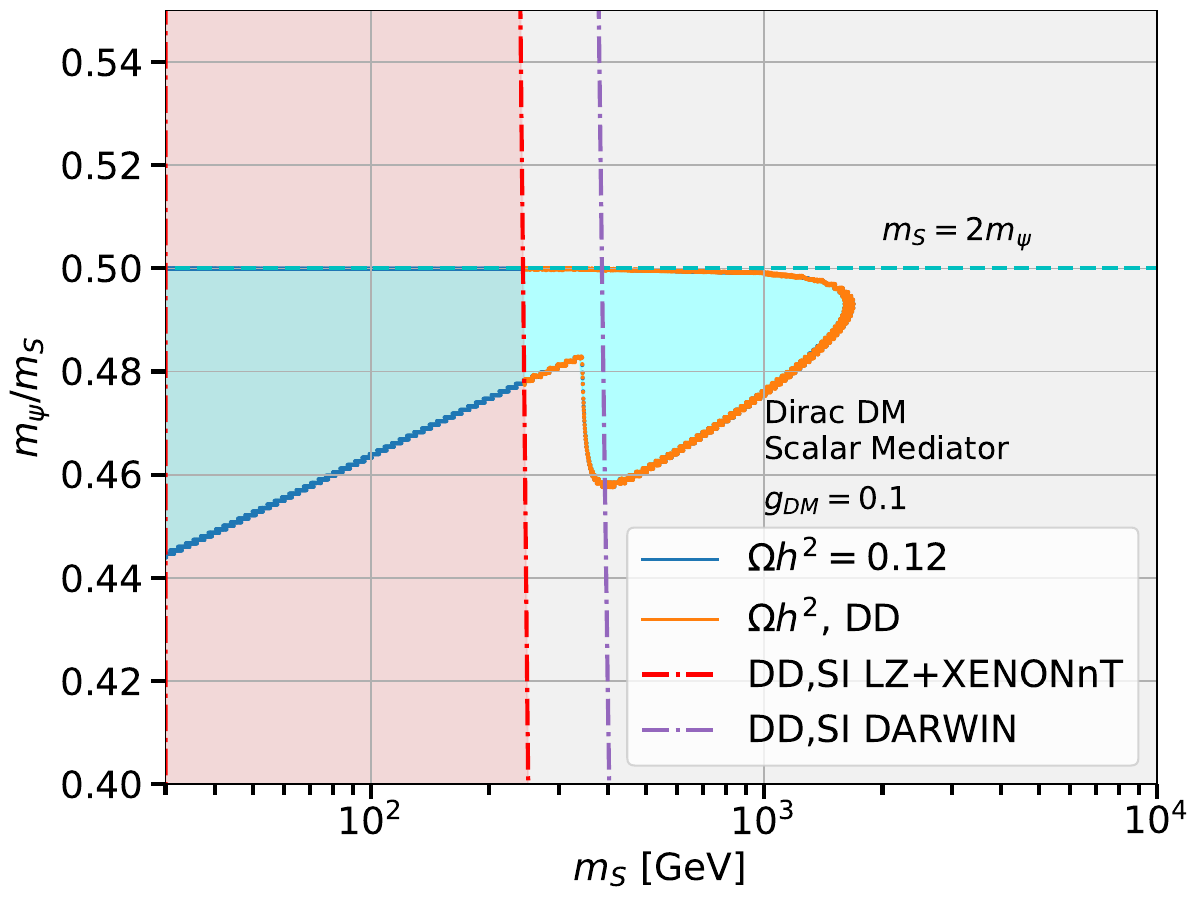}
\caption{The same as in Fig.~\ref{fig:SmedDDM} where we report the $y$ axis as $m_{\psi}/m_S$ in order to zoom in the resonance region.}
\label{fig:SmedDDMrescaling}
\end{figure*}

We begin with the scalar–mediator case, where the DM particle is a Dirac fermion and only CP-even interactions are present.  
The model depends on three parameters: the mediator mass $m_{S}$, the DM mass $m_{\psi}$, and a (universal) coupling $g_{\rm{DM}}$ (defined as $g_{\rm DM}\equiv \lambda=\beta$).

The top panels and the bottom-left panel of Fig.~\ref{fig:SmedDDM} show scans at fixed $g_{\rm{DM}}$, varying $(m_{S},m_{\psi})$.  
For each mass pair values we determine whether one can (i) reproduce the observed relic abundance and (ii) satisfy direct-detection bounds.  
Constraints from indirect detection are not applied because the annihilation is $p$-wave and thus strongly velocity-suppressed today; typical dSph velocities are $v/c\sim 10^{-6}$–$10^{-5}$.

For $g_{\rm DM}=1$, a broad region achieves the observed relic density across the tested mass range ($10$–$10^4$ GeV), consistent with the WIMP expectation that electroweak-sized couplings and weak-scale masses yield the correct freeze-out abundance.  
However, imposing direct-detection limits leaves only a small viable zone, namely $m_{\rm DM}\gtrsim 600~\mathrm{GeV}$ and $m_{\rm med}\gtrsim 1.4~\mathrm{TeV}$, and even there predominantly near the $s$-channel resonance $m_{\rm med}\simeq 2\,m_{\rm DM}$.

Reducing the coupling to $g_{\rm DM}=0.1$ ($0.01$) progressively confines the relic-density–compatible region to the resonance and to $m_{\rm DM}\lesssim 1~\mathrm{TeV}$ ($\lesssim 50~\mathrm{GeV}$).  
After applying direct-detection limits, only a narrow band around $m_{\rm med}\simeq 2\,m_{\rm DM}$ survives with roughly $m_{\rm med}\sim 10^2$–$10^3~\mathrm{GeV}$ (below $50~\mathrm{GeV}$) for $g_{\rm DM}=0.1$ ($0.01$).  
This “resonant funnel” is highlighted in Fig.~\ref{fig:SmedDDMrescaling}, where we plot the mass ratio $m_{\rm med}/m_{\rm DM}$ for $g_{\rm DM}=1$ and $g_{\rm DM}=0.1$ to zoom in on the resonance.
In particular, we note that for the case with $g_{\rm DM}=0.1$ the different between the $m_{\rm med}$ and $2 m_{\rm DM}$ should be less than 10\%.

The trend is further illustrated in the bottom-right panel of Fig.~\ref{fig:SmedDDM}, where we fix $m_{\rm med}=200~\mathrm{GeV}$ and display constraints in the $(g_{\rm DM},m_{\rm DM})$ plane: the allowed region lies very close to resonance and prefers $g_{\rm DM}\lesssim 0.1$.  
Quantitatively, compatibility with LZ requires a that the mediator and DM masses satisfy the following condition
\begin{equation}
   0\lesssim\frac{ (m_{\rm med}-2m_{\rm DM})}{m_{\rm med}}\;\lesssim\;0.08\,, 
\end{equation}
tightening to $\sim 0.05$ when considering DARWIN projected sensitivities.

\subsubsection{Dirac dark matter: Pseudoscalar interactions}
\label{sec:pseudoscalar}

\begin{figure*}[t]
\includegraphics[width=0.49\linewidth]{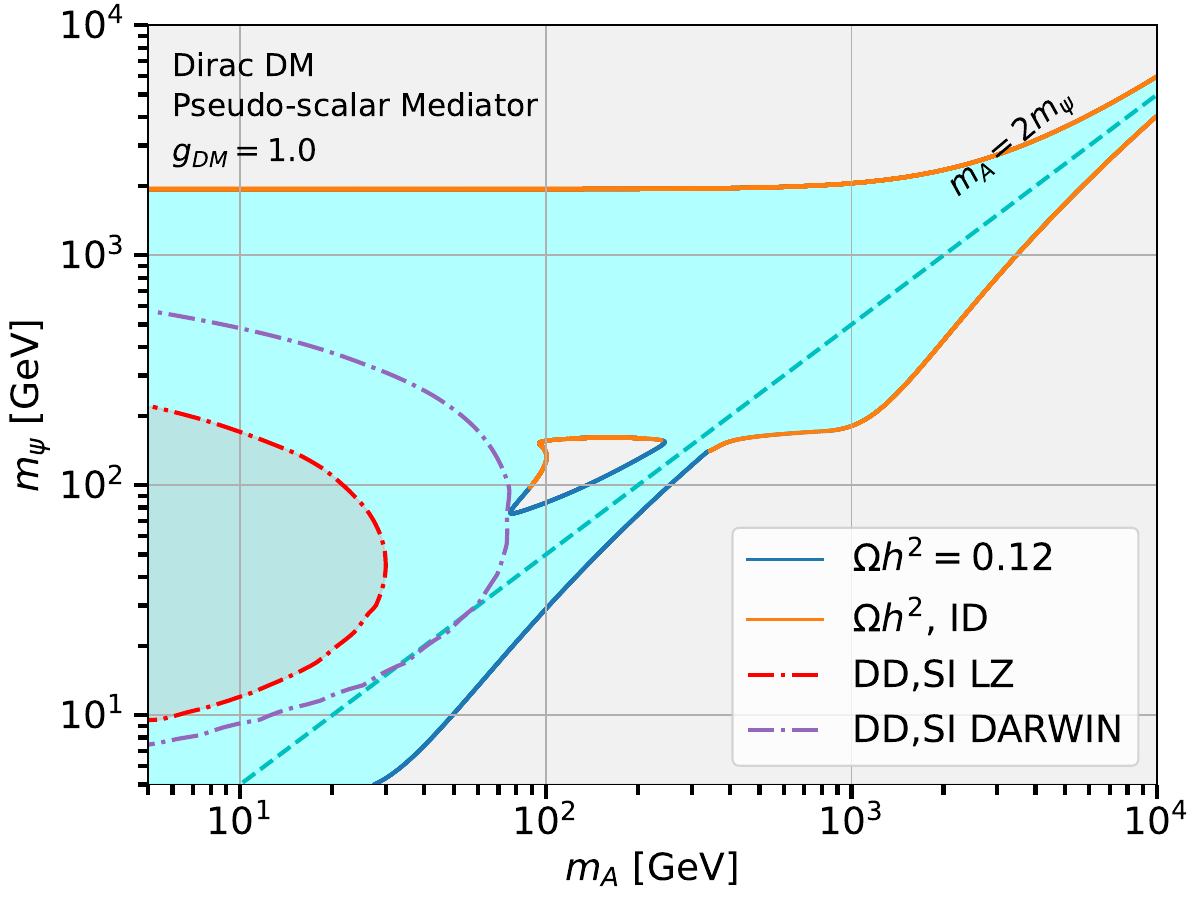}
\includegraphics[width=0.49\linewidth]{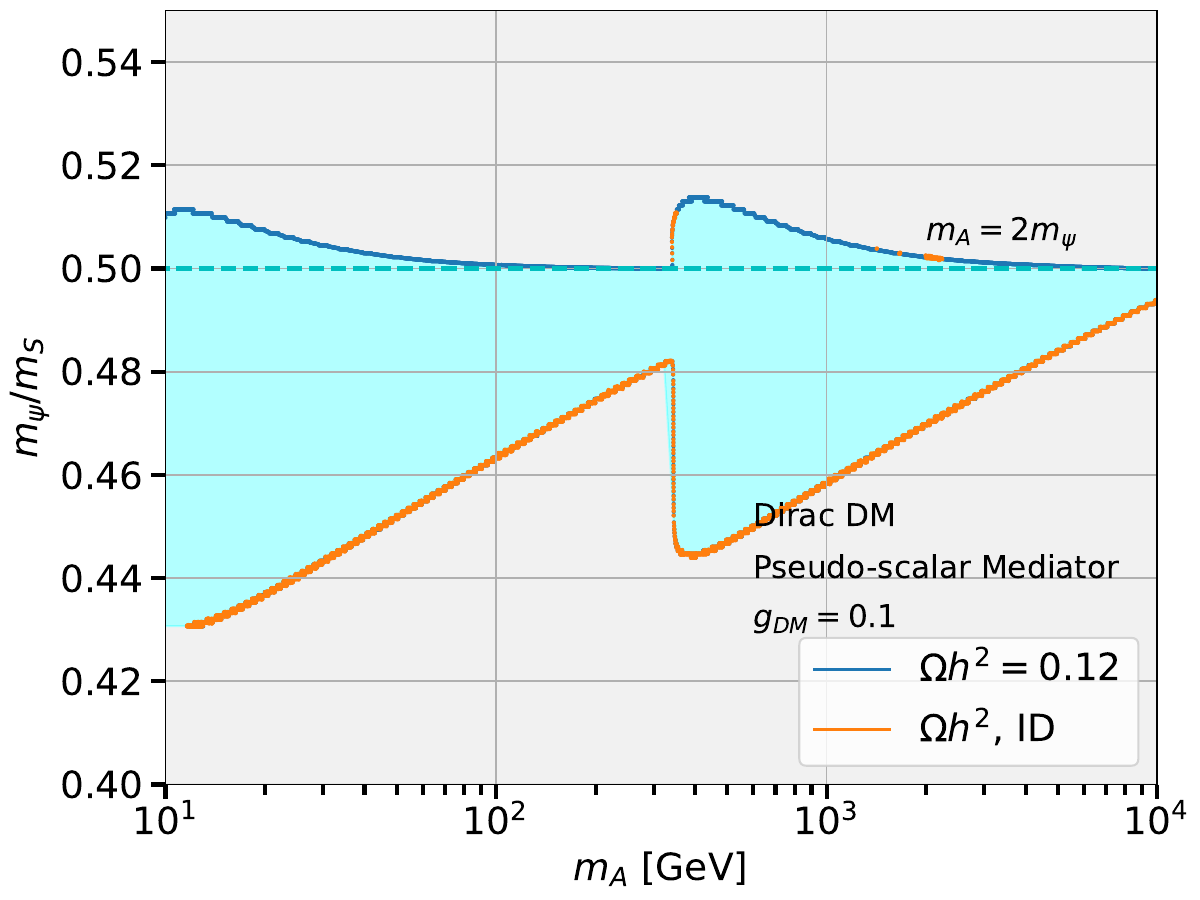}
\includegraphics[width=0.49\linewidth]{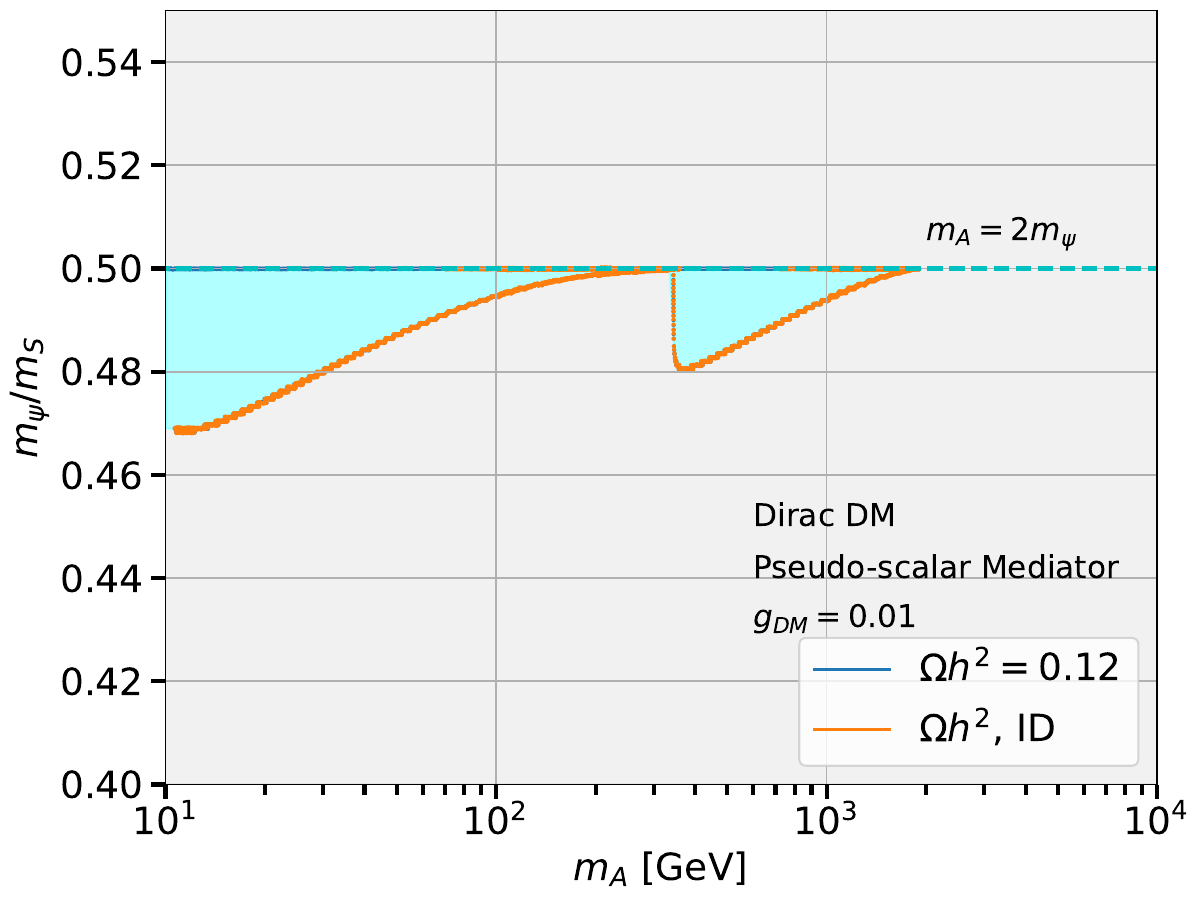}
\includegraphics[width=0.49\linewidth]{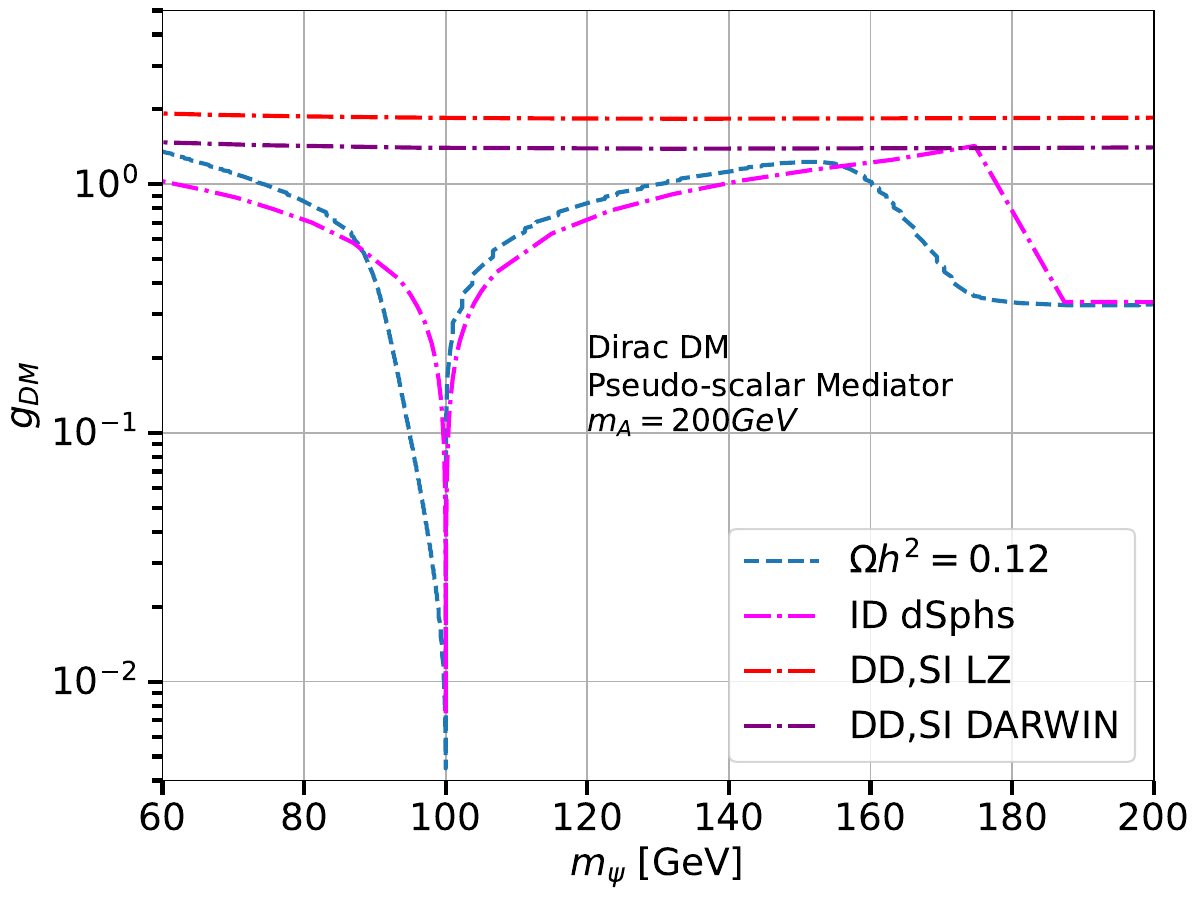}
\caption{Summary of constraints reported as a function of the parameters $(m_{\chi},m_{A})$ and $(g_{DM},m_{\chi})$ for a simplified model with Dirac DM interacting via an s-channel pseudoscalar mediator. 
This figure is the same as Fig.~\ref{fig:SmedDDM} with the addition of indirect-detection constraints from the observation of dSphs that is applied in all the figures. 
In particular, in the first tree figures the orange points represent the parameter space that satisfy relic density, direct and indirect detection while in the bottom right panel we show in the parameter space described by $(g_{DM},m_{\chi})$, the indirect detection upper limits with a pink dot-dashed line.}
\label{fig:PSmedDDM}
\end{figure*}

In this section we present results for Dirac DM coupled to a (CP-odd) pseudoscalar mediator $A$ (see Eq.~\ref{eq:pseudoDM}). 
The model is specified by the mediator mass $m_A$, the DM mass $m_\psi$, and a universal coupling $g_{\rm DM}=\lambda=\beta$.

For pseudoscalar interactions, the tree-level DM–nucleon amplitude is momentum suppressed, leading to a SD 
cross section that scales as $\propto q^{4}$, where $q$ is the momentum transfer. Since $q$ is tiny in WIMP elastic scattering, the rate is strongly suppressed; tree-level limits become relevant only for light mediators 
$m_A \!\lesssim\! \mathrm{GeV}$ as shown in Fig.~\ref{fig:pseudolimits}. As we do not consider that mass range, we omit tree-level direct-detection bounds in the following. 
At loop level a SI contribution arises; as illustrated in Fig.~\ref{fig:pseudolimits}, these loop-induced bounds 
become relevant mainly for $g_{\rm DM}\!\gtrsim\!1$ and for $m_A,m_\psi \!\lesssim\! 100~\mathrm{GeV}$.

In this model the annihilation $\bar\psi\psi\to f\bar f$ via a pseudoscalar is $s$-wave dominated, hence the limits coming from indirect detection are comparatively strong.

Figure~\ref{fig:PSmedDDM} summarizes the constraints. LZ and DARWIN upper limits have visible impact only for 
$g_{\rm DM}\!\gtrsim\!1$ and primarily probe $m_\psi$ below a few hundred GeV. A broad region of parameter space achieves the observed relic abundance; 
however, imposing indirect-detection limits excludes much of this region, leaving $m_{\rm DM}\!\gtrsim\!100~\mathrm{GeV}$ as the 
typical viable range. For smaller couplings ($g_{\rm DM}\!\lesssim\!0.1$), the relic-density–compatible set collapses toward the 
$s$-channel resonance, and compatibility with indirect detection selects preferentially the \emph{left} side of the resonant region, 
$m_\psi \lesssim m_A/2$.
This is due to the following reason.
Near the pole ($m_A\simeq 2m_\psi$) the annihilation rate follows a Breit--Wigner form,
\begin{equation}
\sigma v \;\propto\; 
\frac{g_{\rm DM}^2 g_f^2}{\bigl(4m_\psi^2 - m_A^2\bigr)^2 + m_A^2\Gamma_S^2}\,.
\label{eq:BW}
\end{equation}
At dwarf-spheroidal velocities ($v/c\sim10^{-6}-10^{-5}$), the change of the center of mass energy due the DM velocity $s=4m_\psi^2(1+v^2/4)$) is negligible, so indirect detection probes the intrinsic Breit–Wigner line shape, producing an approximately \emph{symmetric} feature around $m_\psi = m_A/2$ (pink curves) set by $\Gamma_A$.
In contrast, the relic-density condition involves thermal averaging at freeze-out,
\begin{equation}
\langle \sigma v\rangle(x) \;=\; 
\frac{x^{3/2}}{2\sqrt{\pi}}\!
\int_0^\infty\! dv\,v^2\,e^{-x v^2/4}\,
\sigma v,
\label{eq:thermalavg}
\end{equation}
with typical $v_{\rm rms}\sim\sqrt{6/x_f}\approx0.3\,c$. Therefore, the large DM velocity can increase sufficiently the center-of-mass energy so that, 
for $m_\psi < m_A/2$, a fraction of the velocity distribution still samples the resonance ($s$ moves upward). 
Instead, for $m_\psi > m_A/2$ the averaging cannot ``move back'' toward the pole and so the enhancement of the resonant cross section is suddenly lost in that mass region. This one-sided access, possibly reinforced by 
changes in $\Gamma_A$ when $A\!\to\!\bar\psi\psi$ opens, produces the \emph{asymmetric} relic-density curve 
(blue): the required $g_{\rm DM}$ drops more sharply for $m_\psi<m_A/2$ and recovers more slowly for $m_\psi>m_A/2$.

Compared to pure scalar interactions, the pseudoscalar scenario admits a substantially larger viable parameter space because 
direct-detection constraints are far weaker (tree-level $q^4$ suppression; loop-induced SI only). Consequently, mediator and DM masses need not reside at the TeV scale nor exactly on resonance, and $\mathcal{O}(1)$ couplings remain viable.

\subsubsection{Scalar and vector dark matter}
\label{sec:scalarvectorscalar}

\begin{figure*}[t]
\includegraphics[width=0.49\linewidth]{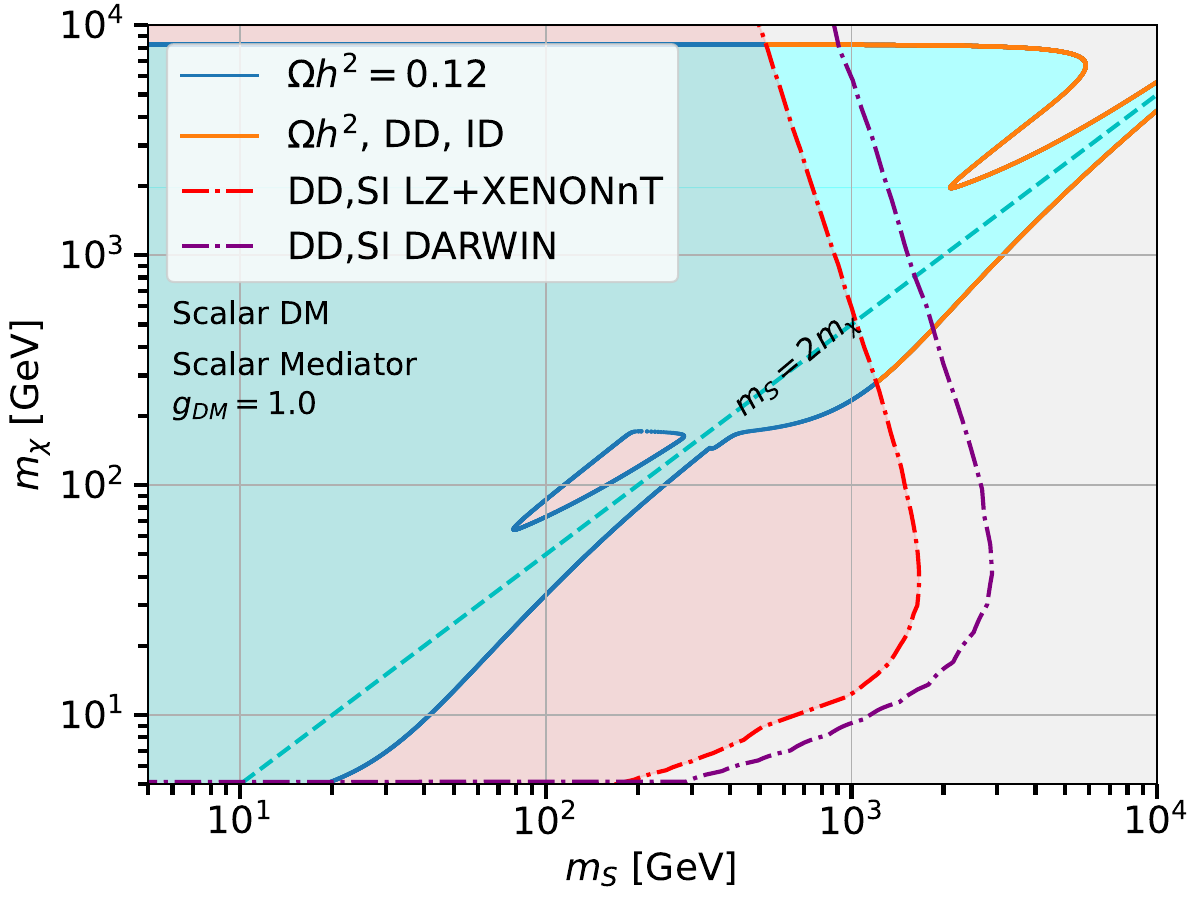}
\includegraphics[width=0.49\linewidth]{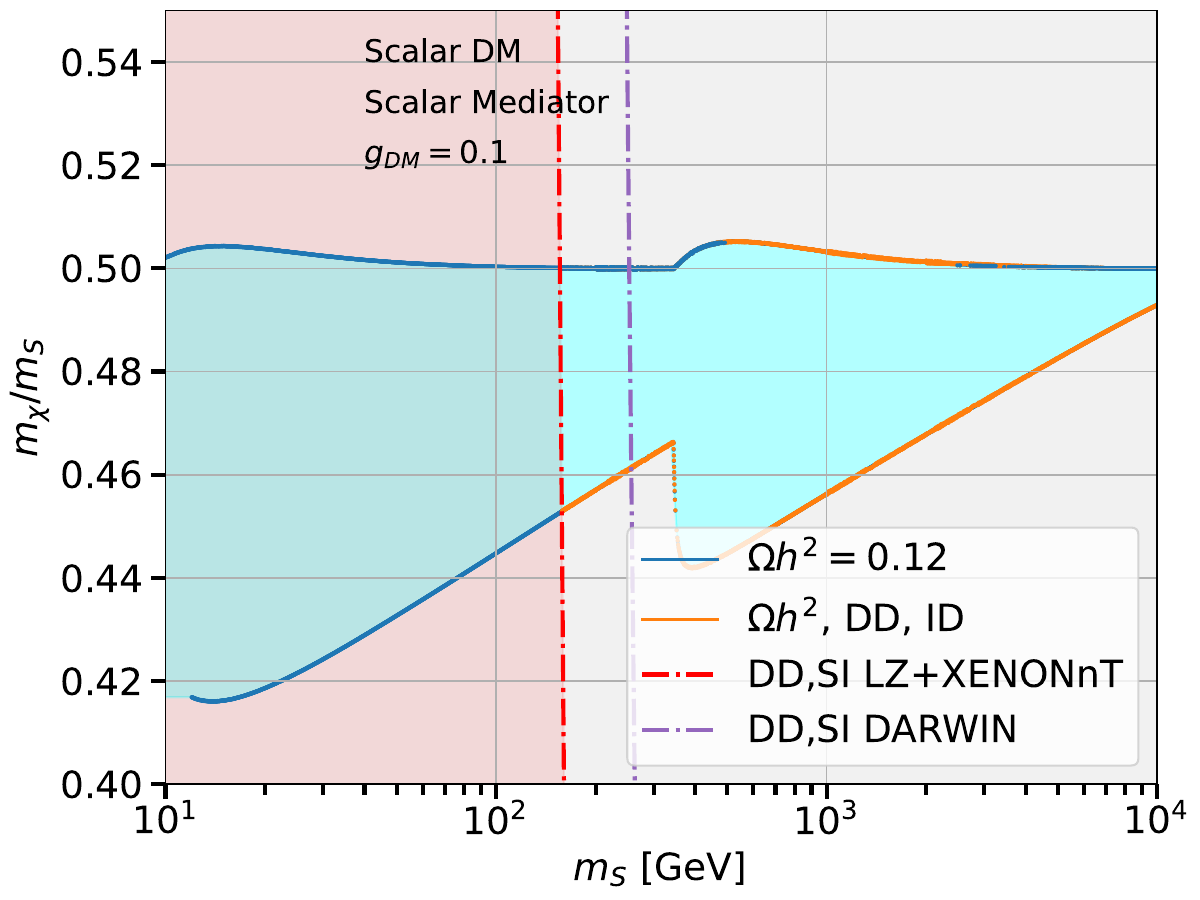}
\includegraphics[width=0.49\linewidth]{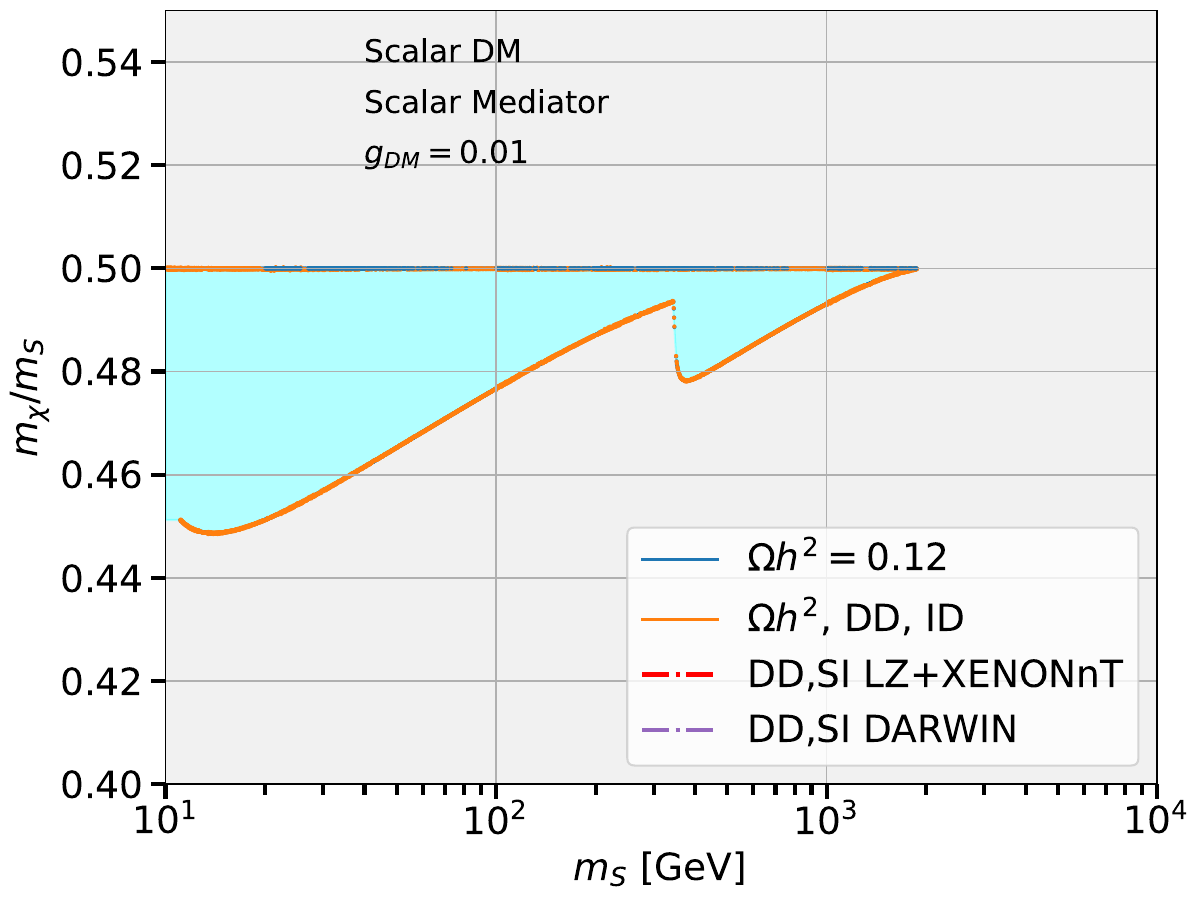}
\includegraphics[width=0.49\linewidth]{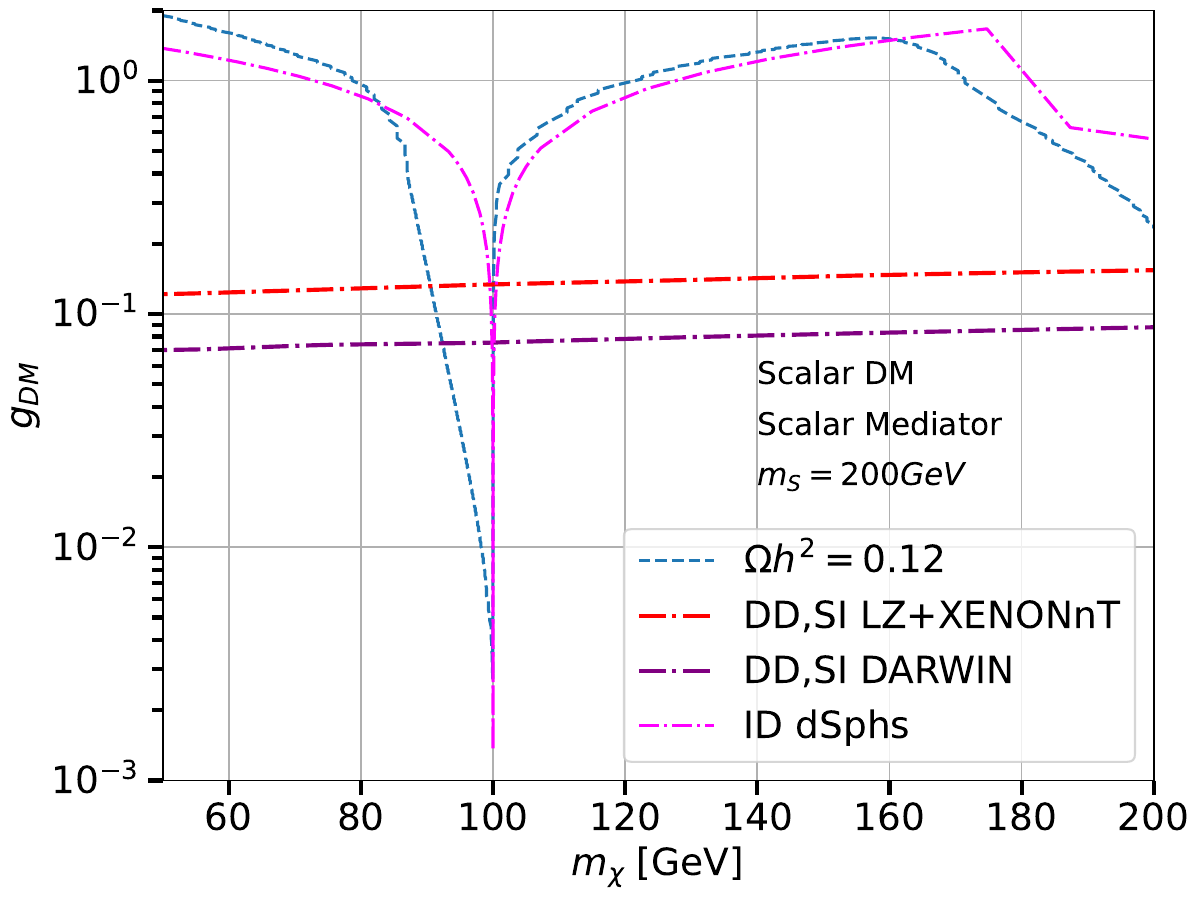}
\caption{Same as previous figures for a simplified model with real scalar DM and a CP-even scalar mediator.}
\label{fig:medDDMscalarscalr}
\end{figure*}

\begin{figure*}[t]
\includegraphics[width=0.49\linewidth]{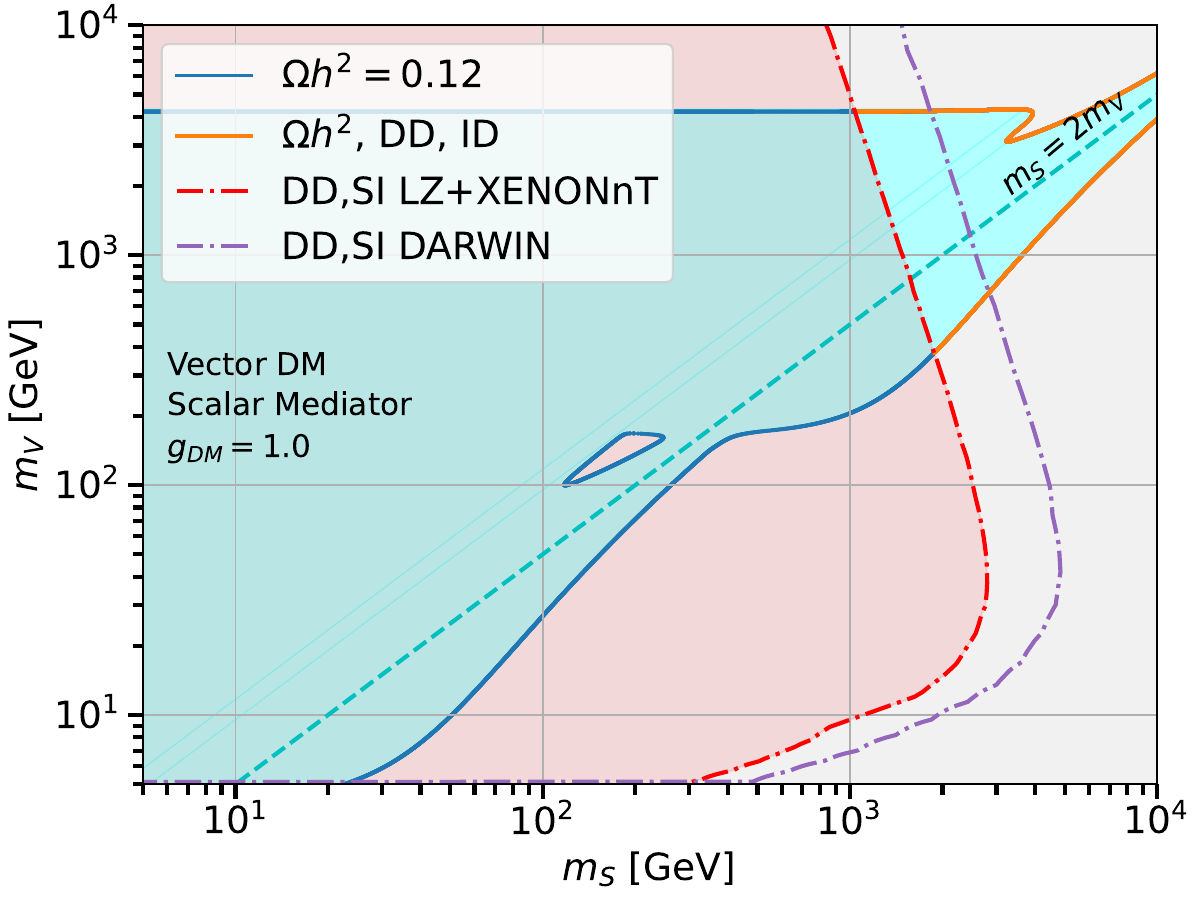}
\includegraphics[width=0.49\linewidth]{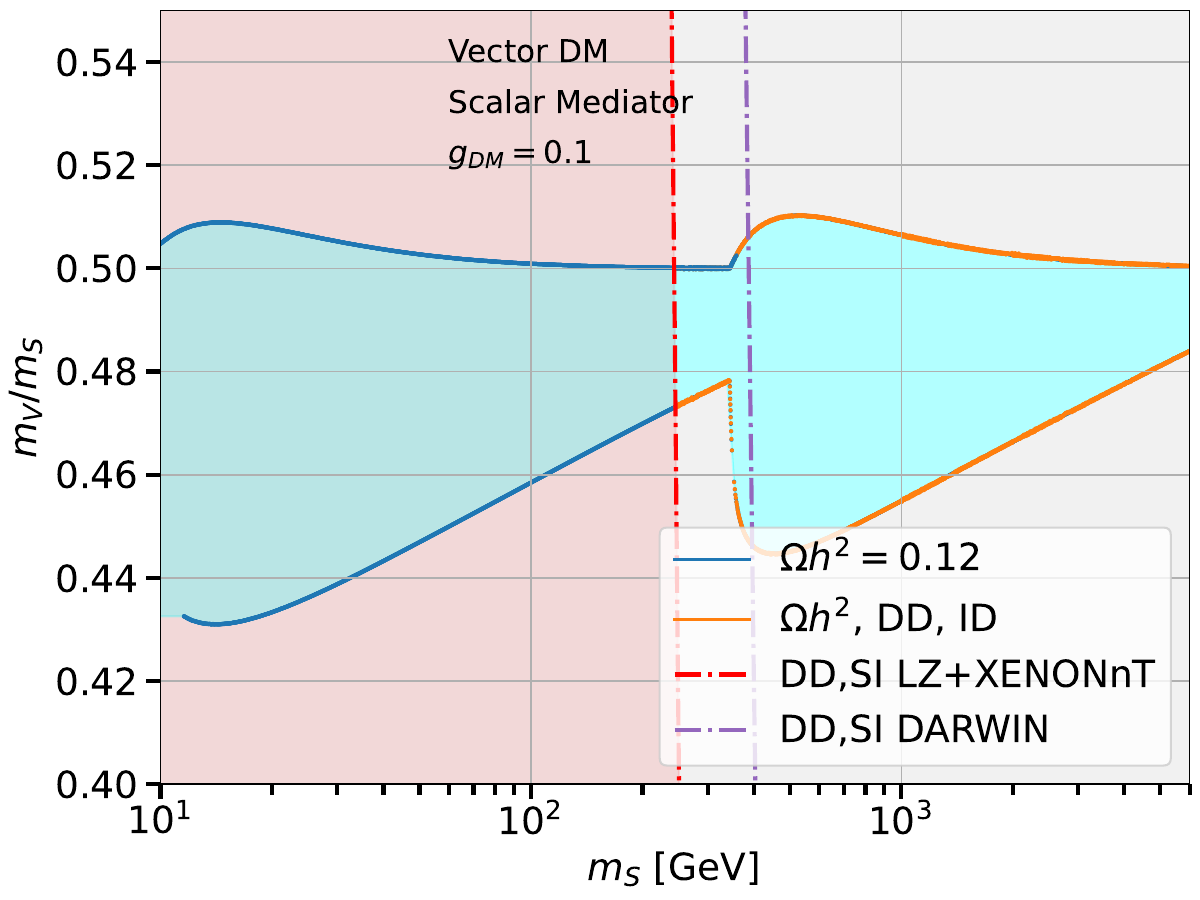}
\includegraphics[width=0.49\linewidth]{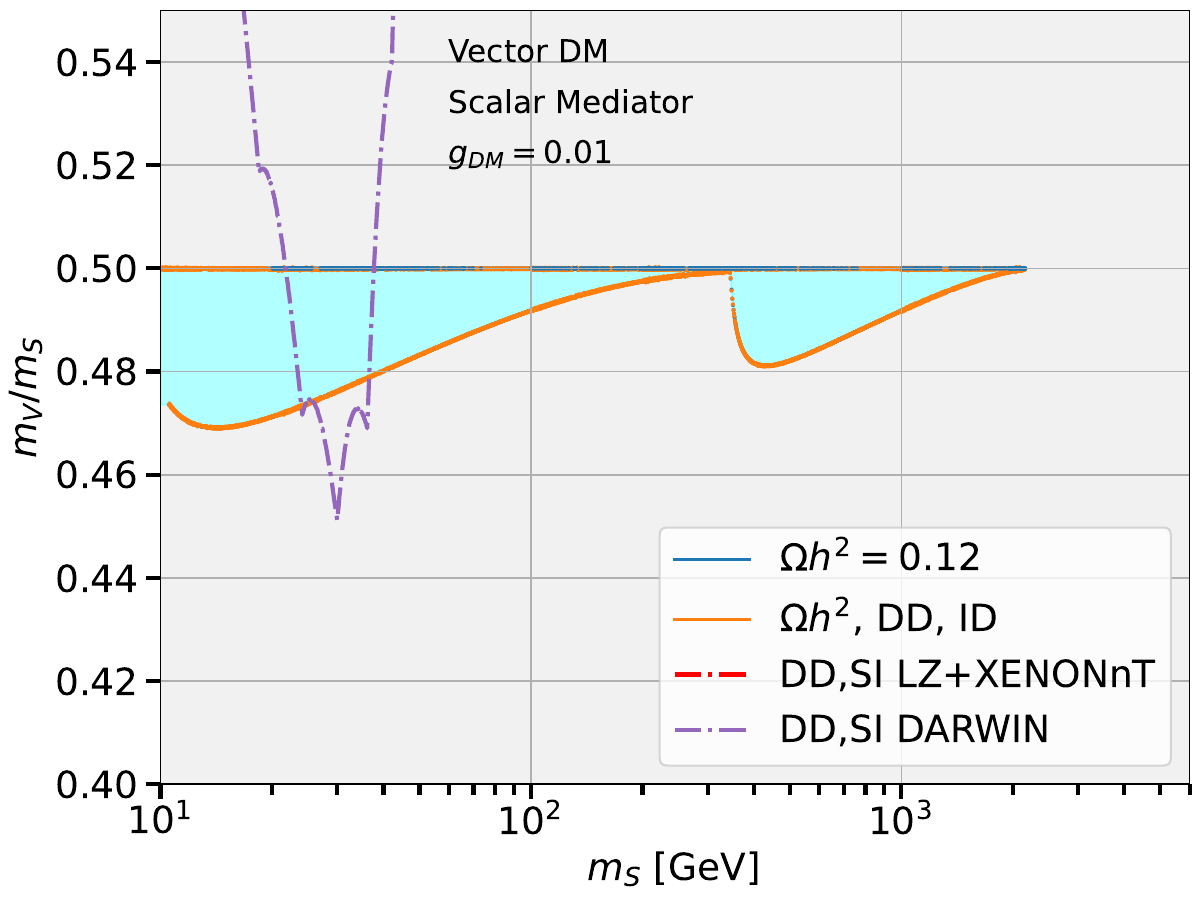}
\includegraphics[width=0.49\linewidth]{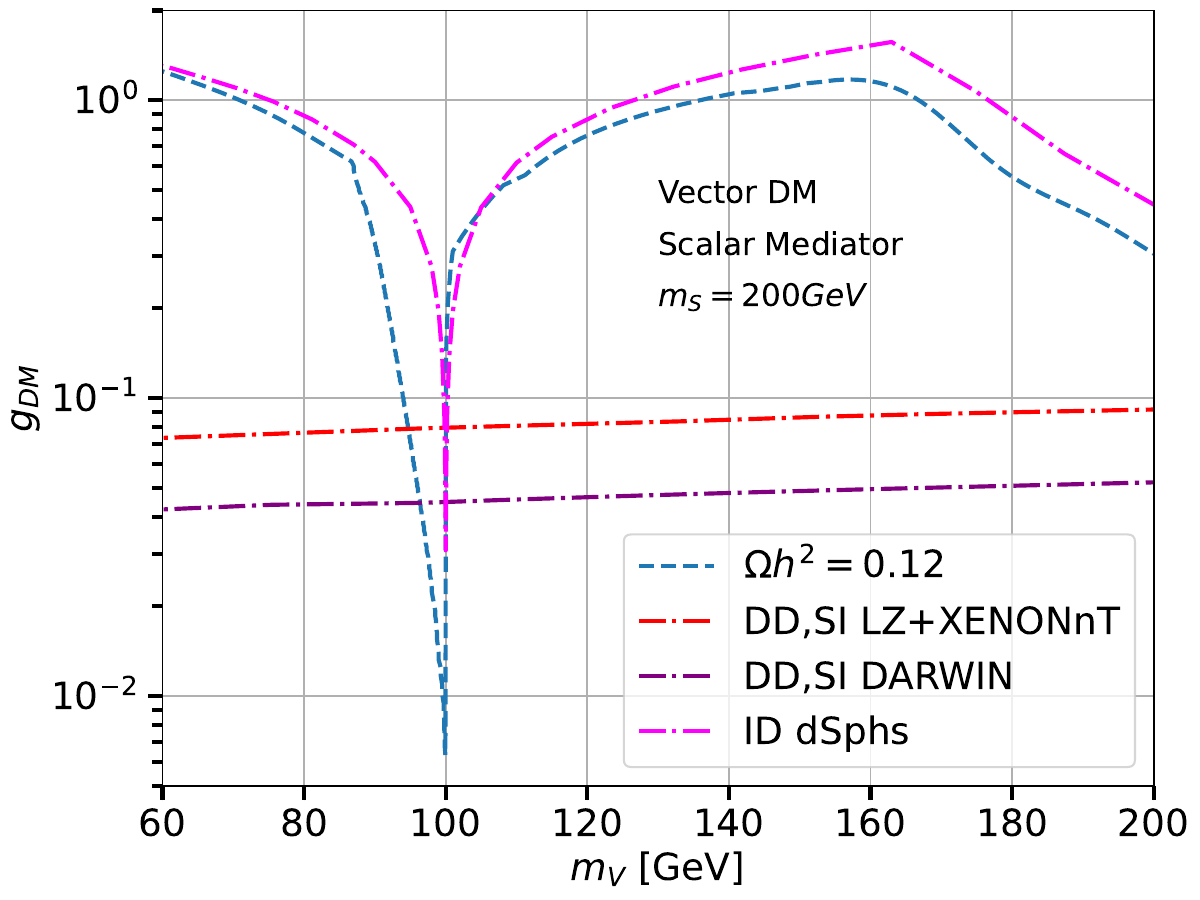}
\caption{Same as previous figures for a simplified model with vector DM and a CP-even scalar mediator.}
\label{fig:scalarmedVDM}
\end{figure*}

Figs.~\ref{fig:medDDMscalarscalr} and \ref{fig:scalarmedVDM} show the constraints for real \emph{scalar} and \emph{vector} DM particles interacting through an $s$-channel CP-even scalar mediator. 
The qualitative picture in both cases closely follows that of the scalar–mediator/Dirac–DM scenario (Sec.~\ref{sec:scalardirac}). 
For $\mathcal{O}(1)$ couplings a broad region of parameter space attains the observed relic abundance; however, current SI direct-detection limits push the viable DM masses to above a few hundred GeV. 
Lower masses are possible for smaller couplings, but only near the $s$-channel resonance: for $g_{\rm DM}=0.1$ ($0.01$) solutions exist for $m_{\rm DM}\!\gtrsim\!100~\mathrm{GeV}$ (down to a few GeV), provided $m_{\rm DM}\simeq m_S/2$.

In these models the annihilation into SM fermions is $s$-wave (velocity independent) but helicity suppressed by the Yukawa factor $\propto m_f^{2}$ (see Eq.~\ref{eq:annihscalar}). 
Consequently, indirect detection can exclude part of the parameter space but vero close to the resonance region and for small values of the couplings. Nevertheless, these bounds are weaker than present SI direct-detection limits, especially for the vector-DM case.

In the bottom-right panels of Figs.~\ref{fig:medDDMscalarscalr} and \ref{fig:scalarmedVDM} we show the case where we fix $m_S=200~\mathrm{GeV}$ and displays the constraints in the $(m_{\rm{DM}},g_{\rm DM})$ plane. 
Consistency with LZ requires $m_\chi$ and $m_V$ to lie extremely close to resonance.
In particular, $m_{\rm{DM}} \in [95,100]~\mathrm{GeV}$, while the DARWIN projection tightens this to $m_\chi \in [97,100]~\mathrm{GeV}$ corresponding to a fractional detuning \(|2m_V-m_S|/m_S\) of only a few percent. 
As reported before, indirect-detection limits from dwarf spheroidals (magenta) are subdominant except near the pole, where the enhanced annihilation rate partially overlaps the relic funnel.

\subsection{Vector mediator}

\subsubsection{Dirac dark matter}

We now consider Dirac DM interacting via a spin-1 mediator $Z'$. The relevant parameters are 
$(m_\psi,\,m_{Z'},\,g_{\rm DM})$, with flavor-universal SM couplings. 
In this setup the DM–nucleus scattering is SI, and the annihilation 
$\bar\psi\psi\!\to\! f\bar f$ is $s$-wave dominated. Consequently, direct-detection limits are typically very strong, 
while indirect-detection constraints are subdominant except near the $s$-channel resonance $m_{Z'}\simeq 2m_\psi$.

\begin{figure*}[t]
\includegraphics[width=0.49\linewidth]{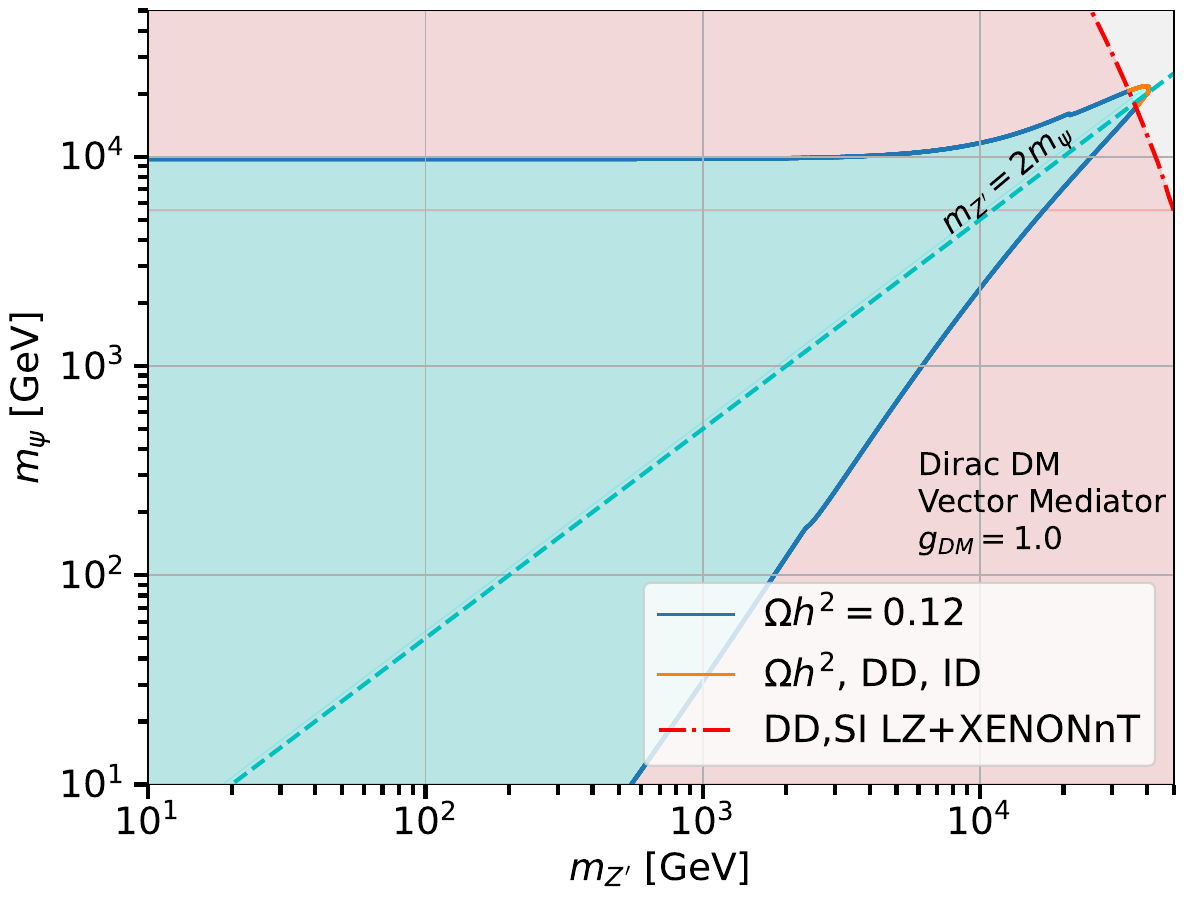}
\includegraphics[width=0.49\linewidth]{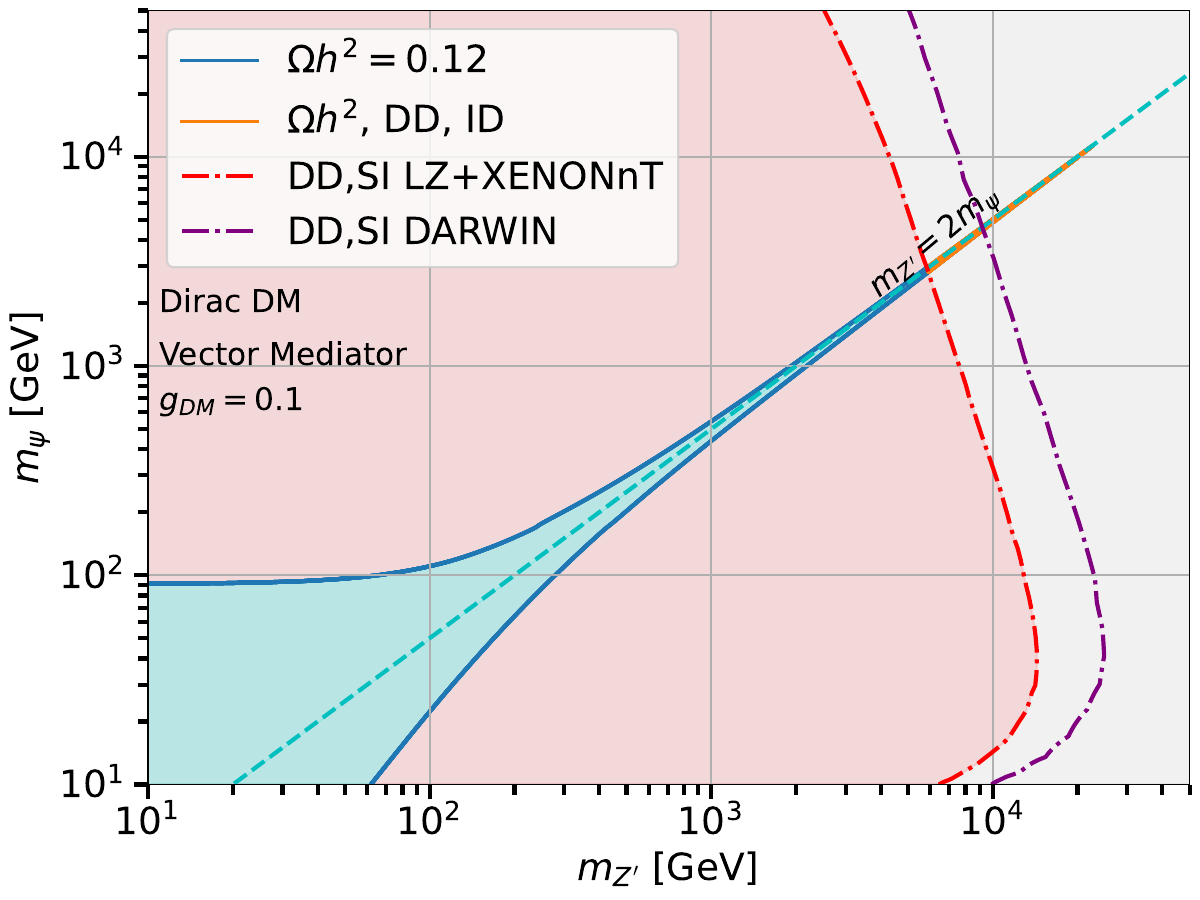}
\includegraphics[width=0.49\linewidth]
{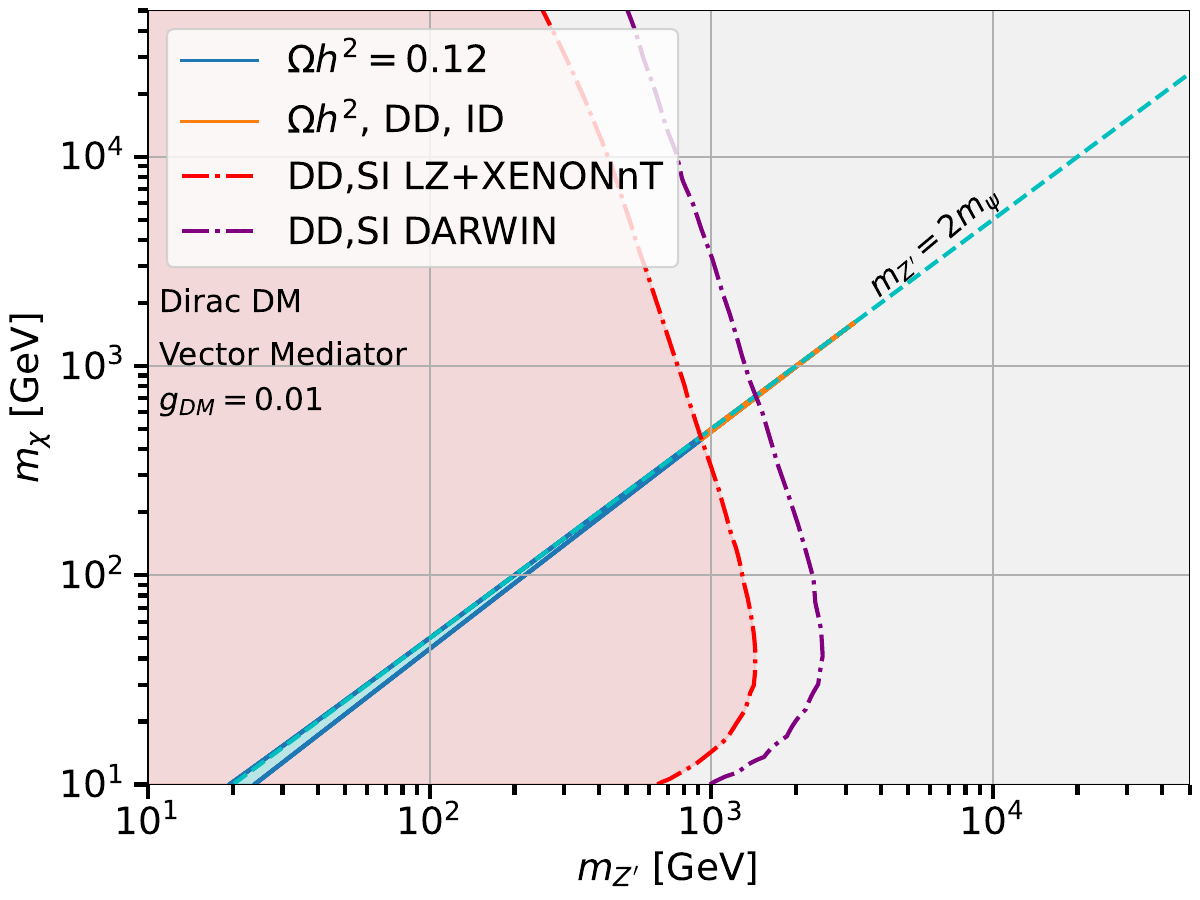}
\includegraphics[width=0.49\linewidth]{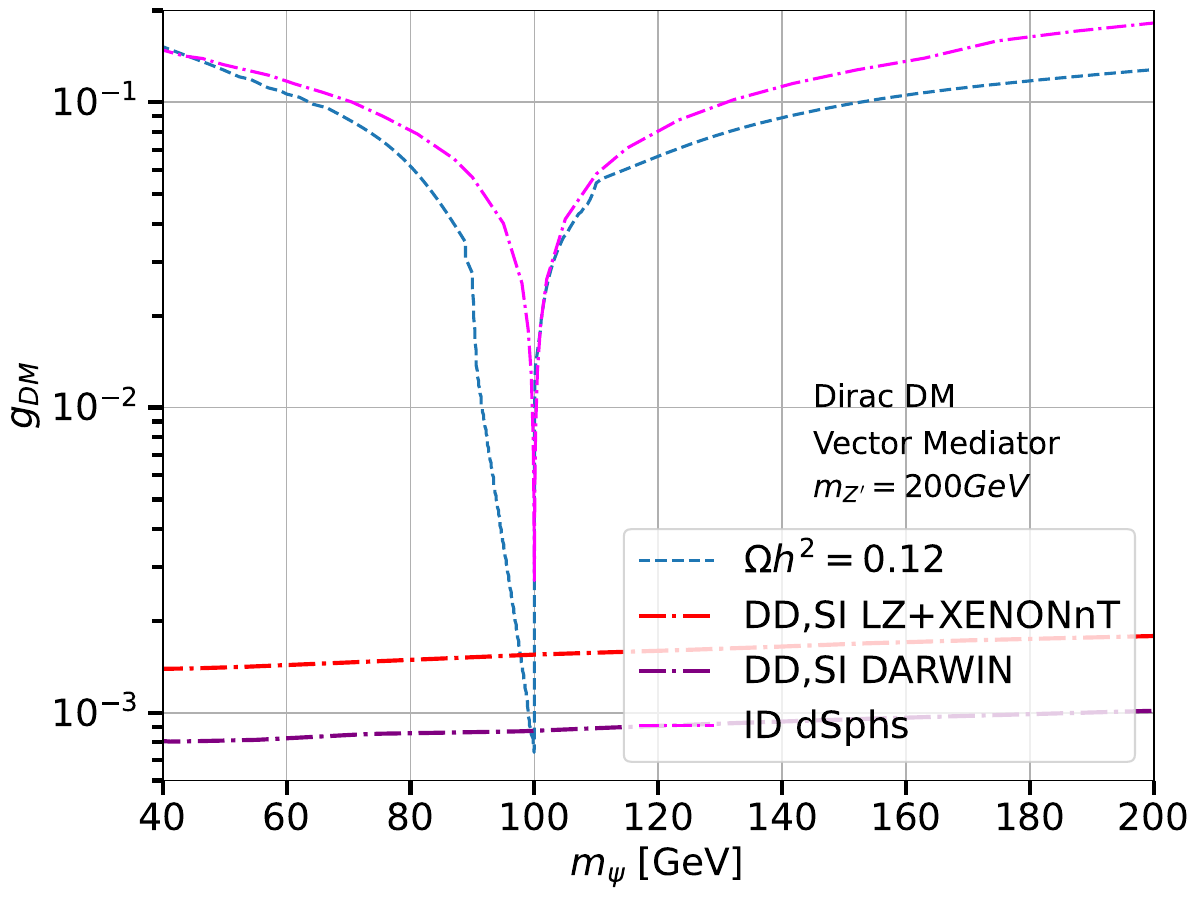}
\caption{The same as Fig.~\ref{fig:SmedDDM} but for vector mediator Dirac DM model.}
\label{fig:medDDMvectorvector}
\end{figure*}

Following the same procedure as in the previous sections, Fig.~\ref{fig:medDDMvectorvector} shows scans at fixed $g_{\rm DM}$ 
in the $(m_{Z'},m_\psi)$ plane (top-left, top-right, bottom-left) and, for fixed $m_{Z'}$, the combined constraints in the 
$(g_{\rm DM},m_\psi)$ plane (bottom-right). 
For $g_{\rm DM}=1.0$, the parameter space is almost entirely excluded by LZ SI bounds; 
projected DARWIN sensitivities would probe essentially the full mass range shown, up to multi-TeV mediator and DM masses. 
As $g_{\rm DM}$ decreases, the SI cross section drops and the relic-density–compatible region collapses toward the 
resonant funnel $m_{Z'}\simeq 2m_\psi$. The surviving points sit increasingly close to resonance as $g_{\rm DM}$ is reduced, 
as seen by comparing the $g_{\rm DM}=1.0$, $0.1$, and $0.01$ panels.

Constraints from dwarf-spheroidal $\gamma$ rays are weaker than SI limits over most of the plane but retain sensitivity in the 
vicinity of $m_{Z'}\simeq 2m_\psi$, where the annihilation rate is enhanced. This behavior is visible in the 
bottom-right panel: fixing $m_{Z'}=200~\mathrm{GeV}$, the SI curve follows an (almost) symmetric dip around 
$m_\psi\simeq m_{Z'}/2$, whereas the relic-density curve is skewed by thermal averaging at freeze-out (see the discussion in the pseudoscalar subsection \ref{sec:pseudoscalar}).

Compared to scalar or pseudoscalar mediators, the vector case is the most constrained by current SI direct detection. 
Viable regions remain predominantly in the resonant corridor with a required fine tuned between $2m_{\rm{DM}}$ and $m_{\rm{med}}$ that is at a few $\%$ level and for sufficiently small $g_{\rm DM}$. The future DARWIN sensitivities will further compress the remaining viable parameter space into a very small corner.
These findings are consistent with the trends seen for scalar mediators (Sec.~\ref{sec:scalarmediator}) and with previous studies (e.g., \cite{Arcadi:2024ukq,chala2015}).

\subsubsection{Dirac dark matter with an axial-vector mediator}

We now consider Dirac DM $\psi$ coupled axially to a spin-1 mediator $Z'$. The parameter set is
$\{m_\psi,\,m_{Z'},\,g_{\rm DM}\}$, with flavor-universal axial couplings to SM fermions. In this model
DM–nucleus scattering is predominantly SD, while annihilation to SM fermions proceeds through an
$s$-wave, so we include relic density, SD direct-detection, and a perturbative-unitarity envelope.

\begin{figure*}[t]
\includegraphics[width=0.49\linewidth]{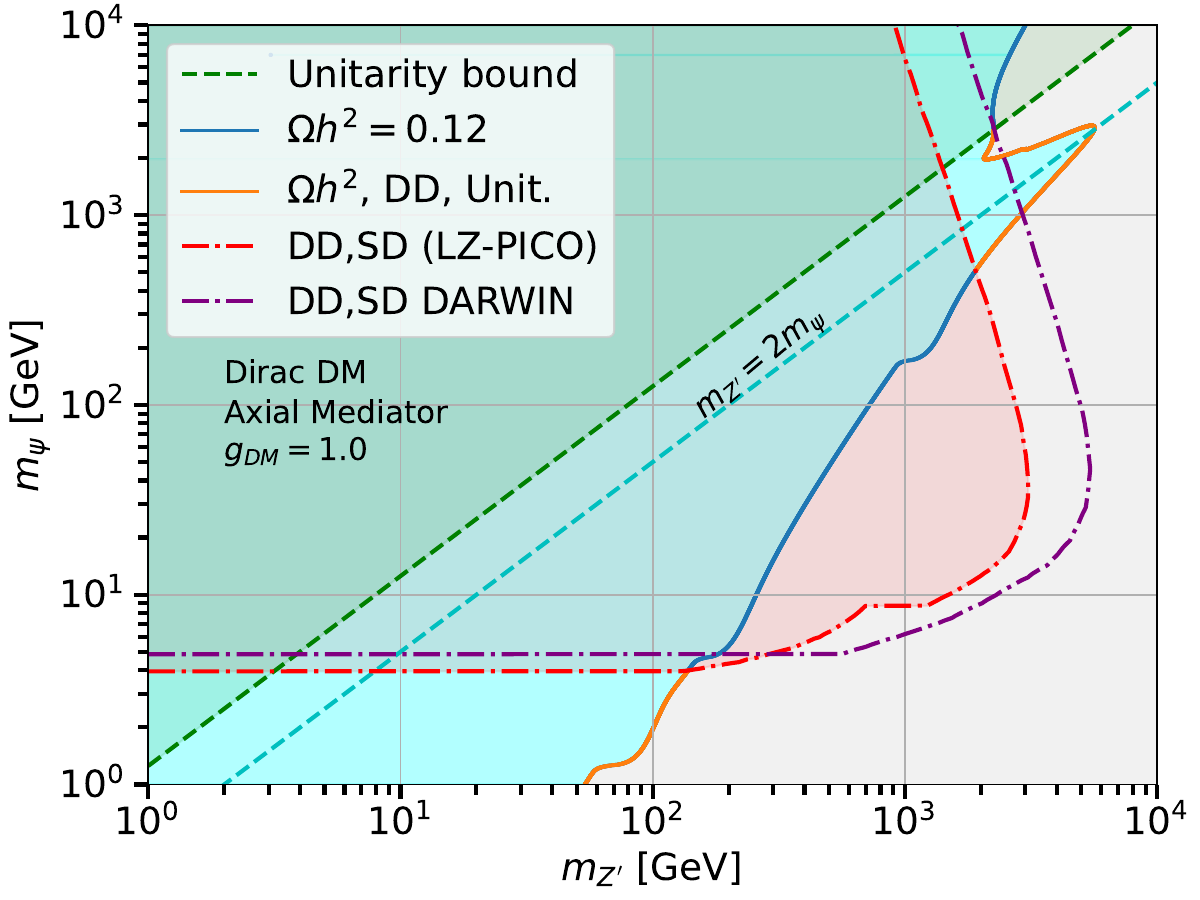}
\includegraphics[width=0.49\linewidth]{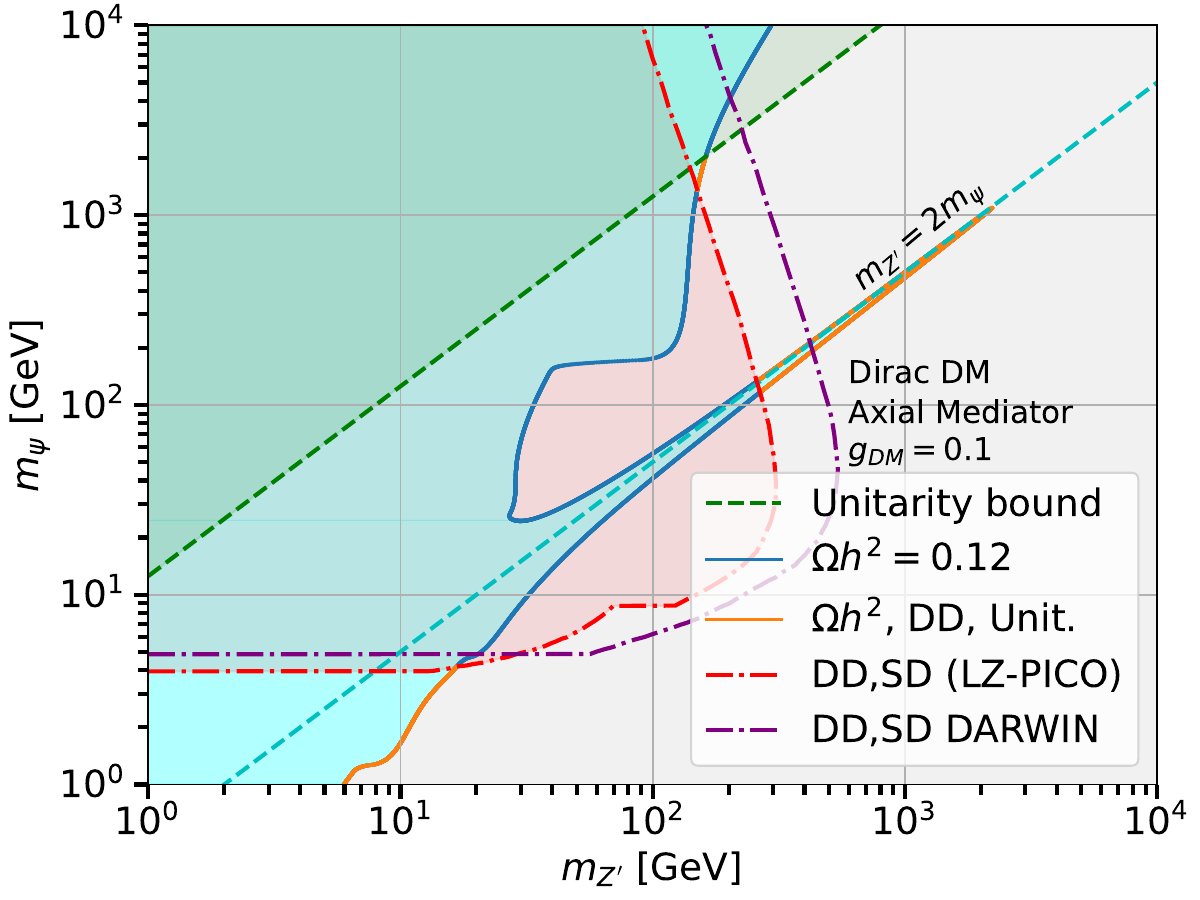}
\includegraphics[width=0.49\linewidth]
{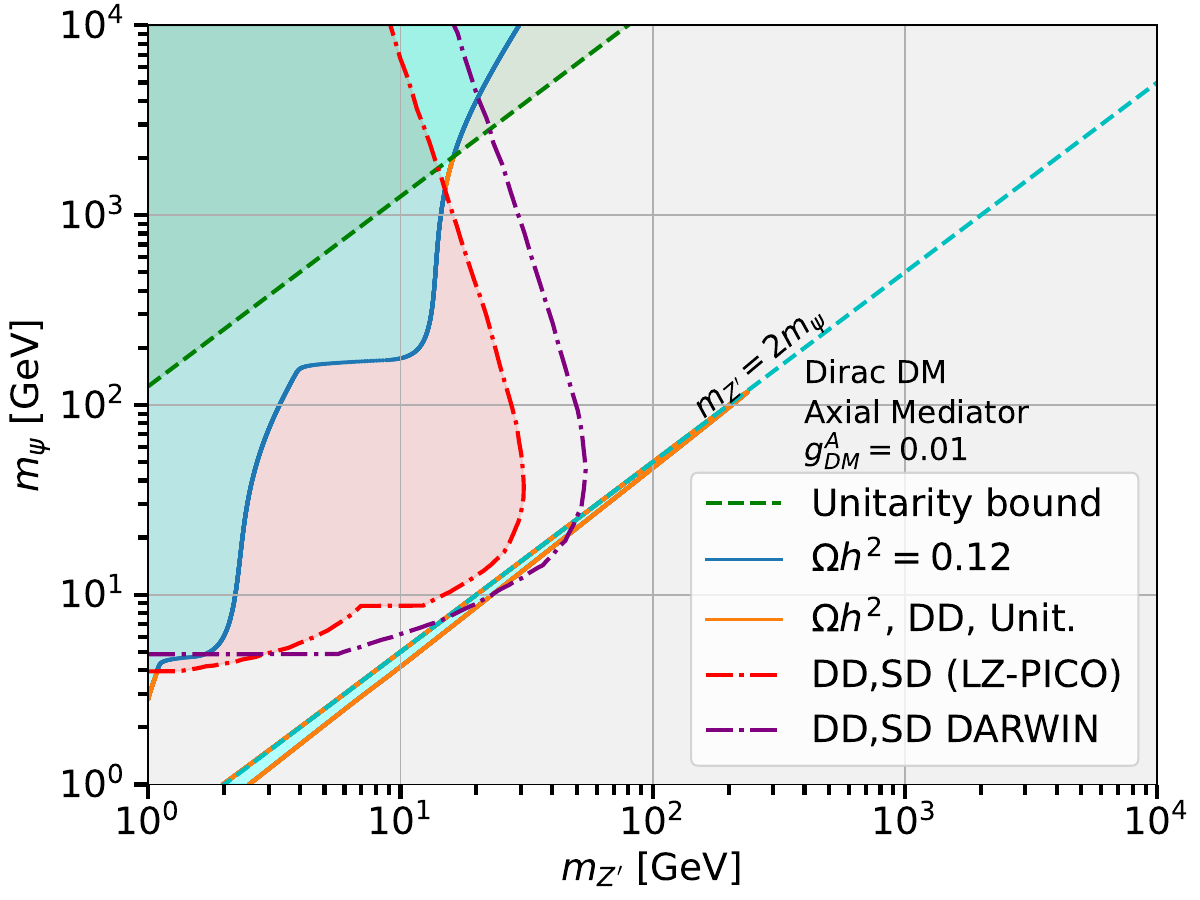}
\includegraphics[width=0.49\linewidth]{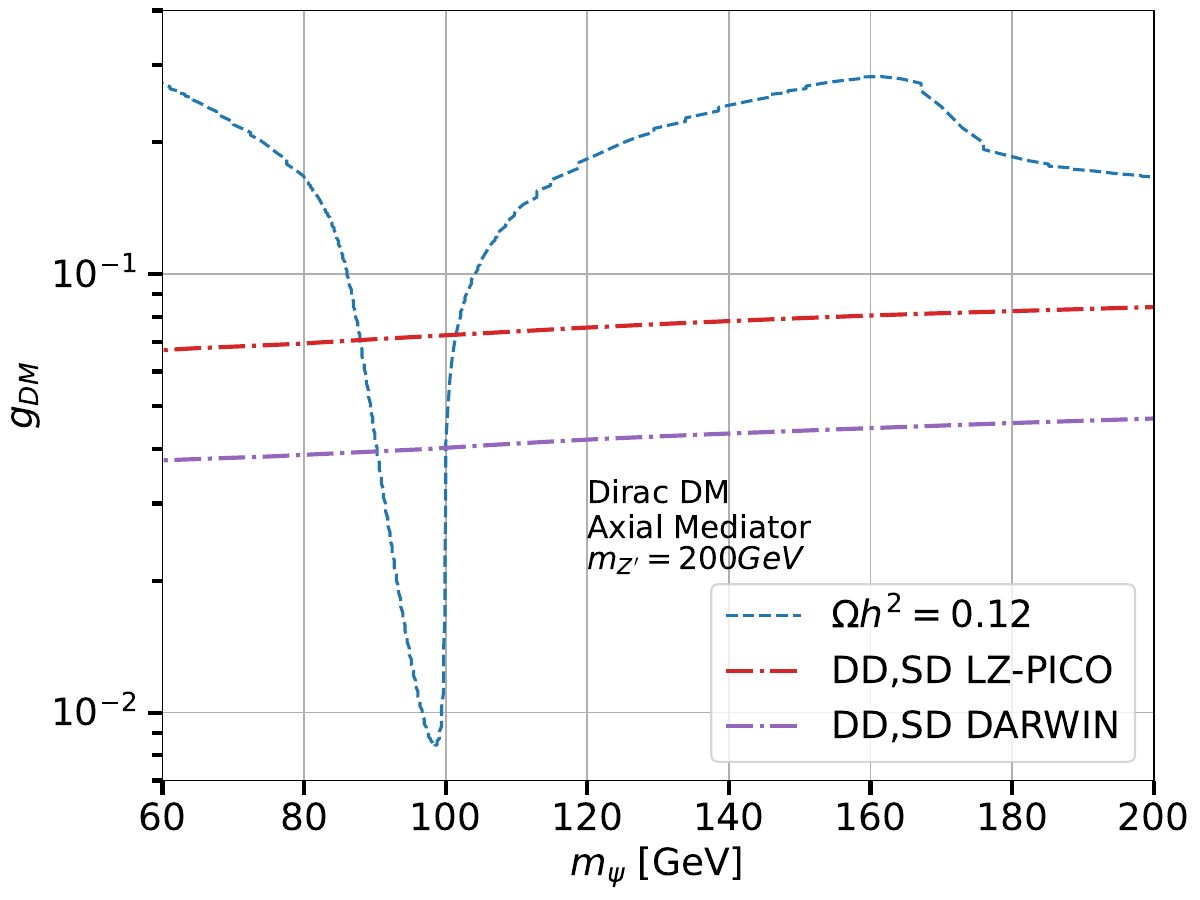}
\caption{The same as Fig.~\ref{fig:SmedDDM} but for axialvector mediator Dirac DM model.
The unitarity bound from eq is shown in green which exclude the parameter space in this region.
The DD exclusion limits are based on SD neutron cross section for this model.
In the $(g^{A}_{DM}=g^{A}_{SM},m_{\chi})$ plane, the unitarity bound is much weaker than DD constraints around the resonance region so that it is not shown in the lower-right panel.}
\label{fig:medDDMaxialvector}
\end{figure*}

Axial interactions with a massive vector lead to amplitudes that grow with energy (via the longitudinal $Z'$),
so partial-wave unitarity bounds the parameter space. Following
Refs.~\cite{PhysRevD.78.096004,PhysRevD.55.3137,BABU2012468}, we overlay a unitarity contour that, in our
normalization, scales approximately as
$m_\psi \propto m_{Z'}^{\,2}/g_{\rm DM}^{\,2}$ (green dashed lines in Fig.~\ref{fig:medDDMaxialvector}).
This provides an upper envelope at large masses and/or large couplings, beyond which the effective field theory description (and
perturbativity) breaks down.

The axial current matches onto the non-relativistic operator $O_4=\mathbf{S}_\psi\!\cdot\!\mathbf{S}_N$
(SD scattering) \cite{Arina:2018zcq}. We use the most constraining SD‐proton limits, combining
LZ \cite{LZ:2024zvo} with PICO-60 \cite{PICO:2019vsc} (red dot–dashed), and show DARWIN‐like SD projections (purple dot–dash–dot). The experimental
searches lose sensitivity below $m_\psi \simeq 3.94~\mathrm{GeV}$, so we keep that low-mass region unshaded.

Figure~\ref{fig:medDDMaxialvector} displays scans at fixed $g_{\rm DM}$ in the $(m_{Z'},m_\psi)$ plane
(left and top-right) and, for fixed $m_{Z'}=200~\mathrm{GeV}$, the $(g_{\rm DM},m_\psi)$ slice (bottom-right).

\begin{itemize}
\item For $g_{\rm DM}=1.0$ the SD limits already exclude most of the plane up to a few hundred of GeV for $m_\psi$, leaving viable
points primarily near the $s$-channel funnel, and below the unitarity envelope.
\item Reducing the coupling to $g_{\rm DM}=0.1$ and $0.01$ weakens SD scattering and opens more parameter space.
Nevertheless, the combination of relic density and SD limits still pushes the solution toward the resonant corridor.
\item In contrast to vector SI mediators, there remains a visible off-resonance band with
$m_\psi \gtrsim m_{Z'}/2$ that attains the correct relic density for each coupling choice; updated LZ+PICO results
now exclude a large fraction of that region (compare to \cite{Arcadi:2024ukq}), especially for $g_{\rm DM}=1.0$ and $0.1$.
For $g_{\rm DM}=0.01$ a significant portion survives all current constraints.
\end{itemize}

For $m_{Z'}=200~\mathrm{GeV}$ (bottom-right) the relic-density curve shows the expected dip around
$m_\psi\simeq m_{Z'}/2$; present and projected SD limits are roughly flat in $m_\psi$ over this range, intersecting
the relic line only away from the exact pole. 
The symmetry/asymmetry of the dip follows the discussion given for the pseudoscalar case, with indirect detection constraints displaying a nearly symmetric Breit–Wigner and the relic curve skewed by thermal averaging.

Compared with the vector  mediator, the axial case is less constrained by direct detection but still substantially restricted by SD searches plus the unitarity envelope. Current data confine most viable solutions to the resonant corridor or to small couplings; future SD improvements (DARWIN-like) will further erode the remaining off-resonance space.

\subsubsection{Scalar dark matter}

We consider a complex scalar DM particle $\chi$ coupled to a spin-1 mediator $Z'$ with coupling $g_{\rm DM}$, and flavor-universal vector couplings to SM quarks. In this setup the dominant annihilation
$\chi\chi^\ast\!\to\! f\bar f$ through an $s$-channel $Z'$ is $p$-wave suppressed, so present-day indirect-detection
constraints are negligible and we focus on the relic density and SI direct detection.

\begin{figure*}[t]
\includegraphics[width=0.49\linewidth]{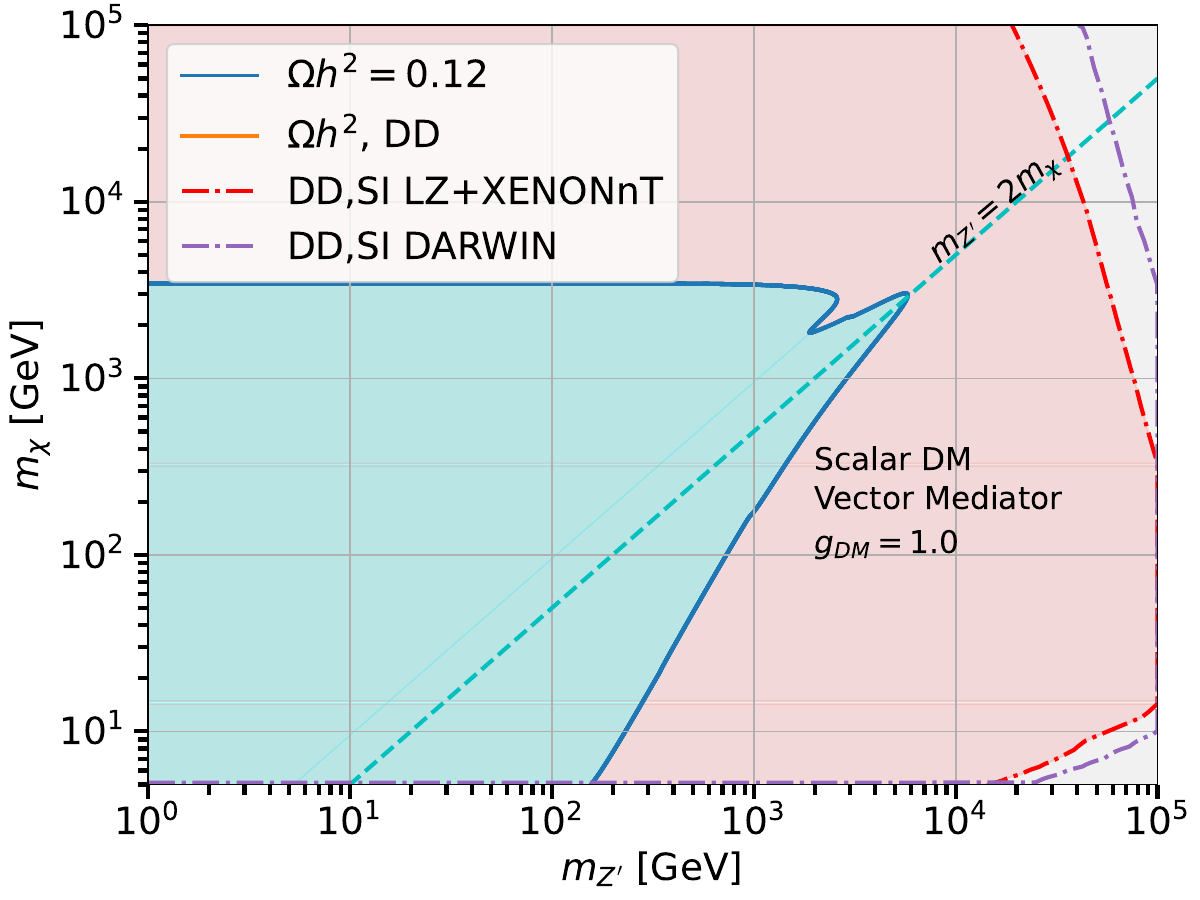}
\includegraphics[width=0.49\linewidth]{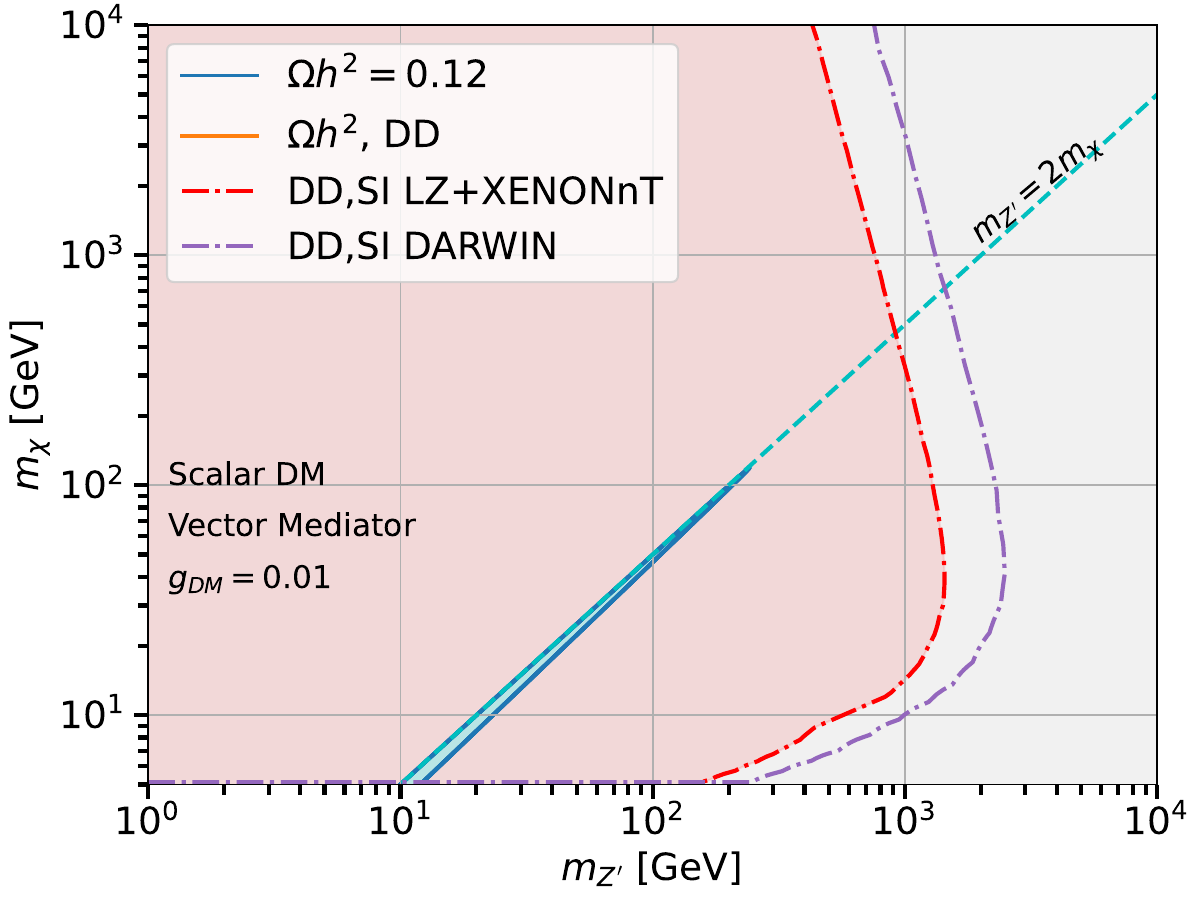}
\includegraphics[width=0.49\linewidth]{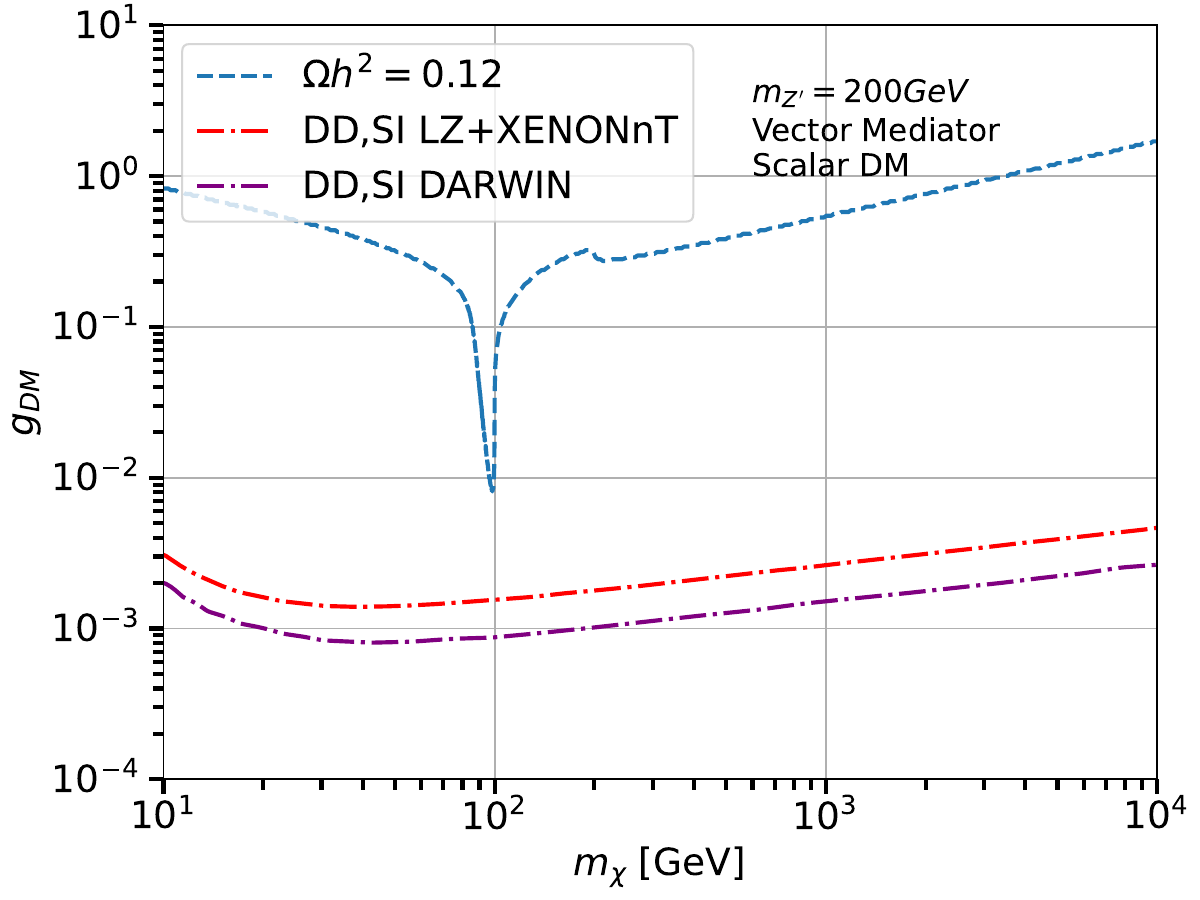}
\caption{The same as Fig.~\ref{fig:SmedDDM} but for vector mediator scalar DM model.}
\label{fig:medDDMvectorscalar}
\end{figure*}

Figure~\ref{fig:medDDMvectorscalar} shows scans in the $(m_{Z'},m_\chi)$ plane at fixed $g_{\rm DM}$ (top panels) and,
in the bottom panel, a scan in $(m_\chi,g_{\rm DM})$ at fixed $m_{Z'}=200~\mathrm{GeV}$.
For $g_{\rm DM}=1$ (top-left), a broad region reproduces $\Omega_{\rm DM}h^2\simeq0.12$ (blue), but
SI direct-detection limits (LZ-XENONnT and DARWIN; red and purple) exclude essentially all of it. Lowering the coupling to $g_{\rm DM}=0.01$ (top-right) pushes the relic-density solution onto the resonant funnel $m_{Z'}\simeq 2m_\chi$, but the required couplings remain well above the nuclear cross section bounds across the mass range shown.

The tension is made explicit in the bottom right panel: for $m_{Z'}=200~\mathrm{GeV}$ the relic-density curve (blue) demands
$g_{\rm DM}\gtrsim 10^{-1}$ away from resonance and $\sim10^{-2}$ at $2m_{\psi}\sim m_{\rm{med}}$, whereas LZ already limits $g_{\rm DM}\lesssim\text{few}\times10^{-3}$ (DARWIN projects even lower). There is no overlap, even at the pole.
Hence, within the mass ranges displayed, the vector-mediator–scalar-DM model is excluded by SI direct detection
once the relic abundance requirement is imposed.

For a vector mediator, the leading SI amplitude is the coherent sum of vector currents over the nucleus,
so the per–nucleon cross section scales as in Eq.~\ref{eq:scalarSI} with $A^2$ coherence and with no $q^2$ or $v^2$
suppression.  The effective proton/neutron couplings are $\mathcal{O}(g)$ combinations of $V_u$ and $V_d$ (e.g.\ $f_p=2V_u+V_d$), so the overall normalization is set directly by the gauge–like couplings.
In contrast, for a CP–even scalar mediator the nucleon coupling proceeds via the scalar form factors and, in Yukawa–aligned setups, is proportional to the quark masses.  After matching to the nucleon one obtains
$g_{SNN}\simeq \beta\, (m_N/v_h)\, f_N$ with $f_N\!\sim\!0.30$ and $m_N/v_h\simeq 4\times10^{-3}$, i.e.\ an intrinsic suppression by $m_N/v_h$ relative to the vector case at fixed couplings and masses.  (There is no leading $q^2$ or $v^2$ suppression here either; the difference is the Yukawa/trace–anomaly normalization.)
Therefore, for comparable $(m_{\rm DM},m_{\rm med},g)$ the vector–mediator prediction for $\sigma^{\rm SI}$ is generically much larger, which explains why direct–detection limits are tighter in the vector case.  Moreover, unlike some Dirac–DM $Z'$ setups where partial $f_p$–$f_n$ cancellations can weaken the nuclear coherence, the scalar–DM scenarios considered here typically add more nearly coherently across protons and neutrons, further strengthening the effective bounds in the vector–mediator case.


\subsection{Dark matter interpretation of the {\it Fermi}-LAT Galactic center excess}
\label{sec:GCE}

The $\gamma$-ray excess observed by \textit{Fermi}–LAT toward the Galactic Center is among the most discussed
possible signals of DM annihilation. It is therefore natural to test whether simplified models can reproduce the GCE while remaining compatible with relic density and direct-detection constraints.

We consider two $s$-wave–dominated benchmarks at present times:
(i) \emph{scalar DM} $\chi$ with a \emph{scalar mediator} $S$; and
(ii) \emph{Dirac DM} $\psi$ with a \emph{vector mediator} $Z'$.
An analogous $s$-wave behavior occurs for vector DM with a scalar mediator.
By contrast, for the coupling choices analyzed here, models with a scalar mediator and Dirac DM, or a vector mediator and scalar DM,
are $p$-wave dominated today and cannot account for the GCE without violating direct-detection limits by orders of magnitude.

In all cases we assume annihilation dominantly to quarks and include only the prompt $\gamma$-ray emission:
secondary components (e.g.\ inverse Compton) are negligible above $E_\gamma \gtrsim 0.3$ GeV \cite{Blanchet:2012vq,Koechler:2025ryv}.
We adopt the same prompt yields and line-of-sight geometry as in \cite{DiMauro:2021qcf}.
The differential flux is
\begin{equation}
\frac{d\Phi_\gamma}{dE_\gamma}
\;=\;
\frac{\langle\sigma v\rangle}{\mathcal{S}\,4\pi\,m_\chi^2}\,
\frac{dN_\gamma}{dE_\gamma}\,
J_{\rm ROI},
\end{equation}
where $J_{\rm ROI}$ is the usual $J$-factor integrated over the region of interest and
$dN_\gamma/dE_\gamma$ is computed with \texttt{MadDM} (using \texttt{Cosmixs} tables) \cite{Arina:2023eic,DiMauro:2024kml},
so that the channel weights reflect the model’s annihilation matrix elements.
The factor $\mathcal{S}$ is $2$ for \emph{self–conjugate} DM (e.g., a real scalar or Majorana fermion) and $4$ for DM with a distinct antiparticle (e.g., a complex scalar or Dirac fermion).\footnote{$\mathcal{S}$ is the \emph{initial–state combinatorial} factor that accounts for how many annihilating pairs exist in a DM fluid of density $\rho(\mathbf{x})$: For \emph{self–conjugate} DM (e.g.\ Majorana, real scalar), the number of pairs in a volume is $n_\chi^2/2$ (the $1/2$ avoids double counting identical pairs), with $n_\chi=\rho/m_\chi$. The annihilation rate density is $\Gamma_{\rm ann}=\tfrac{1}{2}\,\langle\sigma v\rangle\,n_\chi^2
=\tfrac{1}{2}\,\langle\sigma v\rangle\,\frac{\rho^2}{m_\chi^2}.$
For \emph{non–self–conjugate} DM (e.g.\ Dirac, complex scalar) with equal particle/antiparticle densities, $n_\chi=n_{\bar\chi}=\rho/(2m_\chi)$, and the pair number is $n_\chi n_{\bar\chi}$: $\Gamma_{\rm ann}=\langle\sigma v\rangle\,n_\chi n_{\bar\chi}
=\tfrac{1}{4}\,\langle\sigma v\rangle\,\frac{\rho^2}{m_\chi^2}.$}

This factor is \emph{not} part of the usual spin/quantum–number averages in $\langle\sigma v\rangle$. Those averages are purely particle–physics normalizations of the two–body cross section; $\mathcal{S}$ instead encodes the \emph{astrophysical} pair–counting (identical vs.\ distinct initial states) when converting $\langle\sigma v\rangle$ and $\rho$ into an annihilation rate density.

We fit two GCE spectra: (i) Ref.~\cite{DiMauro:2021prd}, ROI $-20^\circ<l<20^\circ$, $-20^\circ<b<20^\circ$;
(ii) Ref.~\cite{Cholis:2022prd}, same ROI with the plane masked ($|b|<2^\circ$).
Both fits include statistical and systematic errors from interstellar-emission modeling.
Free parameters are the DM mass and the velocity-averaged cross section, $(m_\chi,\,\langle\sigma v\rangle)$.

\begin{figure*}[t]
\includegraphics[width=0.49\linewidth]{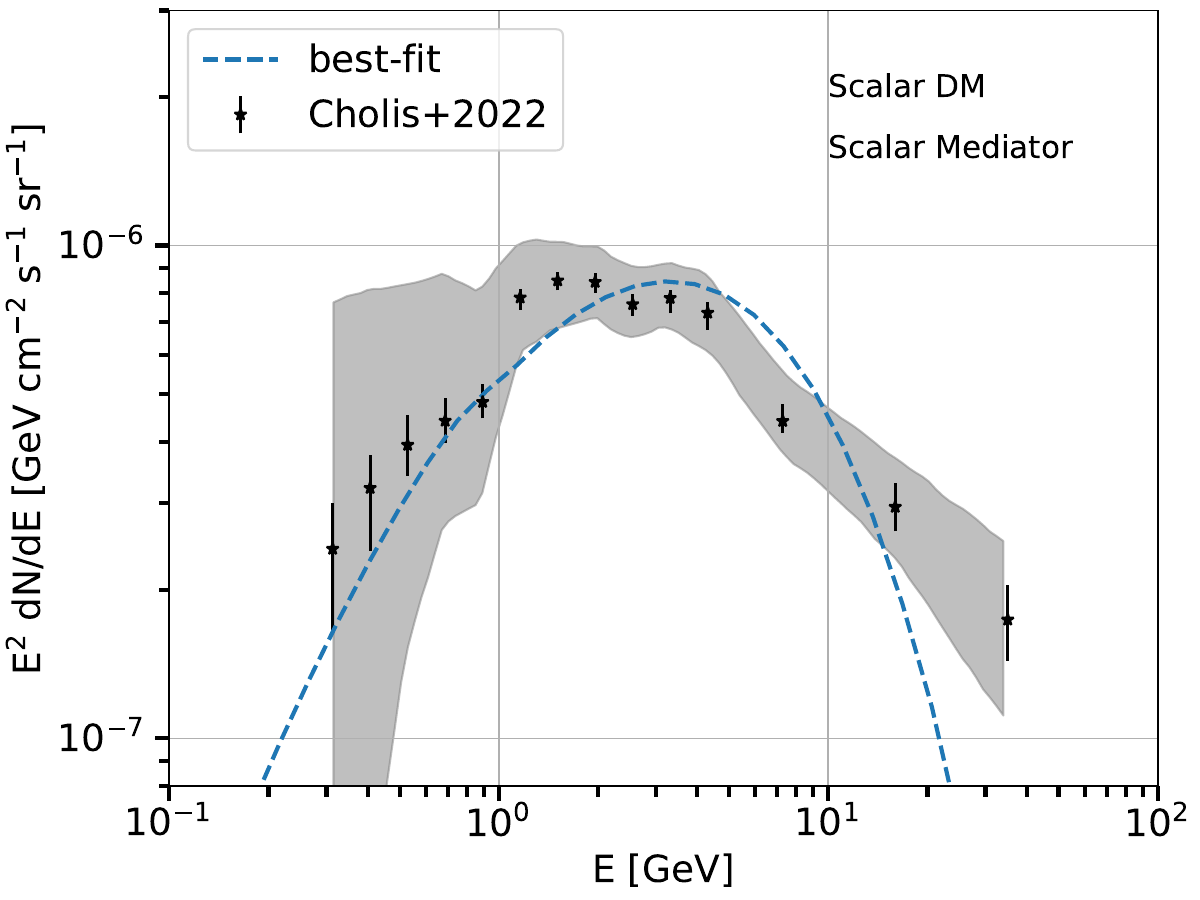}
\includegraphics[width=0.49\linewidth]{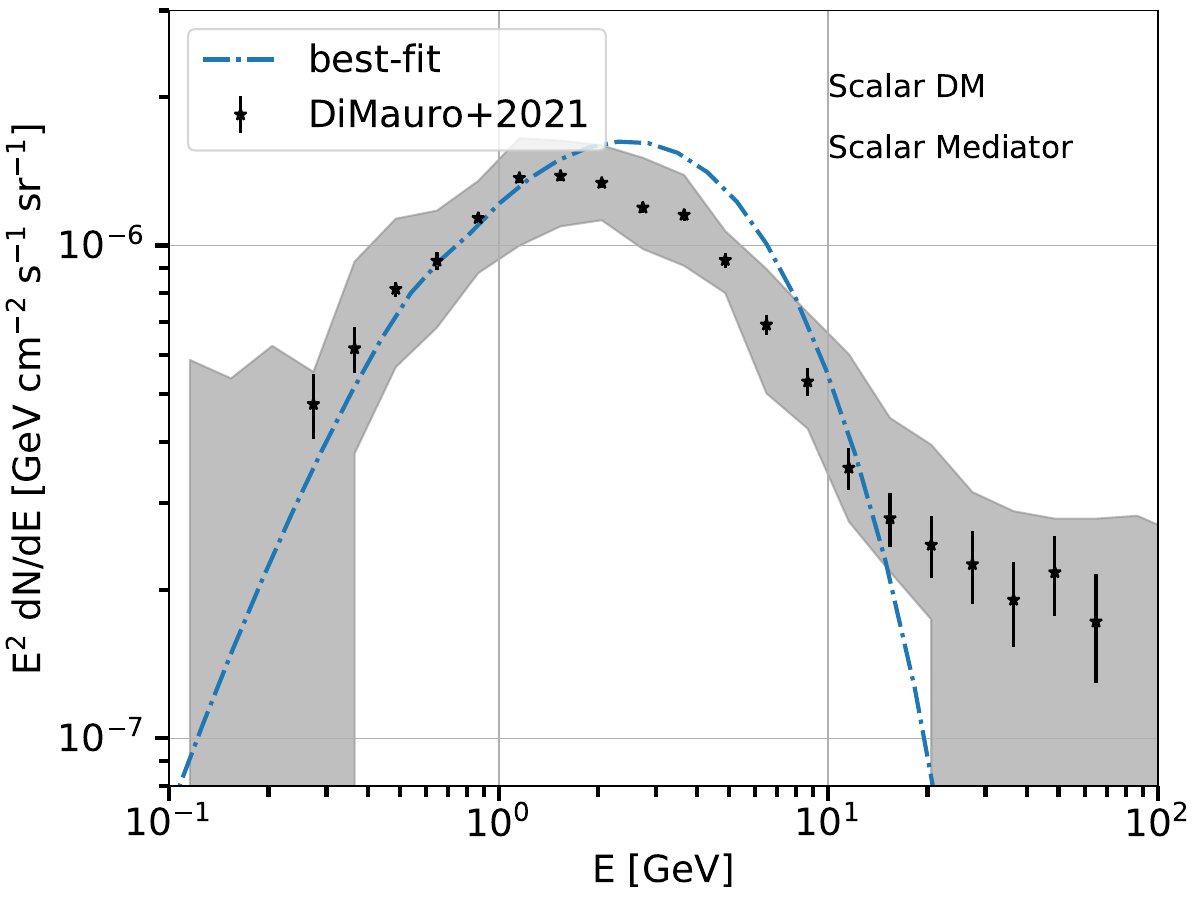}
\caption{Best-fit from DM in the simplified model with scalar DM and spin-0 mediator to the GCE data (black points) reported in \cite{Cholis:2022prd} (left panel) and \cite{DiMauro:2021prd} (right panel).
The gray bands correspond to the statistical and systematic errors for the data, and the blue dashed lines are the theoretical best-fit curves.}
\label{fig:GCEbestfit}
\end{figure*}

Fig.~\ref{fig:GCEbestfit} displays the best fits for which we find we find
\[
m_\chi = (64.5 \pm 6.3)\ \mathrm{GeV}, \,
\langle\sigma v\rangle = (1.7 \pm 0.2)\times 10^{-26}\ \mathrm{cm^3\,s^{-1}}.
\]
when using Cholis+22 data \cite{Cholis:2022prd} and 
\[
m_\chi = (46.5 \pm 0.6)\ \mathrm{GeV}, \,
\langle\sigma v\rangle = (9.6 \pm 0.6)\times 10^{-27}\ \mathrm{cm^3\,s^{-1}}.
\]
when considering Di Mauro+22 \cite{DiMauro:2021prd}.
Uncertainties reflect the fit including systematics due to the choice of the interstellar emission model. An additional normalization uncertainty enters through the $J$-factor:
taking $\rho_\odot = 0.4~\mathrm{GeV\,cm^{-3}}$ with a plausible range $0.3$–$0.5~\mathrm{GeV\,cm^{-3}}$ implies an
overall flux uncertainty of $\mathcal{O}(2)$ (since $\Phi_\gamma\propto \rho_\odot^2$). With this, the preferred
\(\langle\sigma v\rangle\) values are consistent with the canonical thermal value that for \(m_\chi\simeq 40\)–\(70\) GeV is around $2.0-2.3 \times 10^{-26}$ cm$^3$/s
\cite{Steigman:2012nb,BRINGMANN2021136341}.

\begin{figure*}[t]
\includegraphics[width=0.49\linewidth]{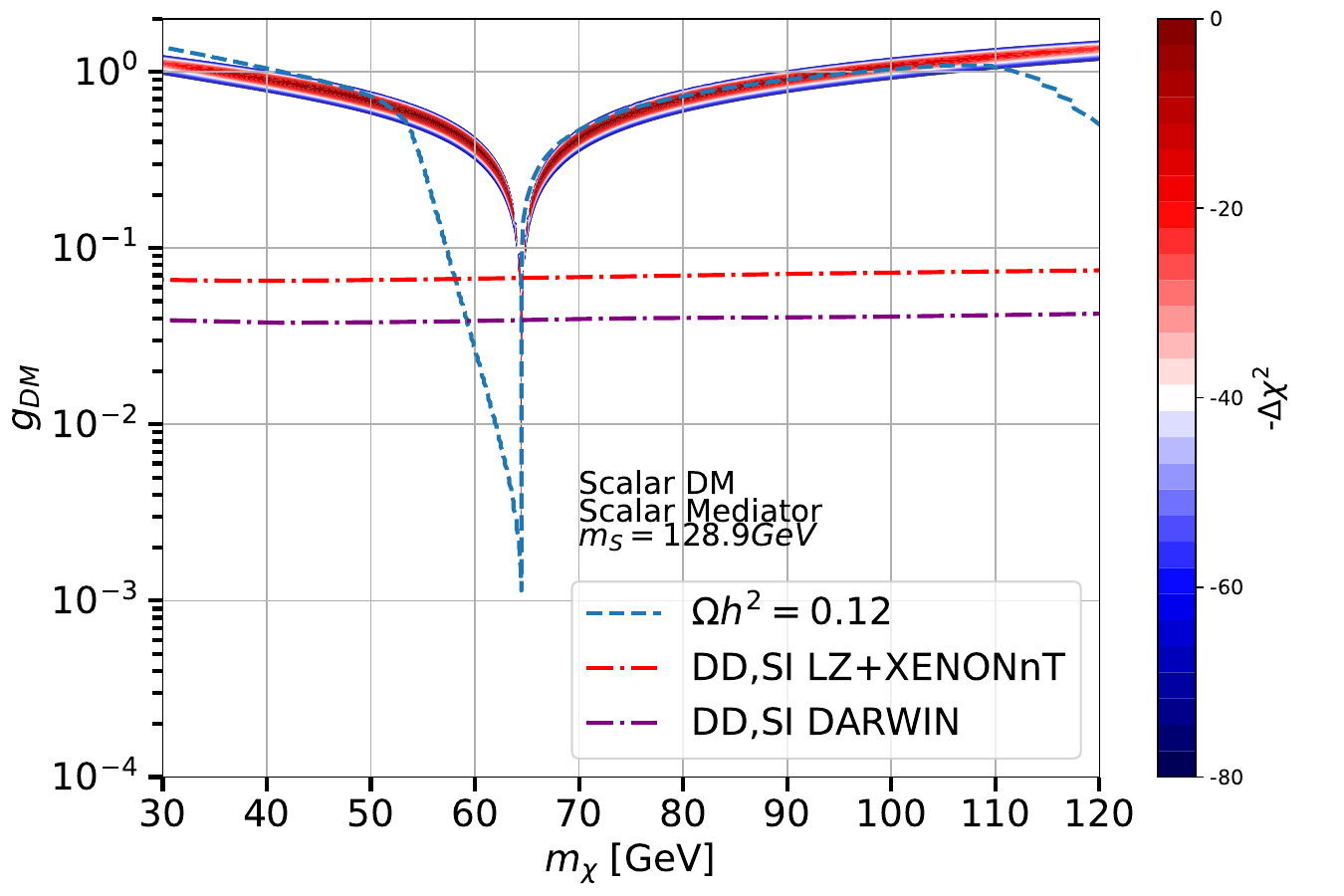}
\includegraphics[width=0.49\linewidth]{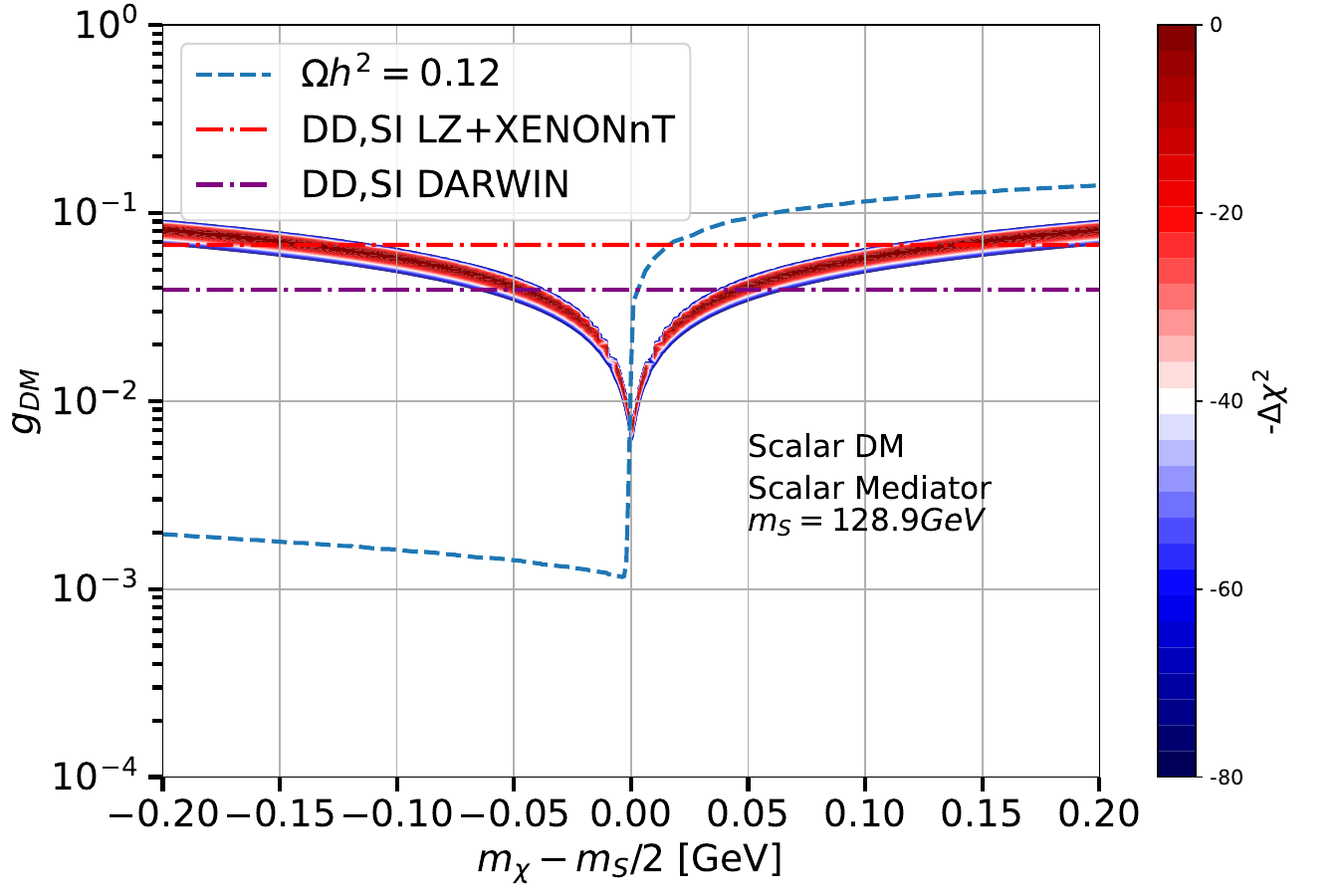}
\includegraphics[width=0.49\linewidth]{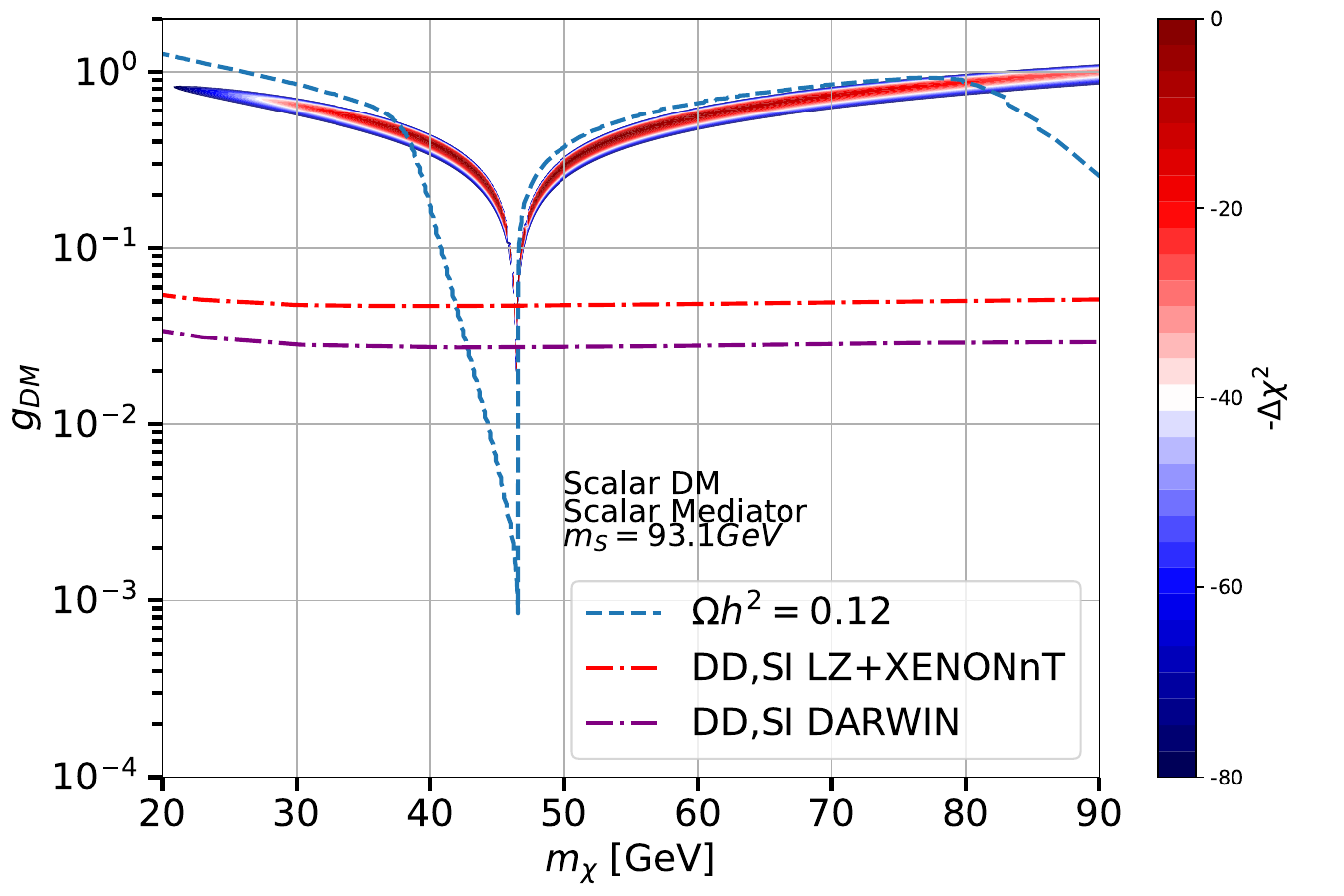}
\includegraphics[width=0.49\linewidth]{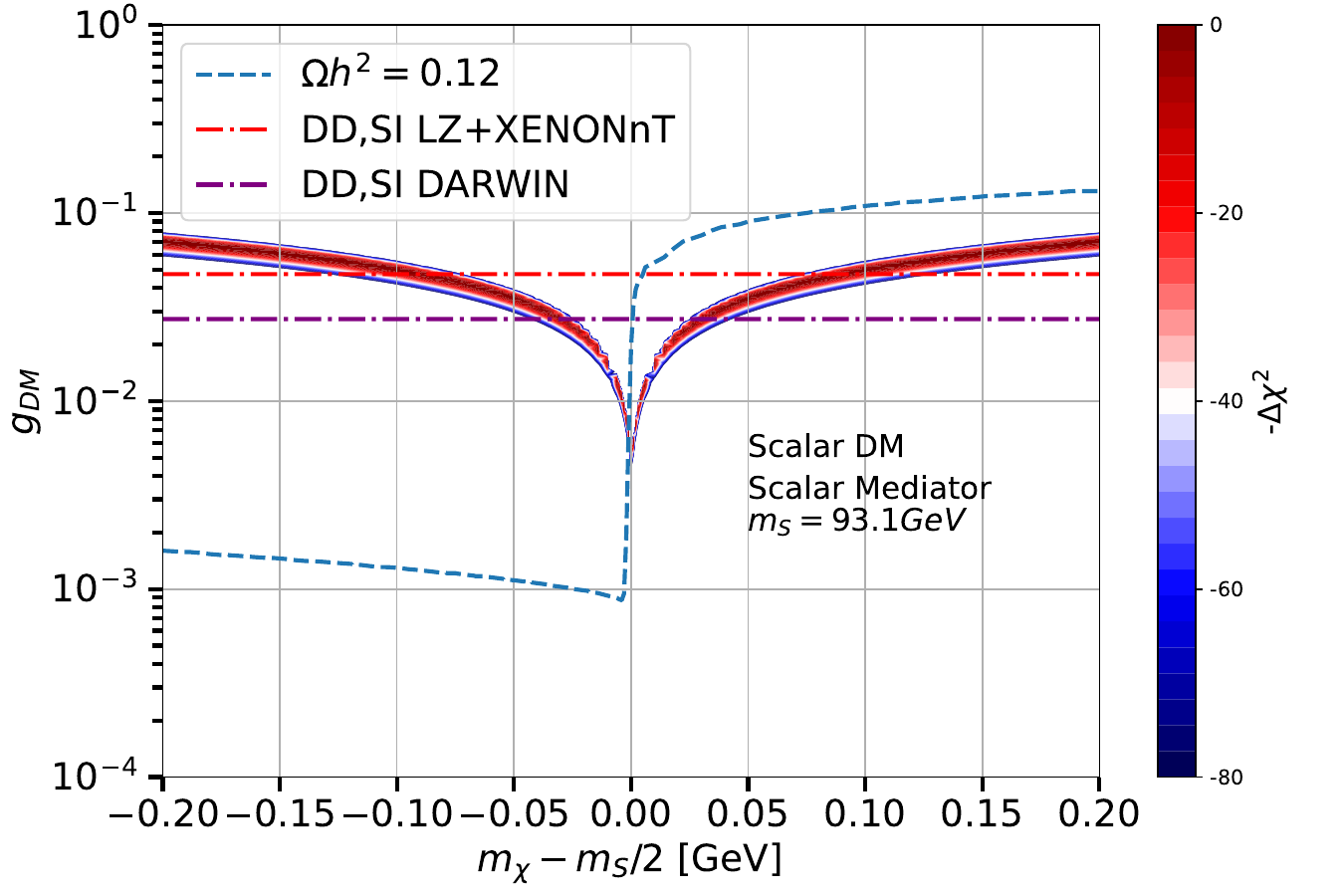}
\caption{Upper-left panel: Contour region for the $\chi^2$ obtained with a fit to the GCE data found in \cite{Cholis:2022prd}.
We also show the direct detection upper limits based on LZ data \cite{LZ:2022ufs} (red dot-dashed line) and projections to DARWIN (purple dot-dashed line).
The blue dashed line corresponds to model parameters that provide the observed DM relic density.
Upper-right panel: Same as the upper-left panel but for the region around resonance $m_{\chi} = \frac{1}{2}m_{S}$.
Lower-left panel: Same as the upper-left panel but the contour regions based on the GCE data in \cite{DiMauro:2021prd}.
Lower-right panel: Same as the lower-left panel but for the region around resonance $m_{\chi} = \frac{1}{2}m_{S}$.}
\label{fig:GCEspin0dmg}
\end{figure*}

Figure~\ref{fig:GCEspin0dmg} shows the $\Delta\chi^2$ profiles versus $m_\chi$ and the coupling \(g_{\rm DM}\). We have fixed the mediator mass as twice the best-fit DM mass.
Away from the $s$-channel pole ($m_S\neq 2m_\chi$), reproducing the GCE requires \(g_{\rm DM}\sim \mathcal{O}(1)\),
which is incompatible with LZ and (even more so) DARWIN SI limits.
Near resonance ($m_S\simeq 2m_\chi$) the annihilation rate is enhanced at fixed masses,
so the GCE can be matched with \(g_{\rm DM}\sim 10^{-2}\), bringing the nuclear cross section below current bounds.
This mirrors the relic-density condition: for fixed masses, \(\langle\sigma v\rangle \propto g_{\rm DM}^4\),
so both the relic contour and the GCE-preferred band track the resonant funnel.

\begin{figure*}[t]
\includegraphics[width=0.49\linewidth]{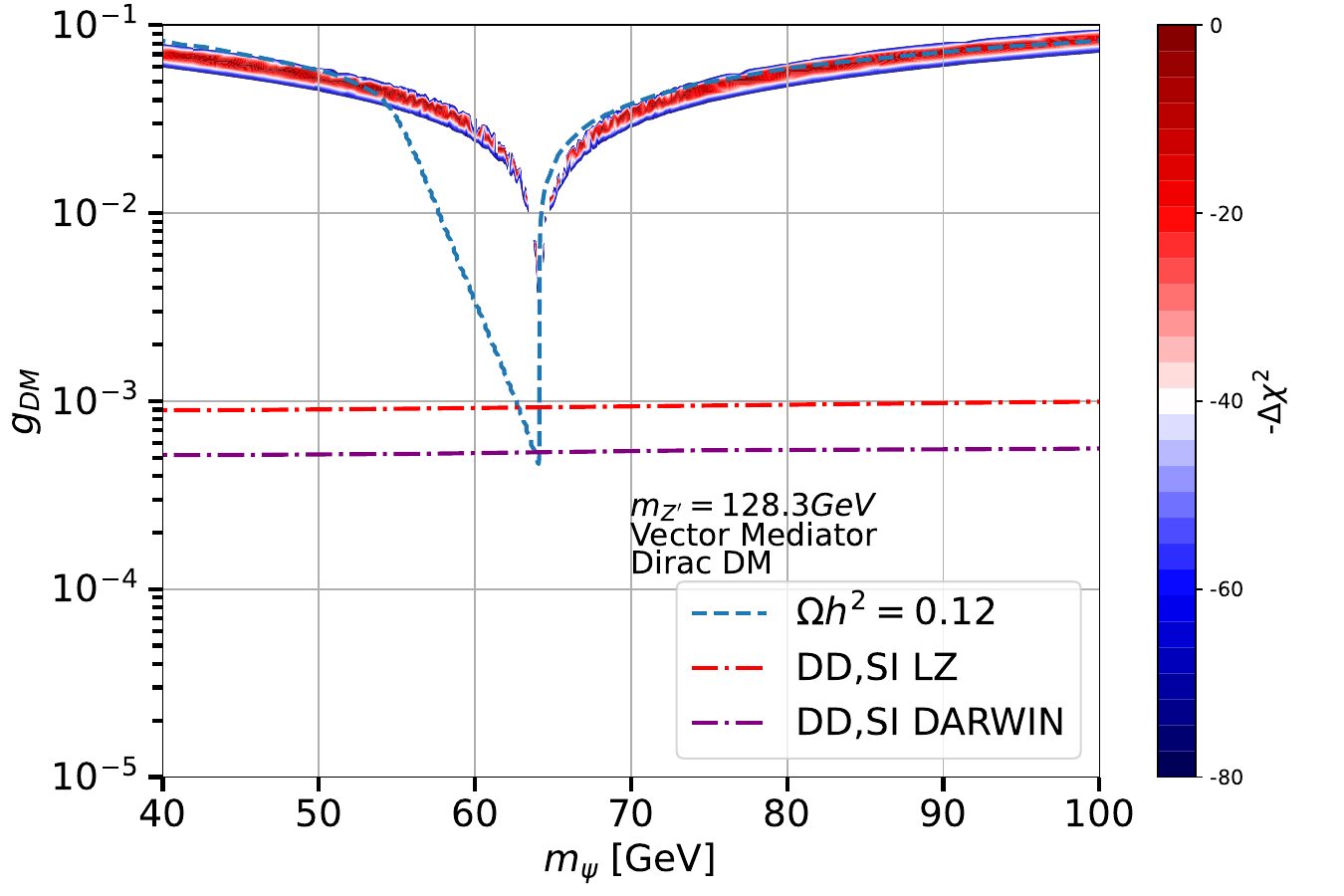}
\includegraphics[width=0.49\linewidth]{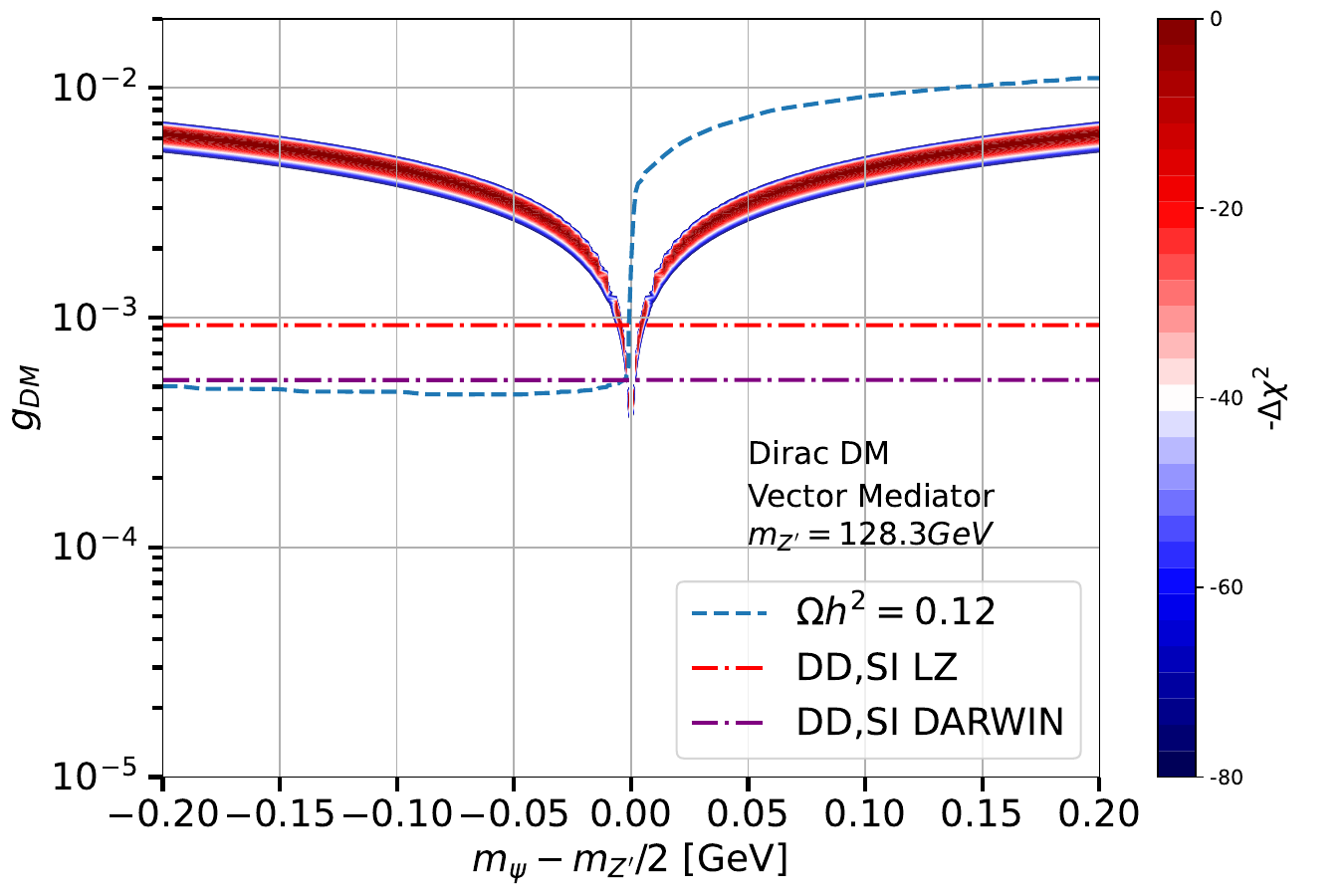}
\includegraphics[width=0.49\linewidth]
{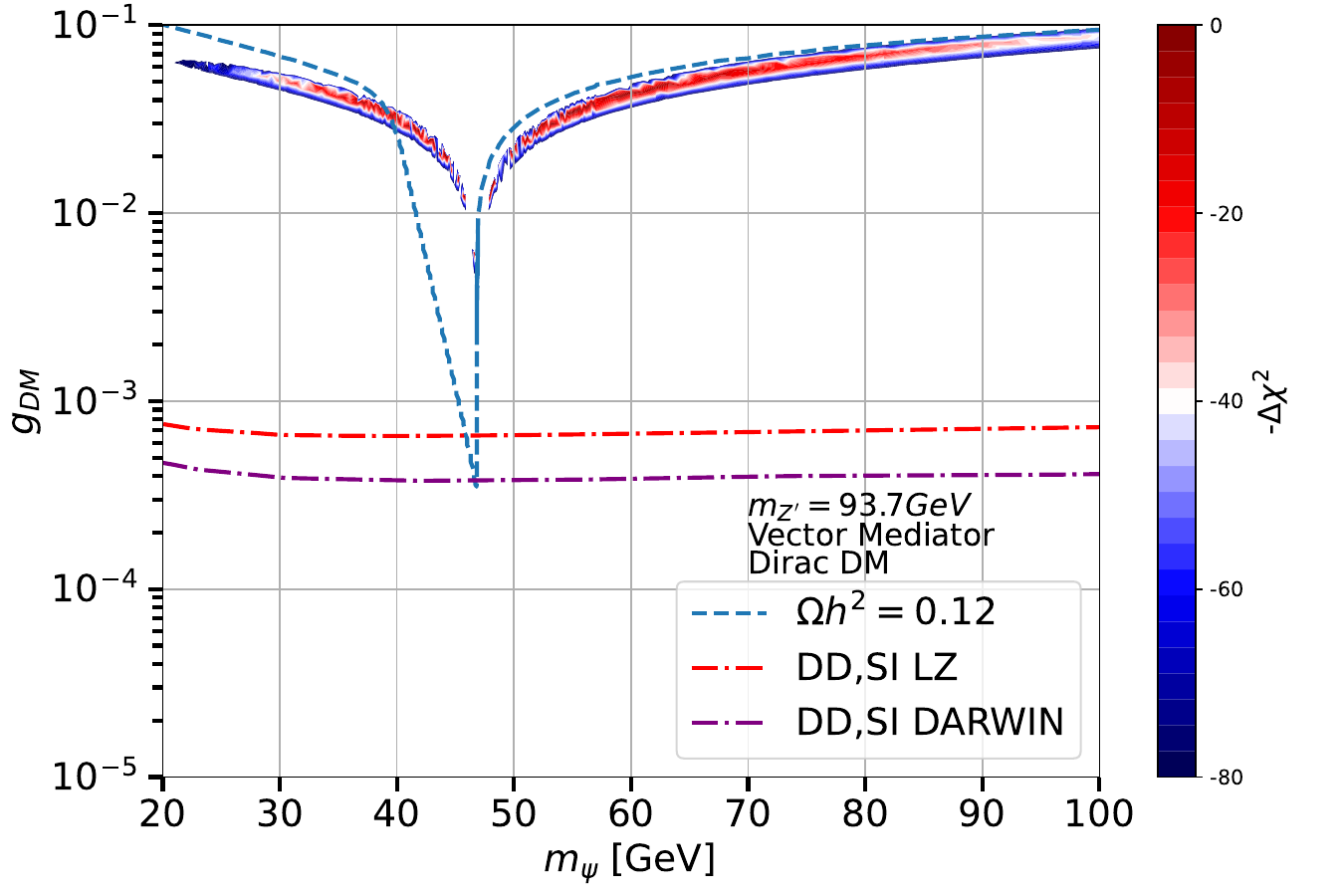}
\includegraphics[width=0.49\linewidth]{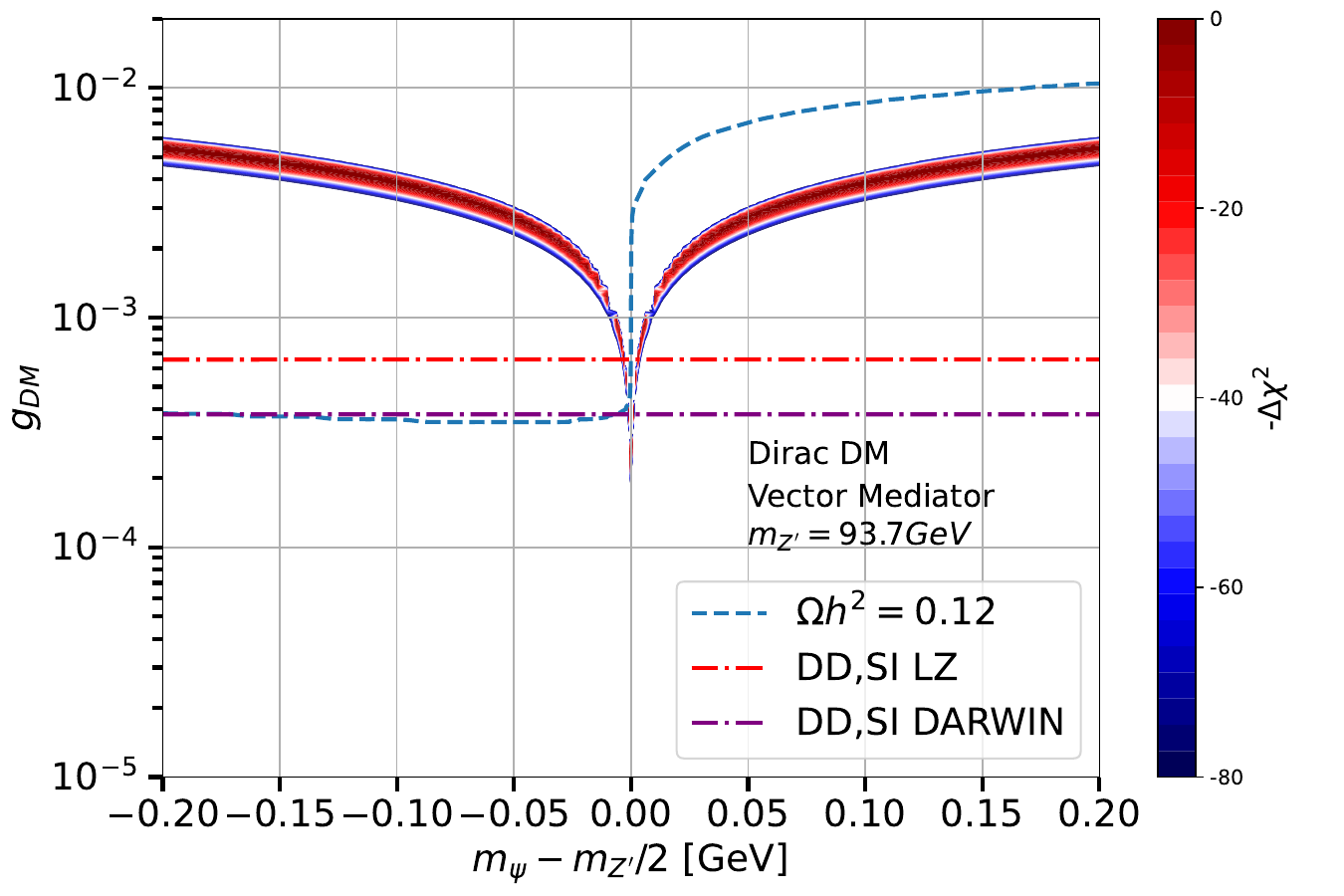}
\caption{The same as Fig.~\ref{fig:GCEspin0dmg} but for vector mediator Dirac DM model.
}
\label{fig:GCEspin1dmg}
\end{figure*}

The corresponding profiles for a $Z'$ mediator and Dirac DM are shown in Fig.~\ref{fig:GCEspin1dmg}.
Because the prompt spectra (and thus the best-fit \((m_\chi,\langle\sigma v\rangle)\)) are very similar in this mass range
(dominated by $b\bar b$), the GCE-preferred regions align with those found above.
However, SI direct-detection bounds are \emph{more} stringent for vector exchange, and the allowed strip near resonance is much narrower:
LZ already excludes most of the band; the DARWIN projection covers essentially all of the parameter space consistent with the GCE,
including very close to the pole.

For {\tt DMsimp} models with $s$-wave DM annihilation cross section into SM pairs, the GCE can be accommodated only within a narrow $s$-channel \textit{resonant funnel}, $m_{\rm med}\simeq 2\,m_{\rm DM}$, where modest couplings ($g_{\rm DM}\sim 10^{-2}$) yield the required $\langle\sigma v\rangle$ while evading SI direct-detection bounds; off resonance, the parameter space is essentially excluded.
Given residual astrophysical systematics and potential contributions from unresolved sources \cite{Macias:2016nev,Bartels:2017vsx,Manconi:2024tgh}, the DM interpretation remains intriguing but not unique; improved GC modeling and complementary searches will be decisive.


\section{Natural origins of resonant dark matter}
\label{sec:natural_resonance}

In this section we present two simple UV setups in which the relation $m_{\rm med}\simeq 2\,m_{\rm DM}$ arises \emph{naturally}, i.e.\ from masses controlled by the
same symmetry-breaking scale and $\mathcal{O}(1)$ couplings, rather than from a tuned choice
of unrelated parameters. The first is the SM Higgs portal with a real
scalar DM field; the second is a dark $U(1)_X$ gauge theory in which both a new gauge boson
$Z'$ and a Dirac DM fermion acquire mass from the VEV of a dark
Higgs field. In both cases, reasonable relations among couplings lead to $m_{\rm med}\!\approx\!2m_{\rm DM}$ without fine tuning, and the resonant enhancement of the $s$-channel annihilation rate then explains why these regions are repeatedly selected by
relic-density and direct-detection complementarity.

\subsection{Example I: Higgs portal with real scalar dark matter}
\label{sec:HP_scalar}

Consider a $\mathcal{Z}_2$-odd real scalar $\chi$ (the DM candidate) coupled to the SM Higgs
doublet $H$ via the renormalizable portal \cite{SILVEIRA1985136,DiMauro:2023tho}:
\begin{align}
\mathcal{L}_{\rm HP} &=
\frac{1}{2}(\partial_\mu \chi)(\partial^\mu \chi)
-\frac{1}{2}\,\mu_\chi^2\,\chi^2 - \frac{\lambda_\chi}{4}\,\chi^4
-\frac{\lambda_{H\chi}}{2}\,\chi^2\,H^\dagger H,
\label{eq:HP_L}
\end{align}
with the usual SM Higgs potential for $H$.
After electroweak symmetry breaking (EWSB),
$H = \big(0,\, (v_h+h)/\sqrt{2}\big)^T$ with $v_h\simeq 246~{\rm GeV}$, one obtains
\begin{align}
m_\chi^2 &= \mu_\chi^2 + \frac{\lambda_{H\chi}}{2}\,v_h^2, &
\mathcal{L} \supset -\,\frac{\lambda_{H\chi}}{2}\,v_h\,h\,\chi^2,
\end{align}
so that the SM Higgs boson $h$ (with $m_h\simeq 125~{\rm GeV}$) mediates $\chi\chi\to {\rm SM}$
in the $s$-channel.

A \emph{resonant} configuration arises for $2m_\chi \simeq m_h$. This happens
\emph{naturally} if the portal dominates the DM mass, $\mu_\chi^2 \ll \lambda_{H\chi}v_h^2/2$,
so that $m_\chi \simeq \sqrt{\lambda_{H\chi}}\,v_h/\sqrt{2}$. The resonance condition then fixes
an $\mathcal{O}(0.1)$ portal coupling,
\begin{align}
2m_\chi \simeq m_h
\quad \Longrightarrow \quad
\lambda_{H\chi} \simeq \frac{m_h^2}{2v_h^2} \approx 0.13,
\end{align}
which is perturbative and does not require tuning beyond the assumption that the DM mass
is chiefly induced by EWSB through the portal. In this limit the same coupling $\lambda_{H\chi}$ controls both
the DM mass and the $h\chi\chi$ vertex, aligning the model toward the resonant strip
favored by relic density with small direct-detection rates (since near resonance much smaller
$\lambda_{H\chi}$ can reproduce the thermal cross section).
In Ref.~\cite{DiMauro:2023tho} the authors have studied this model extensively and demonstrated that it can fit the GCE flux.

\subsection{Example II: Dark $U(1)_X$ with a massive $Z'$ and Dirac dark matter}
\label{sec:U1_resonance}

Consider a new Abelian gauge symmetry $U(1)_X$ with gauge field $X_\mu$ and coupling $g_X$.
The symmetry is spontaneously broken by a complex scalar $\Phi$ of $U(1)_X$ charge $q_\Phi$,
and the DM is a Dirac fermion $\psi$ charged under $U(1)_X$.\footnote{Charge assignments
must satisfy anomaly cancellation; simple consistent choices exist and are not essential
for the present discussion.} The relevant Lagrangian is \cite{Langacker:2007ac,Langacker:2008yv}
\begin{eqnarray}
\mathcal{L}_{U(1)_X} &=&
-\frac{1}{4} X_{\mu\nu} X^{\mu\nu}
+ |D_\mu \Phi|^2 - V(\Phi)
+ \bar{\psi}\, i\slashed{D}\, \psi + \\
&-&\Big( y_\psi\, \Phi\, \bar{\psi}_L \psi_R + \text{h.c.} \Big)
-\frac{\epsilon}{2}\,X_{\mu\nu} B^{\mu\nu} + \nonumber \\
&+& \sum_f \bar{f}\,\gamma^\mu \left( \kappa_V^f - \kappa_A^f \gamma_5 \right) f\, X_\mu, \nonumber
\label{eq:U1_L}
\end{eqnarray}
with $D_\mu = \partial_\mu + i g_X q\, X_\mu$, $B_{\mu\nu}$ the hypercharge field strength,
and a Higgs-like potential $V(\Phi) = -\mu_\Phi^2 |\Phi|^2 + \lambda_\Phi |\Phi|^4$.
When $\Phi$ acquires a VEV, $\langle\Phi\rangle = v_X/\sqrt{2}$, one finds
\begin{align}
m_{Z'} &= g_X\, q_\Phi\, v_X, &
m_\psi &= \frac{y_\psi}{\sqrt{2}}\, v_X, &
m_S &= \sqrt{2\lambda_\Phi}\, v_X,
\label{eq:U1_masses}
\end{align}
where $Z'_\mu \equiv X_\mu$ is the massive gauge boson and $S \equiv \sqrt{2}\,\mathrm{Re}\,\Phi - v_X$
is the dark Higgs.

Because \emph{both} $m_{Z'}$ and $m_\psi$ are set by the same scale $v_X$, their ratio is
coupling-controlled and independent of $v_X$:
\begin{align}
\frac{m_{Z'}}{2 m_\psi} = \frac{g_X\, q_\Phi}{\sqrt{2}\, y_\psi}.
\end{align}
Hence the $Z'$ pole, $m_{Z'} \simeq 2 m_\psi$, is obtained for the \emph{natural} relation
\begin{align}
y_\psi \;\simeq\; \frac{g_X\, q_\Phi}{\sqrt{2}},
\end{align}
which can emerge from $\mathcal{O}(1)$ couplings at the symmetry-breaking scale (and is
radiatively stable up to modest running). In this situation, $\psi\bar{\psi}\to Z'\to f\bar f$
proceeds resonantly in the early Universe while present-day annihilation depends on the
vector/axial structure of the couplings $\kappa_{V,A}^f$.

A closely related possibility is a \emph{scalar} (dark-Higgs) resonance.
Since $m_s = \sqrt{2\lambda_\Phi} v_X$ and $m_\psi = y_\psi v_X/\sqrt{2}$,
\begin{align}
\frac{m_s}{2 m_\psi} = \frac{\sqrt{2\lambda_\Phi}}{2 y_\psi}
\quad\Rightarrow\quad
m_s \simeq 2 m_\psi \;\;\text{for}\;\; \lambda_\Phi \simeq 2 y_\psi^2,
\end{align}
again a mild relation among $\mathcal{O}(1)$ couplings.
Thus the same $U(1)_X$ setup \emph{naturally} realizes either a vector ($Z'$) or scalar ($s$)
resonant mediator, depending on whether $g_X q_\Phi$ or $\sqrt{\lambda_\Phi}$ aligns with $y_\psi$.

In both Examples~\ref{sec:HP_scalar} and \ref{sec:U1_resonance}, the mediator and DM masses are
set by a common symmetry-breaking scale (the SM Higgs VEV or a dark-Higgs VEV). Resonance then
corresponds to simple $\mathcal{O}(1)$ ratios of couplings, rather than to a tuned relation
between unrelated mass parameters. Cosmologically, the resonant enhancement in the
$s$-channel permits the observed relic density to be obtained with small portal/gauge/Yukawa
couplings, which simultaneously suppresses direct-detection rates—precisely the pattern favored by
global constraints in Secs.~\ref{sec:results}--\ref{sec:conclusions}.

\section{Conclusions}
\label{sec:conclusions}

In this work we have revisited DM simplified models \texttt{DMSimps} in the near–resonant regime, confronting them with up-to-date constraints from relic density, direct detection, and indirect detection, and assessing their ability to account
for the \textit{Fermi}-LAT GCE. Our main findings are:

\begin{itemize}
\item \textbf{Resonant viability.} Over most of parameter space \texttt{DMSimps} are severely
constrained by direct detection. The primary surviving region occurs near the $s$-channel resonance,
\mbox{$m_{\rm med}\simeq 2\,m_{\rm DM}$}, where the enhancement of \sigmav\ allows
the observed relic abundance to be achieved with much smaller couplings, thereby
evading current spin indepedent and SI limits. This pattern appears across the spin assignments we studied (see Tab.~\ref{tab:simp_summary_grid_ps_ax_vec}).

\item \textbf{Relic density with kinetic decoupling.} We solved the full Boltzmann equation
including elastic scatterings (``{\tt fBE}'') and compared with the usual treatment that
assumes kinetic equilibrium (``{\tt nBE}''). In the resonance region, the difference can be
significant: for scalar DM we find up to a \emph{factor $\sim$4.5} change in $\Omega_{\rm DM}h^2$ (Fig.~\ref{fig:RD1}, left), translating into $\mathcal{O}(1.5)$ shifts in the couplings that reproduce the Planck value for $\Omega_{\rm{DM}} h^2$ (Fig.~\ref{fig:RD1}, right). We therefore
employed the {\tt fBE} treatment in our scans.

\item \textbf{Spin and coupling systematics.} For a scalar mediator $S$:
scalar and vector DM annihilations are s-wave (with helicity suppression
$\propto m_f^2$), whereas Dirac DM is p-wave and hence weakly constrained by indirect detection today
but strongly by SI direct detection. For a vector mediator $Z'$:
Dirac DM with \emph{pure vector} couplings is $s$-wave and SI, leading to very strong direct detection constraints; \emph{pure axial} couplings yield SI scattering and helicity-/velocity-suppressed annihilation, relaxing direct detection but still confining viable points near resonance.
For a \emph{pseudoscalar} mediator coupled to Dirac DM, tree-level scattering is momentum-suppressed ($\propto q^4$) and SI, while loop-induced SI scattering sets the leading
direct detection bounds for $m_{\rm med}\!\gtrsim\!1$~GeV and $\mathcal{O}(1)$ couplings; annihilation is $s$-wave, so indirect detection can be important (Sec.~\ref{sec:spin0}).

\item \textbf{Global constraints.} In scans where we fix $g_{\rm{DM}}$ and vary masses, or fix $m_{\rm med}$ and vary $(g_{\rm{DM}},m_{\rm DM})$, the interplay of relic density, direct and indirect detection select narrow bands clustered around the resonance with typical couplings $g\sim 10^{-2}$–$10^{-1}$ (model dependent). Outside these bands, either direct detection excludes
(overabundance with too small $g_{\rm{DM}}$ or too large SI/SD cross sections with large $g_{\rm{DM}}$), or the
relic density is not reproduced.

\item \textbf{GCE interpretation.} In the scalar–mediator \emph{scalar}–DM model, the GCE can be fit with $m_\chi \sim 45$–$65$~GeV and $\langle\sigma v\rangle \sim (1$–$2)\times 10^{-26}\,\mathrm{cm^3\,s^{-1}}$ (Fig.~\ref{fig:GCEbestfit}), values compatible with a thermal relic once a plausible $\mathcal{O}(2)$ $J$-factor uncertainty is included. When combined with direct detection and relic density,
the allowed configurations collapse to a near-resonant strip $m_S \simeq 2m_\chi$ with $g\lesssim 10^{-2}$ (Fig.~\ref{fig:GCEspin0dmg}). Models that are $p$-wave dominated at late times (e.g., scalar mediator + Dirac DM; vector mediator + scalar DM) are disfavored as GCE explanations under the same constraints. For a $Z'$ with Dirac DM, SI limits render the
\emph{pure vector} case highly constrained; \emph{pure axial} couplings leave somewhat more room but still prefer the resonant corridor (Fig.~\ref{fig:GCEspin1dmg}).

\item \textbf{Resonant DM models} We also outlined a simple mechanism applied to two UV complete BSMs that naturally favors the mass relation $m_{\rm med}\!\simeq\!2\,m_{\rm DM}$, lending theoretical support to the phenomenologically selected resonant regime. While details are model dependent, such a relation can emerge from dark-sector dynamics or symmetry-breaking patterns that correlate mediator and DM masses.

\item \textbf{Prospects.} Portions of the resonant corridor remain below current SI and SD sensitivities and can persist even as experiments approach the neutrino floor.
Improved $\gamma$-ray systematics (IEM modeling, extended ROI comparisons), deeper dSph limits, and refined nuclear inputs for direct detection will be crucial to further test these scenarios.
In particular, if future experiments detect WIMP DM consistent with a {\tt DMSimps} \emph{resonant} scenario, the mediator mass would be tied to the DM mass as $m_{\rm med}\simeq 2\,m_{\rm DM}$ (up to corrections set by the mediator width).

\end{itemize}


\medskip

\begin{acknowledgments} 
M.D.M. acknowledges support from the research grant {\sl TAsP (Theoretical Astroparticle Physics)} funded by Istituto Nazionale di Fisica Nucleare (INFN) and from the Italian Ministry of University and Research (MUR), PRIN 2022 ``EXSKALIBUR – Euclid-Cross-SKA: Likelihood Inference Building for Universe’s Research'', Grant No. 20222BBYB9, CUP I53D23000610 0006, and from the European Union -- Next Generation EU.
\end{acknowledgments}

\bibliography{paper}

\begin{thebibliography}{86}%
\makeatletter
\providecommand \@ifxundefined [1]{%
 \@ifx{#1\undefined}
}%
\providecommand \@ifnum [1]{%
 \ifnum #1\expandafter \@firstoftwo
 \else \expandafter \@secondoftwo
 \fi
}%
\providecommand \@ifx [1]{%
 \ifx #1\expandafter \@firstoftwo
 \else \expandafter \@secondoftwo
 \fi
}%
\providecommand \natexlab [1]{#1}%
\providecommand \enquote  [1]{``#1''}%
\providecommand \bibnamefont  [1]{#1}%
\providecommand \bibfnamefont [1]{#1}%
\providecommand \citenamefont [1]{#1}%
\providecommand \href@noop [0]{\@secondoftwo}%
\providecommand \href [0]{\begingroup \@sanitize@url \@href}%
\providecommand \@href[1]{\@@startlink{#1}\@@href}%
\providecommand \@@href[1]{\endgroup#1\@@endlink}%
\providecommand \@sanitize@url [0]{\catcode `\\12\catcode `\$12\catcode
  `\&12\catcode `\#12\catcode `\^12\catcode `\_12\catcode `\%12\relax}%
\providecommand \@@startlink[1]{}%
\providecommand \@@endlink[0]{}%
\providecommand \url  [0]{\begingroup\@sanitize@url \@url }%
\providecommand \@url [1]{\endgroup\@href {#1}{\urlprefix }}%
\providecommand \urlprefix  [0]{URL }%
\providecommand \Eprint [0]{\href }%
\providecommand \doibase [0]{http://dx.doi.org/}%
\providecommand \selectlanguage [0]{\@gobble}%
\providecommand \bibinfo  [0]{\@secondoftwo}%
\providecommand \bibfield  [0]{\@secondoftwo}%
\providecommand \translation [1]{[#1]}%
\providecommand \BibitemOpen [0]{}%
\providecommand \bibitemStop [0]{}%
\providecommand \bibitemNoStop [0]{.\EOS\space}%
\providecommand \EOS [0]{\spacefactor3000\relax}%
\providecommand \BibitemShut  [1]{\csname bibitem#1\endcsname}%
\let\auto@bib@innerbib\@empty
\bibitem [{\citenamefont {Bertone}\ and\ \citenamefont
  {Hooper}(2018)}]{Bertone:2016nfn}%
  \BibitemOpen
  \bibfield  {author} {\bibinfo {author} {\bibfnamefont {G.}~\bibnamefont
  {Bertone}}\ and\ \bibinfo {author} {\bibfnamefont {D.}~\bibnamefont
  {Hooper}},\ }\href {\doibase 10.1103/RevModPhys.90.045002} {\bibfield
  {journal} {\bibinfo  {journal} {Rev. Mod. Phys.}\ }\textbf {\bibinfo {volume}
  {90}},\ \bibinfo {pages} {045002} (\bibinfo {year} {2018})},\ \Eprint
  {http://arxiv.org/abs/1605.04909} {arXiv:1605.04909 [astro-ph.CO]}
  \BibitemShut {NoStop}%
\bibitem [{\citenamefont {Silk}\ \emph {et~al.}(2010)\citenamefont {Silk} \emph
  {et~al.}}]{Bertone:2010zza}%
  \BibitemOpen
  \bibfield  {author} {\bibinfo {author} {\bibfnamefont {J.}~\bibnamefont
  {Silk}} \emph {et~al.},\ }\href {\doibase 10.1017/CBO9780511770739} {\emph
  {\bibinfo {title} {{Particle Dark Matter: Observations, Models and
  Searches}}}},\ edited by\ \bibinfo {editor} {\bibfnamefont {G.}~\bibnamefont
  {Bertone}}\ (\bibinfo  {publisher} {Cambridge Univ. Press},\ \bibinfo
  {address} {Cambridge},\ \bibinfo {year} {2010})\BibitemShut {NoStop}%
\bibitem [{\citenamefont {Cirelli}\ \emph {et~al.}(2024)\citenamefont
  {Cirelli}, \citenamefont {Strumia},\ and\ \citenamefont
  {Zupan}}]{Cirelli:2024ssz}%
  \BibitemOpen
  \bibfield  {author} {\bibinfo {author} {\bibfnamefont {M.}~\bibnamefont
  {Cirelli}}, \bibinfo {author} {\bibfnamefont {A.}~\bibnamefont {Strumia}}, \
  and\ \bibinfo {author} {\bibfnamefont {J.}~\bibnamefont {Zupan}},\
  }\href@noop {} {\  (\bibinfo {year} {2024})},\ \Eprint
  {http://arxiv.org/abs/2406.01705} {arXiv:2406.01705 [hep-ph]} \BibitemShut
  {NoStop}%
\bibitem [{\citenamefont {Aghanim}\ \emph {et~al.}(2018)\citenamefont {Aghanim}
  \emph {et~al.}}]{Aghanim:2018eyx}%
  \BibitemOpen
  \bibfield  {author} {\bibinfo {author} {\bibfnamefont {N.}~\bibnamefont
  {Aghanim}} \emph {et~al.} (\bibinfo {collaboration} {Planck}),\ }\href@noop
  {} {\  (\bibinfo {year} {2018})},\ \Eprint {http://arxiv.org/abs/1807.06209}
  {arXiv:1807.06209 [astro-ph.CO]} \BibitemShut {NoStop}%
\bibitem [{\citenamefont {Schumann}(2019)}]{Schumann:2019eaa}%
  \BibitemOpen
  \bibfield  {author} {\bibinfo {author} {\bibfnamefont {M.}~\bibnamefont
  {Schumann}},\ }\href {\doibase 10.1088/1361-6471/ab2ea5} {\bibfield
  {journal} {\bibinfo  {journal} {J. Phys. G}\ }\textbf {\bibinfo {volume}
  {46}},\ \bibinfo {pages} {103003} (\bibinfo {year} {2019})},\ \Eprint
  {http://arxiv.org/abs/1903.03026} {arXiv:1903.03026 [astro-ph.CO]}
  \BibitemShut {NoStop}%
\bibitem [{\citenamefont {Boveia}\ and\ \citenamefont
  {Doglioni}(2018)}]{Boveia:2018yeb}%
  \BibitemOpen
  \bibfield  {author} {\bibinfo {author} {\bibfnamefont {A.}~\bibnamefont
  {Boveia}}\ and\ \bibinfo {author} {\bibfnamefont {C.}~\bibnamefont
  {Doglioni}},\ }\href {\doibase 10.1146/annurev-nucl-101917-021008} {\bibfield
   {journal} {\bibinfo  {journal} {Ann. Rev. Nucl. Part. Sci.}\ }\textbf
  {\bibinfo {volume} {68}},\ \bibinfo {pages} {429} (\bibinfo {year} {2018})},\
  \Eprint {http://arxiv.org/abs/1810.12238} {arXiv:1810.12238 [hep-ex]}
  \BibitemShut {NoStop}%
\bibitem [{\citenamefont {Gaskins}(2016)}]{Gaskins:2016cha}%
  \BibitemOpen
  \bibfield  {author} {\bibinfo {author} {\bibfnamefont {J.~M.}\ \bibnamefont
  {Gaskins}},\ }\href {\doibase 10.1080/00107514.2016.1175160} {\bibfield
  {journal} {\bibinfo  {journal} {Contemp. Phys.}\ }\textbf {\bibinfo {volume}
  {57}},\ \bibinfo {pages} {496} (\bibinfo {year} {2016})},\ \Eprint
  {http://arxiv.org/abs/1604.00014} {arXiv:1604.00014 [astro-ph.HE]}
  \BibitemShut {NoStop}%
\bibitem [{\citenamefont {Goodenough}\ and\ \citenamefont
  {Hooper}(2009)}]{Goodenough:2009gk}%
  \BibitemOpen
  \bibfield  {author} {\bibinfo {author} {\bibfnamefont {L.}~\bibnamefont
  {Goodenough}}\ and\ \bibinfo {author} {\bibfnamefont {D.}~\bibnamefont
  {Hooper}},\ }\href@noop {} {\  (\bibinfo {year} {2009})},\ \Eprint
  {http://arxiv.org/abs/0910.2998} {arXiv:0910.2998 [hep-ph]} \BibitemShut
  {NoStop}%
\bibitem [{\citenamefont {Hooper}\ and\ \citenamefont
  {Goodenough}(2011)}]{Hooper:2010mq}%
  \BibitemOpen
  \bibfield  {author} {\bibinfo {author} {\bibfnamefont {D.}~\bibnamefont
  {Hooper}}\ and\ \bibinfo {author} {\bibfnamefont {L.}~\bibnamefont
  {Goodenough}},\ }\href {\doibase 10.1016/j.physletb.2011.02.029} {\bibfield
  {journal} {\bibinfo  {journal} {Phys. Lett.}\ }\textbf {\bibinfo {volume}
  {B697}},\ \bibinfo {pages} {412} (\bibinfo {year} {2011})},\ \Eprint
  {http://arxiv.org/abs/1010.2752} {arXiv:1010.2752 [hep-ph]} \BibitemShut
  {NoStop}%
\bibitem [{\citenamefont {Boyarsky}\ \emph {et~al.}(2011)\citenamefont
  {Boyarsky}, \citenamefont {Malyshev},\ and\ \citenamefont
  {Ruchayskiy}}]{Boyarsky:2010dr}%
  \BibitemOpen
  \bibfield  {author} {\bibinfo {author} {\bibfnamefont {A.}~\bibnamefont
  {Boyarsky}}, \bibinfo {author} {\bibfnamefont {D.}~\bibnamefont {Malyshev}},
  \ and\ \bibinfo {author} {\bibfnamefont {O.}~\bibnamefont {Ruchayskiy}},\
  }\href {\doibase 10.1016/j.physletb.2011.10.014} {\bibfield  {journal}
  {\bibinfo  {journal} {Phys. Lett.}\ }\textbf {\bibinfo {volume} {B705}},\
  \bibinfo {pages} {165} (\bibinfo {year} {2011})},\ \Eprint
  {http://arxiv.org/abs/1012.5839} {arXiv:1012.5839 [hep-ph]} \BibitemShut
  {NoStop}%
\bibitem [{\citenamefont {Hooper}\ and\ \citenamefont
  {Linden}(2011)}]{Hooper:2011ti}%
  \BibitemOpen
  \bibfield  {author} {\bibinfo {author} {\bibfnamefont {D.}~\bibnamefont
  {Hooper}}\ and\ \bibinfo {author} {\bibfnamefont {T.}~\bibnamefont
  {Linden}},\ }\href {\doibase 10.1103/PhysRevD.84.123005} {\bibfield
  {journal} {\bibinfo  {journal} {Phys. Rev.}\ }\textbf {\bibinfo {volume}
  {D84}},\ \bibinfo {pages} {123005} (\bibinfo {year} {2011})},\ \Eprint
  {http://arxiv.org/abs/1110.0006} {arXiv:1110.0006 [astro-ph.HE]} \BibitemShut
  {NoStop}%
\bibitem [{\citenamefont {Abazajian}\ and\ \citenamefont
  {Kaplinghat}(2012)}]{Abazajian:2012pn}%
  \BibitemOpen
  \bibfield  {author} {\bibinfo {author} {\bibfnamefont {K.~N.}\ \bibnamefont
  {Abazajian}}\ and\ \bibinfo {author} {\bibfnamefont {M.}~\bibnamefont
  {Kaplinghat}},\ }\href {\doibase 10.1103/PhysRevD.86.083511,
  10.1103/PhysRevD.87.129902} {\bibfield  {journal} {\bibinfo  {journal} {Phys.
  Rev.}\ }\textbf {\bibinfo {volume} {D86}},\ \bibinfo {pages} {083511}
  (\bibinfo {year} {2012})},\ \bibinfo {note} {[Erratum: Phys.
  Rev.D87,129902(2013)]},\ \Eprint {http://arxiv.org/abs/1207.6047}
  {arXiv:1207.6047 [astro-ph.HE]} \BibitemShut {NoStop}%
\bibitem [{\citenamefont {Gordon}\ and\ \citenamefont
  {Macias}(2013)}]{Gordon:2013vta}%
  \BibitemOpen
  \bibfield  {author} {\bibinfo {author} {\bibfnamefont {C.}~\bibnamefont
  {Gordon}}\ and\ \bibinfo {author} {\bibfnamefont {O.}~\bibnamefont
  {Macias}},\ }\href {\doibase 10.1103/PhysRevD.88.083521,
  10.1103/PhysRevD.89.049901} {\bibfield  {journal} {\bibinfo  {journal} {Phys.
  Rev.}\ }\textbf {\bibinfo {volume} {D88}},\ \bibinfo {pages} {083521}
  (\bibinfo {year} {2013})},\ \bibinfo {note} {[Erratum: Phys.
  Rev.D89,no.4,049901(2014)]},\ \Eprint {http://arxiv.org/abs/1306.5725}
  {arXiv:1306.5725 [astro-ph.HE]} \BibitemShut {NoStop}%
\bibitem [{\citenamefont {Abazajian}\ \emph {et~al.}(2014)\citenamefont
  {Abazajian}, \citenamefont {Canac}, \citenamefont {Horiuchi},\ and\
  \citenamefont {Kaplinghat}}]{Abazajian:2014fta}%
  \BibitemOpen
  \bibfield  {author} {\bibinfo {author} {\bibfnamefont {K.~N.}\ \bibnamefont
  {Abazajian}}, \bibinfo {author} {\bibfnamefont {N.}~\bibnamefont {Canac}},
  \bibinfo {author} {\bibfnamefont {S.}~\bibnamefont {Horiuchi}}, \ and\
  \bibinfo {author} {\bibfnamefont {M.}~\bibnamefont {Kaplinghat}},\ }\href
  {\doibase 10.1103/PhysRevD.90.023526} {\bibfield  {journal} {\bibinfo
  {journal} {Phys. Rev.}\ }\textbf {\bibinfo {volume} {D90}},\ \bibinfo {pages}
  {023526} (\bibinfo {year} {2014})},\ \Eprint {http://arxiv.org/abs/1402.4090}
  {arXiv:1402.4090 [astro-ph.HE]} \BibitemShut {NoStop}%
\bibitem [{\citenamefont {Daylan}\ \emph {et~al.}(2016)\citenamefont {Daylan},
  \citenamefont {Finkbeiner}, \citenamefont {Hooper}, \citenamefont {Linden},
  \citenamefont {Portillo}, \citenamefont {Rodd},\ and\ \citenamefont
  {Slatyer}}]{Daylan:2014rsa}%
  \BibitemOpen
  \bibfield  {author} {\bibinfo {author} {\bibfnamefont {T.}~\bibnamefont
  {Daylan}}, \bibinfo {author} {\bibfnamefont {D.~P.}\ \bibnamefont
  {Finkbeiner}}, \bibinfo {author} {\bibfnamefont {D.}~\bibnamefont {Hooper}},
  \bibinfo {author} {\bibfnamefont {T.}~\bibnamefont {Linden}}, \bibinfo
  {author} {\bibfnamefont {S.~K.~N.}\ \bibnamefont {Portillo}}, \bibinfo
  {author} {\bibfnamefont {N.~L.}\ \bibnamefont {Rodd}}, \ and\ \bibinfo
  {author} {\bibfnamefont {T.~R.}\ \bibnamefont {Slatyer}},\ }\href {\doibase
  10.1016/j.dark.2015.12.005} {\bibfield  {journal} {\bibinfo  {journal} {Phys.
  Dark Univ.}\ }\textbf {\bibinfo {volume} {12}},\ \bibinfo {pages} {1}
  (\bibinfo {year} {2016})},\ \Eprint {http://arxiv.org/abs/1402.6703}
  {arXiv:1402.6703 [astro-ph.HE]} \BibitemShut {NoStop}%
\bibitem [{\citenamefont {Calore}\ \emph
  {et~al.}(2015{\natexlab{a}})\citenamefont {Calore}, \citenamefont {Cholis},
  \citenamefont {McCabe},\ and\ \citenamefont {Weniger}}]{Calore:2014nla}%
  \BibitemOpen
  \bibfield  {author} {\bibinfo {author} {\bibfnamefont {F.}~\bibnamefont
  {Calore}}, \bibinfo {author} {\bibfnamefont {I.}~\bibnamefont {Cholis}},
  \bibinfo {author} {\bibfnamefont {C.}~\bibnamefont {McCabe}}, \ and\ \bibinfo
  {author} {\bibfnamefont {C.}~\bibnamefont {Weniger}},\ }\href {\doibase
  10.1103/PhysRevD.91.063003} {\bibfield  {journal} {\bibinfo  {journal} {Phys.
  Rev. D}\ }\textbf {\bibinfo {volume} {91}},\ \bibinfo {pages} {063003}
  (\bibinfo {year} {2015}{\natexlab{a}})},\ \Eprint
  {http://arxiv.org/abs/1411.4647} {arXiv:1411.4647 [hep-ph]} \BibitemShut
  {NoStop}%
\bibitem [{\citenamefont {Calore}\ \emph
  {et~al.}(2015{\natexlab{b}})\citenamefont {Calore}, \citenamefont {Cholis},\
  and\ \citenamefont {Weniger}}]{Calore:2014xka}%
  \BibitemOpen
  \bibfield  {author} {\bibinfo {author} {\bibfnamefont {F.}~\bibnamefont
  {Calore}}, \bibinfo {author} {\bibfnamefont {I.}~\bibnamefont {Cholis}}, \
  and\ \bibinfo {author} {\bibfnamefont {C.}~\bibnamefont {Weniger}},\ }\href
  {\doibase 10.1088/1475-7516/2015/03/038} {\bibfield  {journal} {\bibinfo
  {journal} {JCAP}\ }\textbf {\bibinfo {volume} {1503}},\ \bibinfo {pages}
  {038} (\bibinfo {year} {2015}{\natexlab{b}})},\ \Eprint
  {http://arxiv.org/abs/1409.0042} {arXiv:1409.0042 [astro-ph.CO]} \BibitemShut
  {NoStop}%
\bibitem [{\citenamefont {Ajello}\ \emph {et~al.}(2016)\citenamefont {Ajello}
  \emph {et~al.}}]{TheFermi-LAT:2015kwa}%
  \BibitemOpen
  \bibfield  {author} {\bibinfo {author} {\bibfnamefont {M.}~\bibnamefont
  {Ajello}} \emph {et~al.} (\bibinfo {collaboration} {Fermi-LAT}),\ }\href
  {\doibase 10.3847/0004-637X/819/1/44} {\bibfield  {journal} {\bibinfo
  {journal} {Astrophys. J.}\ }\textbf {\bibinfo {volume} {819}},\ \bibinfo
  {pages} {44} (\bibinfo {year} {2016})},\ \Eprint
  {http://arxiv.org/abs/1511.02938} {arXiv:1511.02938 [astro-ph.HE]}
  \BibitemShut {NoStop}%
\bibitem [{\citenamefont {Ackermann}\ \emph {et~al.}(2017)\citenamefont
  {Ackermann} \emph {et~al.}}]{TheFermi-LAT:2017vmf}%
  \BibitemOpen
  \bibfield  {author} {\bibinfo {author} {\bibfnamefont {M.}~\bibnamefont
  {Ackermann}} \emph {et~al.} (\bibinfo {collaboration} {Fermi-LAT}),\ }\href
  {\doibase 10.3847/1538-4357/aa6cab} {\bibfield  {journal} {\bibinfo
  {journal} {Astrophys. J.}\ }\textbf {\bibinfo {volume} {840}},\ \bibinfo
  {pages} {43} (\bibinfo {year} {2017})},\ \Eprint
  {http://arxiv.org/abs/1704.03910} {arXiv:1704.03910 [astro-ph.HE]}
  \BibitemShut {NoStop}%
\bibitem [{\citenamefont {Di~Mauro}\ \emph {et~al.}(2019)\citenamefont
  {Di~Mauro}, \citenamefont {Hou}, \citenamefont {Eckner}, \citenamefont
  {Zaharijas},\ and\ \citenamefont {Charles}}]{DiMauro:2019frs}%
  \BibitemOpen
  \bibfield  {author} {\bibinfo {author} {\bibfnamefont {M.}~\bibnamefont
  {Di~Mauro}}, \bibinfo {author} {\bibfnamefont {X.}~\bibnamefont {Hou}},
  \bibinfo {author} {\bibfnamefont {C.}~\bibnamefont {Eckner}}, \bibinfo
  {author} {\bibfnamefont {G.}~\bibnamefont {Zaharijas}}, \ and\ \bibinfo
  {author} {\bibfnamefont {E.}~\bibnamefont {Charles}},\ }\href {\doibase
  10.1103/PhysRevD.99.123027} {\bibfield  {journal} {\bibinfo  {journal} {Phys.
  Rev. D}\ }\textbf {\bibinfo {volume} {99}},\ \bibinfo {pages} {123027}
  (\bibinfo {year} {2019})},\ \Eprint {http://arxiv.org/abs/1904.10977}
  {arXiv:1904.10977 [astro-ph.HE]} \BibitemShut {NoStop}%
\bibitem [{\citenamefont {Di~Mauro}(2021{\natexlab{a}})}]{DiMauro:2021raz}%
  \BibitemOpen
  \bibfield  {author} {\bibinfo {author} {\bibfnamefont {M.}~\bibnamefont
  {Di~Mauro}},\ }\href {\doibase 10.1103/PhysRevD.103.063029} {\bibfield
  {journal} {\bibinfo  {journal} {Phys. Rev. D}\ }\textbf {\bibinfo {volume}
  {103}},\ \bibinfo {pages} {063029} (\bibinfo {year} {2021}{\natexlab{a}})},\
  \Eprint {http://arxiv.org/abs/2101.04694} {arXiv:2101.04694 [astro-ph.HE]}
  \BibitemShut {NoStop}%
\bibitem [{\citenamefont {Cholis}\ \emph
  {et~al.}(2022{\natexlab{a}})\citenamefont {Cholis}, \citenamefont {Zhong},
  \citenamefont {McDermott},\ and\ \citenamefont
  {Surdutovich}}]{Cholis:2021rpp}%
  \BibitemOpen
  \bibfield  {author} {\bibinfo {author} {\bibfnamefont {I.}~\bibnamefont
  {Cholis}}, \bibinfo {author} {\bibfnamefont {Y.-M.}\ \bibnamefont {Zhong}},
  \bibinfo {author} {\bibfnamefont {S.~D.}\ \bibnamefont {McDermott}}, \ and\
  \bibinfo {author} {\bibfnamefont {J.~P.}\ \bibnamefont {Surdutovich}},\
  }\href {\doibase 10.1103/PhysRevD.105.103023} {\bibfield  {journal} {\bibinfo
   {journal} {Phys. Rev. D}\ }\textbf {\bibinfo {volume} {105}},\ \bibinfo
  {pages} {103023} (\bibinfo {year} {2022}{\natexlab{a}})},\ \Eprint
  {http://arxiv.org/abs/2112.09706} {arXiv:2112.09706 [astro-ph.HE]}
  \BibitemShut {NoStop}%
\bibitem [{\citenamefont {Di~Mauro}\ \emph {et~al.}(2016)\citenamefont
  {Di~Mauro}, \citenamefont {Donato}, \citenamefont {Fornengo},\ and\
  \citenamefont {Vittino}}]{DiMauro:2015jxa}%
  \BibitemOpen
  \bibfield  {author} {\bibinfo {author} {\bibfnamefont {M.}~\bibnamefont
  {Di~Mauro}}, \bibinfo {author} {\bibfnamefont {F.}~\bibnamefont {Donato}},
  \bibinfo {author} {\bibfnamefont {N.}~\bibnamefont {Fornengo}}, \ and\
  \bibinfo {author} {\bibfnamefont {A.}~\bibnamefont {Vittino}},\ }\href
  {\doibase 10.1088/1475-7516/2016/05/031} {\bibfield  {journal} {\bibinfo
  {journal} {JCAP}\ }\textbf {\bibinfo {volume} {05}},\ \bibinfo {pages} {031}
  (\bibinfo {year} {2016})},\ \Eprint {http://arxiv.org/abs/1507.07001}
  {arXiv:1507.07001 [astro-ph.HE]} \BibitemShut {NoStop}%
\bibitem [{\citenamefont {Charles}\ \emph {et~al.}(2016)\citenamefont {Charles}
  \emph {et~al.}}]{Fermi-LAT:2016afa}%
  \BibitemOpen
  \bibfield  {author} {\bibinfo {author} {\bibfnamefont {E.}~\bibnamefont
  {Charles}} \emph {et~al.} (\bibinfo {collaboration} {Fermi-LAT}),\ }\href
  {\doibase 10.1016/j.physrep.2016.05.001} {\bibfield  {journal} {\bibinfo
  {journal} {Phys. Rept.}\ }\textbf {\bibinfo {volume} {636}},\ \bibinfo
  {pages} {1} (\bibinfo {year} {2016})},\ \Eprint
  {http://arxiv.org/abs/1605.02016} {arXiv:1605.02016 [astro-ph.HE]}
  \BibitemShut {NoStop}%
\bibitem [{\citenamefont {Abercrombie}\ \emph {et~al.}(2020)\citenamefont
  {Abercrombie}, \citenamefont {Akchurin}, \citenamefont {Akilli},
  \citenamefont {Maestre}, \citenamefont {Allen}, \citenamefont {Gonzalez},
  \citenamefont {Andrea}, \citenamefont {Arbey}, \citenamefont {Azuelos},
  \citenamefont {Azzi}, \citenamefont {Backović}, \citenamefont {Bai},
  \citenamefont {Banerjee}, \citenamefont {Beacham}, \citenamefont {Belyaev},
  \citenamefont {Boveia}, \citenamefont {Brennan}, \citenamefont {Buchmueller},
  \citenamefont {Buckley}, \citenamefont {Busoni}, \citenamefont {Buttignol},
  \citenamefont {Cacciapaglia}, \citenamefont {Caputo}, \citenamefont
  {Carpenter}, \citenamefont {Castro}, \citenamefont {Ceballos}, \citenamefont
  {Cheng}, \citenamefont {Chou}, \citenamefont {Gonzalez}, \citenamefont
  {Cowden}, \citenamefont {D’Eramo}, \citenamefont {{De Cosa}}, \citenamefont
  {{De Gruttola}}, \citenamefont {{De Roeck}}, \citenamefont {{De Simone}},
  \citenamefont {Deandrea}, \citenamefont {Demiragli}, \citenamefont
  {DiFranzo}, \citenamefont {Doglioni}, \citenamefont {{du Pree}},
  \citenamefont {Erbacher}, \citenamefont {Erdmann}, \citenamefont {Fischer},
  \citenamefont {Flaecher}, \citenamefont {Fox}, \citenamefont {Fuks},
  \citenamefont {Genest}, \citenamefont {Gomber}, \citenamefont {Goudelis},
  \citenamefont {Gramling}, \citenamefont {Gunion}, \citenamefont {Hahn},
  \citenamefont {Haisch}, \citenamefont {Harnik}, \citenamefont {Harris},
  \citenamefont {Hoepfner}, \citenamefont {Hoh}, \citenamefont {Hsu},
  \citenamefont {Hsu}, \citenamefont {Iiyama}, \citenamefont {Ippolito},
  \citenamefont {Jacques}, \citenamefont {Ju}, \citenamefont {Kahlhoefer},
  \citenamefont {Kalogeropoulos}, \citenamefont {Kaplan}, \citenamefont
  {Kashif}, \citenamefont {Khoze}, \citenamefont {Khurana}, \citenamefont
  {Kotov}, \citenamefont {Kovalskyi}, \citenamefont {Kulkarni}, \citenamefont
  {Kunori}, \citenamefont {Kutzner}, \citenamefont {Lee}, \citenamefont {Lee},
  \citenamefont {Liew}, \citenamefont {Lin}, \citenamefont {Lowette},
  \citenamefont {Madar}, \citenamefont {Malik}, \citenamefont {Maltoni},
  \citenamefont {Perez}, \citenamefont {Mattelaer}, \citenamefont {Mawatari},
  \citenamefont {McCabe}, \citenamefont {Megy}, \citenamefont {Morgante},
  \citenamefont {Mrenna}, \citenamefont {Moon}, \citenamefont {Narayanan},
  \citenamefont {Nelson}, \citenamefont {Novaes}, \citenamefont {Padeken},
  \citenamefont {Pani}, \citenamefont {Papucci}, \citenamefont {Paulini},
  \citenamefont {Paus}, \citenamefont {Pazzini}, \citenamefont {Penning},
  \citenamefont {Peskin}, \citenamefont {Pinna}, \citenamefont {Procura},
  \citenamefont {Qazi}, \citenamefont {Racco}, \citenamefont {Re},
  \citenamefont {Riotto}, \citenamefont {Rizzo}, \citenamefont {Roehrig},
  \citenamefont {Salek}, \citenamefont {Pineda}, \citenamefont {Sarkar},
  \citenamefont {Schmidt}, \citenamefont {Schramm}, \citenamefont {Shepherd},
  \citenamefont {Singh}, \citenamefont {Soffi}, \citenamefont {Srimanobhas},
  \citenamefont {Sung}, \citenamefont {Tait}, \citenamefont
  {Theveneaux-Pelzer}, \citenamefont {Thomas}, \citenamefont {Tosi},
  \citenamefont {Trocino}, \citenamefont {Undleeb}, \citenamefont {Vichi},
  \citenamefont {Wang}, \citenamefont {Wang}, \citenamefont {Wang},
  \citenamefont {Whallon}, \citenamefont {Worm}, \citenamefont {Wu},
  \citenamefont {Wu}, \citenamefont {Yang}, \citenamefont {Yang}, \citenamefont
  {Yu}, \citenamefont {Zaldivar}, \citenamefont {Zanetti}, \citenamefont
  {Zhang},\ and\ \citenamefont {Zucchetta}}]{ABERCROMBIE2020100371}%
  \BibitemOpen
  \bibfield  {author} {\bibinfo {author} {\bibfnamefont {D.}~\bibnamefont
  {Abercrombie}}, \bibinfo {author} {\bibfnamefont {N.}~\bibnamefont
  {Akchurin}}, \bibinfo {author} {\bibfnamefont {E.}~\bibnamefont {Akilli}},
  \bibinfo {author} {\bibfnamefont {J.~A.}\ \bibnamefont {Maestre}}, \bibinfo
  {author} {\bibfnamefont {B.}~\bibnamefont {Allen}}, \bibinfo {author}
  {\bibfnamefont {B.~A.}\ \bibnamefont {Gonzalez}}, \bibinfo {author}
  {\bibfnamefont {J.}~\bibnamefont {Andrea}}, \bibinfo {author} {\bibfnamefont
  {A.}~\bibnamefont {Arbey}}, \bibinfo {author} {\bibfnamefont
  {G.}~\bibnamefont {Azuelos}}, \bibinfo {author} {\bibfnamefont
  {P.}~\bibnamefont {Azzi}}, \bibinfo {author} {\bibfnamefont {M.}~\bibnamefont
  {Backović}}, \bibinfo {author} {\bibfnamefont {Y.}~\bibnamefont {Bai}},
  \bibinfo {author} {\bibfnamefont {S.}~\bibnamefont {Banerjee}}, \bibinfo
  {author} {\bibfnamefont {J.}~\bibnamefont {Beacham}}, \bibinfo {author}
  {\bibfnamefont {A.}~\bibnamefont {Belyaev}}, \bibinfo {author} {\bibfnamefont
  {A.}~\bibnamefont {Boveia}}, \bibinfo {author} {\bibfnamefont {A.~J.}\
  \bibnamefont {Brennan}}, \bibinfo {author} {\bibfnamefont {O.}~\bibnamefont
  {Buchmueller}}, \bibinfo {author} {\bibfnamefont {M.~R.}\ \bibnamefont
  {Buckley}}, \bibinfo {author} {\bibfnamefont {G.}~\bibnamefont {Busoni}},
  \bibinfo {author} {\bibfnamefont {M.}~\bibnamefont {Buttignol}}, \bibinfo
  {author} {\bibfnamefont {G.}~\bibnamefont {Cacciapaglia}}, \bibinfo {author}
  {\bibfnamefont {R.}~\bibnamefont {Caputo}}, \bibinfo {author} {\bibfnamefont
  {L.}~\bibnamefont {Carpenter}}, \bibinfo {author} {\bibfnamefont {N.~F.}\
  \bibnamefont {Castro}}, \bibinfo {author} {\bibfnamefont {G.~G.}\
  \bibnamefont {Ceballos}}, \bibinfo {author} {\bibfnamefont {Y.}~\bibnamefont
  {Cheng}}, \bibinfo {author} {\bibfnamefont {J.~P.}\ \bibnamefont {Chou}},
  \bibinfo {author} {\bibfnamefont {A.~C.}\ \bibnamefont {Gonzalez}}, \bibinfo
  {author} {\bibfnamefont {C.}~\bibnamefont {Cowden}}, \bibinfo {author}
  {\bibfnamefont {F.}~\bibnamefont {D’Eramo}}, \bibinfo {author}
  {\bibfnamefont {A.}~\bibnamefont {{De Cosa}}}, \bibinfo {author}
  {\bibfnamefont {M.}~\bibnamefont {{De Gruttola}}}, \bibinfo {author}
  {\bibfnamefont {A.}~\bibnamefont {{De Roeck}}}, \bibinfo {author}
  {\bibfnamefont {A.}~\bibnamefont {{De Simone}}}, \bibinfo {author}
  {\bibfnamefont {A.}~\bibnamefont {Deandrea}}, \bibinfo {author}
  {\bibfnamefont {Z.}~\bibnamefont {Demiragli}}, \bibinfo {author}
  {\bibfnamefont {A.}~\bibnamefont {DiFranzo}}, \bibinfo {author}
  {\bibfnamefont {C.}~\bibnamefont {Doglioni}}, \bibinfo {author}
  {\bibfnamefont {T.}~\bibnamefont {{du Pree}}}, \bibinfo {author}
  {\bibfnamefont {R.}~\bibnamefont {Erbacher}}, \bibinfo {author}
  {\bibfnamefont {J.}~\bibnamefont {Erdmann}}, \bibinfo {author} {\bibfnamefont
  {C.}~\bibnamefont {Fischer}}, \bibinfo {author} {\bibfnamefont
  {H.}~\bibnamefont {Flaecher}}, \bibinfo {author} {\bibfnamefont {P.~J.}\
  \bibnamefont {Fox}}, \bibinfo {author} {\bibfnamefont {B.}~\bibnamefont
  {Fuks}}, \bibinfo {author} {\bibfnamefont {M.-H.}\ \bibnamefont {Genest}},
  \bibinfo {author} {\bibfnamefont {B.}~\bibnamefont {Gomber}}, \bibinfo
  {author} {\bibfnamefont {A.}~\bibnamefont {Goudelis}}, \bibinfo {author}
  {\bibfnamefont {J.}~\bibnamefont {Gramling}}, \bibinfo {author}
  {\bibfnamefont {J.}~\bibnamefont {Gunion}}, \bibinfo {author} {\bibfnamefont
  {K.}~\bibnamefont {Hahn}}, \bibinfo {author} {\bibfnamefont {U.}~\bibnamefont
  {Haisch}}, \bibinfo {author} {\bibfnamefont {R.}~\bibnamefont {Harnik}},
  \bibinfo {author} {\bibfnamefont {P.~C.}\ \bibnamefont {Harris}}, \bibinfo
  {author} {\bibfnamefont {K.}~\bibnamefont {Hoepfner}}, \bibinfo {author}
  {\bibfnamefont {S.~Y.}\ \bibnamefont {Hoh}}, \bibinfo {author} {\bibfnamefont
  {D.~G.}\ \bibnamefont {Hsu}}, \bibinfo {author} {\bibfnamefont {S.-C.}\
  \bibnamefont {Hsu}}, \bibinfo {author} {\bibfnamefont {Y.}~\bibnamefont
  {Iiyama}}, \bibinfo {author} {\bibfnamefont {V.}~\bibnamefont {Ippolito}},
  \bibinfo {author} {\bibfnamefont {T.}~\bibnamefont {Jacques}}, \bibinfo
  {author} {\bibfnamefont {X.}~\bibnamefont {Ju}}, \bibinfo {author}
  {\bibfnamefont {F.}~\bibnamefont {Kahlhoefer}}, \bibinfo {author}
  {\bibfnamefont {A.}~\bibnamefont {Kalogeropoulos}}, \bibinfo {author}
  {\bibfnamefont {L.~S.}\ \bibnamefont {Kaplan}}, \bibinfo {author}
  {\bibfnamefont {L.}~\bibnamefont {Kashif}}, \bibinfo {author} {\bibfnamefont
  {V.~V.}\ \bibnamefont {Khoze}}, \bibinfo {author} {\bibfnamefont
  {R.}~\bibnamefont {Khurana}}, \bibinfo {author} {\bibfnamefont
  {K.}~\bibnamefont {Kotov}}, \bibinfo {author} {\bibfnamefont
  {D.}~\bibnamefont {Kovalskyi}}, \bibinfo {author} {\bibfnamefont
  {S.}~\bibnamefont {Kulkarni}}, \bibinfo {author} {\bibfnamefont
  {S.}~\bibnamefont {Kunori}}, \bibinfo {author} {\bibfnamefont
  {V.}~\bibnamefont {Kutzner}}, \bibinfo {author} {\bibfnamefont {H.~M.}\
  \bibnamefont {Lee}}, \bibinfo {author} {\bibfnamefont {S.-W.}\ \bibnamefont
  {Lee}}, \bibinfo {author} {\bibfnamefont {S.~P.}\ \bibnamefont {Liew}},
  \bibinfo {author} {\bibfnamefont {T.}~\bibnamefont {Lin}}, \bibinfo {author}
  {\bibfnamefont {S.}~\bibnamefont {Lowette}}, \bibinfo {author} {\bibfnamefont
  {R.}~\bibnamefont {Madar}}, \bibinfo {author} {\bibfnamefont
  {S.}~\bibnamefont {Malik}}, \bibinfo {author} {\bibfnamefont
  {F.}~\bibnamefont {Maltoni}}, \bibinfo {author} {\bibfnamefont {M.~M.}\
  \bibnamefont {Perez}}, \bibinfo {author} {\bibfnamefont {O.}~\bibnamefont
  {Mattelaer}}, \bibinfo {author} {\bibfnamefont {K.}~\bibnamefont {Mawatari}},
  \bibinfo {author} {\bibfnamefont {C.}~\bibnamefont {McCabe}}, \bibinfo
  {author} {\bibfnamefont {T.}~\bibnamefont {Megy}}, \bibinfo {author}
  {\bibfnamefont {E.}~\bibnamefont {Morgante}}, \bibinfo {author}
  {\bibfnamefont {S.}~\bibnamefont {Mrenna}}, \bibinfo {author} {\bibfnamefont
  {C.-S.}\ \bibnamefont {Moon}}, \bibinfo {author} {\bibfnamefont {S.~M.}\
  \bibnamefont {Narayanan}}, \bibinfo {author} {\bibfnamefont {A.}~\bibnamefont
  {Nelson}}, \bibinfo {author} {\bibfnamefont {S.~F.}\ \bibnamefont {Novaes}},
  \bibinfo {author} {\bibfnamefont {K.~O.}\ \bibnamefont {Padeken}}, \bibinfo
  {author} {\bibfnamefont {P.}~\bibnamefont {Pani}}, \bibinfo {author}
  {\bibfnamefont {M.}~\bibnamefont {Papucci}}, \bibinfo {author} {\bibfnamefont
  {M.}~\bibnamefont {Paulini}}, \bibinfo {author} {\bibfnamefont
  {C.}~\bibnamefont {Paus}}, \bibinfo {author} {\bibfnamefont {J.}~\bibnamefont
  {Pazzini}}, \bibinfo {author} {\bibfnamefont {B.}~\bibnamefont {Penning}},
  \bibinfo {author} {\bibfnamefont {M.~E.}\ \bibnamefont {Peskin}}, \bibinfo
  {author} {\bibfnamefont {D.}~\bibnamefont {Pinna}}, \bibinfo {author}
  {\bibfnamefont {M.}~\bibnamefont {Procura}}, \bibinfo {author} {\bibfnamefont
  {S.~F.}\ \bibnamefont {Qazi}}, \bibinfo {author} {\bibfnamefont
  {D.}~\bibnamefont {Racco}}, \bibinfo {author} {\bibfnamefont
  {E.}~\bibnamefont {Re}}, \bibinfo {author} {\bibfnamefont {A.}~\bibnamefont
  {Riotto}}, \bibinfo {author} {\bibfnamefont {T.~G.}\ \bibnamefont {Rizzo}},
  \bibinfo {author} {\bibfnamefont {R.}~\bibnamefont {Roehrig}}, \bibinfo
  {author} {\bibfnamefont {D.}~\bibnamefont {Salek}}, \bibinfo {author}
  {\bibfnamefont {A.~S.}\ \bibnamefont {Pineda}}, \bibinfo {author}
  {\bibfnamefont {S.}~\bibnamefont {Sarkar}}, \bibinfo {author} {\bibfnamefont
  {A.}~\bibnamefont {Schmidt}}, \bibinfo {author} {\bibfnamefont {S.~R.}\
  \bibnamefont {Schramm}}, \bibinfo {author} {\bibfnamefont {W.}~\bibnamefont
  {Shepherd}}, \bibinfo {author} {\bibfnamefont {G.}~\bibnamefont {Singh}},
  \bibinfo {author} {\bibfnamefont {L.}~\bibnamefont {Soffi}}, \bibinfo
  {author} {\bibfnamefont {N.}~\bibnamefont {Srimanobhas}}, \bibinfo {author}
  {\bibfnamefont {K.}~\bibnamefont {Sung}}, \bibinfo {author} {\bibfnamefont
  {T.~M.}\ \bibnamefont {Tait}}, \bibinfo {author} {\bibfnamefont
  {T.}~\bibnamefont {Theveneaux-Pelzer}}, \bibinfo {author} {\bibfnamefont
  {M.}~\bibnamefont {Thomas}}, \bibinfo {author} {\bibfnamefont
  {M.}~\bibnamefont {Tosi}}, \bibinfo {author} {\bibfnamefont {D.}~\bibnamefont
  {Trocino}}, \bibinfo {author} {\bibfnamefont {S.}~\bibnamefont {Undleeb}},
  \bibinfo {author} {\bibfnamefont {A.}~\bibnamefont {Vichi}}, \bibinfo
  {author} {\bibfnamefont {F.}~\bibnamefont {Wang}}, \bibinfo {author}
  {\bibfnamefont {L.-T.}\ \bibnamefont {Wang}}, \bibinfo {author}
  {\bibfnamefont {R.-J.}\ \bibnamefont {Wang}}, \bibinfo {author}
  {\bibfnamefont {N.}~\bibnamefont {Whallon}}, \bibinfo {author} {\bibfnamefont
  {S.}~\bibnamefont {Worm}}, \bibinfo {author} {\bibfnamefont {M.}~\bibnamefont
  {Wu}}, \bibinfo {author} {\bibfnamefont {S.~L.}\ \bibnamefont {Wu}}, \bibinfo
  {author} {\bibfnamefont {H.}~\bibnamefont {Yang}}, \bibinfo {author}
  {\bibfnamefont {Y.}~\bibnamefont {Yang}}, \bibinfo {author} {\bibfnamefont
  {S.-S.}\ \bibnamefont {Yu}}, \bibinfo {author} {\bibfnamefont
  {B.}~\bibnamefont {Zaldivar}}, \bibinfo {author} {\bibfnamefont
  {M.}~\bibnamefont {Zanetti}}, \bibinfo {author} {\bibfnamefont
  {Z.}~\bibnamefont {Zhang}}, \ and\ \bibinfo {author} {\bibfnamefont
  {A.}~\bibnamefont {Zucchetta}},\ }\href {\doibase
  https://doi.org/10.1016/j.dark.2019.100371} {\bibfield  {journal} {\bibinfo
  {journal} {Physics of the Dark Universe}\ }\textbf {\bibinfo {volume} {27}},\
  \bibinfo {pages} {100371} (\bibinfo {year} {2020})}\BibitemShut {NoStop}%
\bibitem [{\citenamefont {Aad}\ \emph {et~al.}(2024)\citenamefont {Aad} \emph
  {et~al.}}]{ATLAS:2024kpy}%
  \BibitemOpen
  \bibfield  {author} {\bibinfo {author} {\bibfnamefont {G.}~\bibnamefont
  {Aad}} \emph {et~al.} (\bibinfo {collaboration} {ATLAS}),\ }\href {\doibase
  10.1140/epjc/s10052-024-13215-5} {\bibfield  {journal} {\bibinfo  {journal}
  {Eur. Phys. J. C}\ }\textbf {\bibinfo {volume} {84}},\ \bibinfo {pages}
  {1102} (\bibinfo {year} {2024})},\ \Eprint {http://arxiv.org/abs/2404.15930}
  {arXiv:2404.15930 [hep-ex]} \BibitemShut {NoStop}%
\bibitem [{\citenamefont {Cline}\ \emph {et~al.}(2013)\citenamefont {Cline},
  \citenamefont {Kainulainen}, \citenamefont {Scott},\ and\ \citenamefont
  {Weniger}}]{Cline:2013gha}%
  \BibitemOpen
  \bibfield  {author} {\bibinfo {author} {\bibfnamefont {J.~M.}\ \bibnamefont
  {Cline}}, \bibinfo {author} {\bibfnamefont {K.}~\bibnamefont {Kainulainen}},
  \bibinfo {author} {\bibfnamefont {P.}~\bibnamefont {Scott}}, \ and\ \bibinfo
  {author} {\bibfnamefont {C.}~\bibnamefont {Weniger}},\ }\href {\doibase
  10.1103/PhysRevD.88.055025} {\bibfield  {journal} {\bibinfo  {journal} {Phys.
  Rev. D}\ }\textbf {\bibinfo {volume} {88}},\ \bibinfo {pages} {055025}
  (\bibinfo {year} {2013})},\ \bibinfo {note} {[Erratum: Phys.Rev.D 92, 039906
  (2015)]},\ \Eprint {http://arxiv.org/abs/1306.4710} {arXiv:1306.4710
  [hep-ph]} \BibitemShut {NoStop}%
\bibitem [{\citenamefont {Arcadi}\ \emph {et~al.}(2015)\citenamefont {Arcadi},
  \citenamefont {Mambrini},\ and\ \citenamefont {Richard}}]{Arcadi:2014lta}%
  \BibitemOpen
  \bibfield  {author} {\bibinfo {author} {\bibfnamefont {G.}~\bibnamefont
  {Arcadi}}, \bibinfo {author} {\bibfnamefont {Y.}~\bibnamefont {Mambrini}}, \
  and\ \bibinfo {author} {\bibfnamefont {F.}~\bibnamefont {Richard}},\ }\href
  {\doibase 10.1088/1475-7516/2015/03/018} {\bibfield  {journal} {\bibinfo
  {journal} {JCAP}\ }\textbf {\bibinfo {volume} {03}},\ \bibinfo {pages} {018}
  (\bibinfo {year} {2015})},\ \Eprint {http://arxiv.org/abs/1411.2985}
  {arXiv:1411.2985 [hep-ph]} \BibitemShut {NoStop}%
\bibitem [{\citenamefont {Beniwal}\ \emph {et~al.}(2016)\citenamefont
  {Beniwal}, \citenamefont {Rajec}, \citenamefont {Savage}, \citenamefont
  {Scott}, \citenamefont {Weniger}, \citenamefont {White},\ and\ \citenamefont
  {Williams}}]{Beniwal:2015sdl}%
  \BibitemOpen
  \bibfield  {author} {\bibinfo {author} {\bibfnamefont {A.}~\bibnamefont
  {Beniwal}}, \bibinfo {author} {\bibfnamefont {F.}~\bibnamefont {Rajec}},
  \bibinfo {author} {\bibfnamefont {C.}~\bibnamefont {Savage}}, \bibinfo
  {author} {\bibfnamefont {P.}~\bibnamefont {Scott}}, \bibinfo {author}
  {\bibfnamefont {C.}~\bibnamefont {Weniger}}, \bibinfo {author} {\bibfnamefont
  {M.}~\bibnamefont {White}}, \ and\ \bibinfo {author} {\bibfnamefont {A.~G.}\
  \bibnamefont {Williams}},\ }\href {\doibase 10.1103/PhysRevD.93.115016}
  {\bibfield  {journal} {\bibinfo  {journal} {Phys. Rev. D}\ }\textbf {\bibinfo
  {volume} {93}},\ \bibinfo {pages} {115016} (\bibinfo {year} {2016})},\
  \Eprint {http://arxiv.org/abs/1512.06458} {arXiv:1512.06458 [hep-ph]}
  \BibitemShut {NoStop}%
\bibitem [{\citenamefont {Arcadi}\ \emph {et~al.}(2020)\citenamefont {Arcadi},
  \citenamefont {Djouadi},\ and\ \citenamefont {Raidal}}]{Arcadi:2019lka}%
  \BibitemOpen
  \bibfield  {author} {\bibinfo {author} {\bibfnamefont {G.}~\bibnamefont
  {Arcadi}}, \bibinfo {author} {\bibfnamefont {A.}~\bibnamefont {Djouadi}}, \
  and\ \bibinfo {author} {\bibfnamefont {M.}~\bibnamefont {Raidal}},\ }\href
  {\doibase 10.1016/j.physrep.2019.11.003} {\bibfield  {journal} {\bibinfo
  {journal} {Phys. Rept.}\ }\textbf {\bibinfo {volume} {842}},\ \bibinfo
  {pages} {1} (\bibinfo {year} {2020})},\ \Eprint
  {http://arxiv.org/abs/1903.03616} {arXiv:1903.03616 [hep-ph]} \BibitemShut
  {NoStop}%
\bibitem [{\citenamefont {Arcadi}\ \emph {et~al.}(2024)\citenamefont {Arcadi},
  \citenamefont {Cabo-Almeida}, \citenamefont {Dutra}, \citenamefont {Ghosh},
  \citenamefont {Lindner}, \citenamefont {Mambrini}, \citenamefont {Neto},
  \citenamefont {Pierre}, \citenamefont {Profumo},\ and\ \citenamefont
  {Queiroz}}]{Arcadi:2024ukq}%
  \BibitemOpen
  \bibfield  {author} {\bibinfo {author} {\bibfnamefont {G.}~\bibnamefont
  {Arcadi}}, \bibinfo {author} {\bibfnamefont {D.}~\bibnamefont
  {Cabo-Almeida}}, \bibinfo {author} {\bibfnamefont {M.}~\bibnamefont {Dutra}},
  \bibinfo {author} {\bibfnamefont {P.}~\bibnamefont {Ghosh}}, \bibinfo
  {author} {\bibfnamefont {M.}~\bibnamefont {Lindner}}, \bibinfo {author}
  {\bibfnamefont {Y.}~\bibnamefont {Mambrini}}, \bibinfo {author}
  {\bibfnamefont {J.~P.}\ \bibnamefont {Neto}}, \bibinfo {author}
  {\bibfnamefont {M.}~\bibnamefont {Pierre}}, \bibinfo {author} {\bibfnamefont
  {S.}~\bibnamefont {Profumo}}, \ and\ \bibinfo {author} {\bibfnamefont
  {F.~S.}\ \bibnamefont {Queiroz}},\ }\href@noop {} {\  (\bibinfo {year}
  {2024})},\ \Eprint {http://arxiv.org/abs/2403.15860} {arXiv:2403.15860
  [hep-ph]} \BibitemShut {NoStop}%
\bibitem [{\citenamefont {Di~Mauro}\ \emph {et~al.}(2023)\citenamefont
  {Di~Mauro}, \citenamefont {Arina}, \citenamefont {Fornengo}, \citenamefont
  {Heisig},\ and\ \citenamefont {Massaro}}]{DiMauro:2023tho}%
  \BibitemOpen
  \bibfield  {author} {\bibinfo {author} {\bibfnamefont {M.}~\bibnamefont
  {Di~Mauro}}, \bibinfo {author} {\bibfnamefont {C.}~\bibnamefont {Arina}},
  \bibinfo {author} {\bibfnamefont {N.}~\bibnamefont {Fornengo}}, \bibinfo
  {author} {\bibfnamefont {J.}~\bibnamefont {Heisig}}, \ and\ \bibinfo {author}
  {\bibfnamefont {D.}~\bibnamefont {Massaro}},\ }\href {\doibase
  10.1103/PhysRevD.108.095008} {\bibfield  {journal} {\bibinfo  {journal}
  {Phys. Rev. D}\ }\textbf {\bibinfo {volume} {108}},\ \bibinfo {pages}
  {095008} (\bibinfo {year} {2023})},\ \Eprint
  {http://arxiv.org/abs/2305.11937} {arXiv:2305.11937 [hep-ph]} \BibitemShut
  {NoStop}%
\bibitem [{\citenamefont {Abdallah}\ \emph {et~al.}(2015)\citenamefont
  {Abdallah}, \citenamefont {Araujo}, \citenamefont {Arbey}, \citenamefont
  {Ashkenazi}, \citenamefont {Belyaev}, \citenamefont {Berger}, \citenamefont
  {Boehm}, \citenamefont {Boveia}, \citenamefont {Brennan}, \citenamefont
  {Brooke}, \citenamefont {Buchmueller}, \citenamefont {Buckley}, \citenamefont
  {Busoni}, \citenamefont {Calibbi}, \citenamefont {Chauhan}, \citenamefont
  {Daci}, \citenamefont {Davies}, \citenamefont {{De Bruyn}}, \citenamefont
  {{De Jong}}, \citenamefont {{De Roeck}}, \citenamefont {{de Vries}},
  \citenamefont {{Del Re}}, \citenamefont {{De Simone}}, \citenamefont {{Di
  Simone}}, \citenamefont {Doglioni}, \citenamefont {Dolan}, \citenamefont
  {Dreiner}, \citenamefont {Ellis}, \citenamefont {Eno}, \citenamefont
  {Etzion}, \citenamefont {Fairbairn}, \citenamefont {Feldstein}, \citenamefont
  {Flaecher}, \citenamefont {Feng}, \citenamefont {Fox}, \citenamefont
  {Genest}, \citenamefont {Gouskos}, \citenamefont {Gramling}, \citenamefont
  {Haisch}, \citenamefont {Harnik}, \citenamefont {Hibbs}, \citenamefont {Hoh},
  \citenamefont {Hopkins}, \citenamefont {Ippolito}, \citenamefont {Jacques},
  \citenamefont {Kahlhoefer}, \citenamefont {Khoze}, \citenamefont {Kirk},
  \citenamefont {Korn}, \citenamefont {Kotov}, \citenamefont {Kunori},
  \citenamefont {Landsberg}, \citenamefont {Liem}, \citenamefont {Lin},
  \citenamefont {Lowette}, \citenamefont {Lucas}, \citenamefont {Malgeri},
  \citenamefont {Malik}, \citenamefont {McCabe}, \citenamefont {Mete},
  \citenamefont {Morgante}, \citenamefont {Mrenna}, \citenamefont {Nakahama},
  \citenamefont {Newbold}, \citenamefont {Nordstrom}, \citenamefont {Pani},
  \citenamefont {Papucci}, \citenamefont {Pataraia}, \citenamefont {Penning},
  \citenamefont {Pinna}, \citenamefont {Polesello}, \citenamefont {Racco},
  \citenamefont {Re}, \citenamefont {Riotto}, \citenamefont {Rizzo},
  \citenamefont {Salek}, \citenamefont {Sarkar}, \citenamefont {Schramm},
  \citenamefont {Skubic}, \citenamefont {Slone}, \citenamefont {Smirnov},
  \citenamefont {Soreq}, \citenamefont {Sumner}, \citenamefont {Tait},
  \citenamefont {Thomas}, \citenamefont {Tomalin}, \citenamefont {Tunnell},
  \citenamefont {Vichi}, \citenamefont {Volansky}, \citenamefont {Weiner},
  \citenamefont {West}, \citenamefont {Wielers}, \citenamefont {Worm},
  \citenamefont {Yavin}, \citenamefont {Zaldivar}, \citenamefont {Zhou},\ and\
  \citenamefont {Zurek}}]{ABDALLAH20158}%
  \BibitemOpen
  \bibfield  {author} {\bibinfo {author} {\bibfnamefont {J.}~\bibnamefont
  {Abdallah}}, \bibinfo {author} {\bibfnamefont {H.}~\bibnamefont {Araujo}},
  \bibinfo {author} {\bibfnamefont {A.}~\bibnamefont {Arbey}}, \bibinfo
  {author} {\bibfnamefont {A.}~\bibnamefont {Ashkenazi}}, \bibinfo {author}
  {\bibfnamefont {A.}~\bibnamefont {Belyaev}}, \bibinfo {author} {\bibfnamefont
  {J.}~\bibnamefont {Berger}}, \bibinfo {author} {\bibfnamefont
  {C.}~\bibnamefont {Boehm}}, \bibinfo {author} {\bibfnamefont
  {A.}~\bibnamefont {Boveia}}, \bibinfo {author} {\bibfnamefont
  {A.}~\bibnamefont {Brennan}}, \bibinfo {author} {\bibfnamefont
  {J.}~\bibnamefont {Brooke}}, \bibinfo {author} {\bibfnamefont
  {O.}~\bibnamefont {Buchmueller}}, \bibinfo {author} {\bibfnamefont
  {M.}~\bibnamefont {Buckley}}, \bibinfo {author} {\bibfnamefont
  {G.}~\bibnamefont {Busoni}}, \bibinfo {author} {\bibfnamefont
  {L.}~\bibnamefont {Calibbi}}, \bibinfo {author} {\bibfnamefont
  {S.}~\bibnamefont {Chauhan}}, \bibinfo {author} {\bibfnamefont
  {N.}~\bibnamefont {Daci}}, \bibinfo {author} {\bibfnamefont {G.}~\bibnamefont
  {Davies}}, \bibinfo {author} {\bibfnamefont {I.}~\bibnamefont {{De Bruyn}}},
  \bibinfo {author} {\bibfnamefont {P.}~\bibnamefont {{De Jong}}}, \bibinfo
  {author} {\bibfnamefont {A.}~\bibnamefont {{De Roeck}}}, \bibinfo {author}
  {\bibfnamefont {K.}~\bibnamefont {{de Vries}}}, \bibinfo {author}
  {\bibfnamefont {D.}~\bibnamefont {{Del Re}}}, \bibinfo {author}
  {\bibfnamefont {A.}~\bibnamefont {{De Simone}}}, \bibinfo {author}
  {\bibfnamefont {A.}~\bibnamefont {{Di Simone}}}, \bibinfo {author}
  {\bibfnamefont {C.}~\bibnamefont {Doglioni}}, \bibinfo {author}
  {\bibfnamefont {M.}~\bibnamefont {Dolan}}, \bibinfo {author} {\bibfnamefont
  {H.~K.}\ \bibnamefont {Dreiner}}, \bibinfo {author} {\bibfnamefont
  {J.}~\bibnamefont {Ellis}}, \bibinfo {author} {\bibfnamefont
  {S.}~\bibnamefont {Eno}}, \bibinfo {author} {\bibfnamefont {E.}~\bibnamefont
  {Etzion}}, \bibinfo {author} {\bibfnamefont {M.}~\bibnamefont {Fairbairn}},
  \bibinfo {author} {\bibfnamefont {B.}~\bibnamefont {Feldstein}}, \bibinfo
  {author} {\bibfnamefont {H.}~\bibnamefont {Flaecher}}, \bibinfo {author}
  {\bibfnamefont {E.}~\bibnamefont {Feng}}, \bibinfo {author} {\bibfnamefont
  {P.}~\bibnamefont {Fox}}, \bibinfo {author} {\bibfnamefont {M.-H.}\
  \bibnamefont {Genest}}, \bibinfo {author} {\bibfnamefont {L.}~\bibnamefont
  {Gouskos}}, \bibinfo {author} {\bibfnamefont {J.}~\bibnamefont {Gramling}},
  \bibinfo {author} {\bibfnamefont {U.}~\bibnamefont {Haisch}}, \bibinfo
  {author} {\bibfnamefont {R.}~\bibnamefont {Harnik}}, \bibinfo {author}
  {\bibfnamefont {A.}~\bibnamefont {Hibbs}}, \bibinfo {author} {\bibfnamefont
  {S.}~\bibnamefont {Hoh}}, \bibinfo {author} {\bibfnamefont {W.}~\bibnamefont
  {Hopkins}}, \bibinfo {author} {\bibfnamefont {V.}~\bibnamefont {Ippolito}},
  \bibinfo {author} {\bibfnamefont {T.}~\bibnamefont {Jacques}}, \bibinfo
  {author} {\bibfnamefont {F.}~\bibnamefont {Kahlhoefer}}, \bibinfo {author}
  {\bibfnamefont {V.~V.}\ \bibnamefont {Khoze}}, \bibinfo {author}
  {\bibfnamefont {R.}~\bibnamefont {Kirk}}, \bibinfo {author} {\bibfnamefont
  {A.}~\bibnamefont {Korn}}, \bibinfo {author} {\bibfnamefont {K.}~\bibnamefont
  {Kotov}}, \bibinfo {author} {\bibfnamefont {S.}~\bibnamefont {Kunori}},
  \bibinfo {author} {\bibfnamefont {G.}~\bibnamefont {Landsberg}}, \bibinfo
  {author} {\bibfnamefont {S.}~\bibnamefont {Liem}}, \bibinfo {author}
  {\bibfnamefont {T.}~\bibnamefont {Lin}}, \bibinfo {author} {\bibfnamefont
  {S.}~\bibnamefont {Lowette}}, \bibinfo {author} {\bibfnamefont
  {R.}~\bibnamefont {Lucas}}, \bibinfo {author} {\bibfnamefont
  {L.}~\bibnamefont {Malgeri}}, \bibinfo {author} {\bibfnamefont
  {S.}~\bibnamefont {Malik}}, \bibinfo {author} {\bibfnamefont
  {C.}~\bibnamefont {McCabe}}, \bibinfo {author} {\bibfnamefont {A.~S.}\
  \bibnamefont {Mete}}, \bibinfo {author} {\bibfnamefont {E.}~\bibnamefont
  {Morgante}}, \bibinfo {author} {\bibfnamefont {S.}~\bibnamefont {Mrenna}},
  \bibinfo {author} {\bibfnamefont {Y.}~\bibnamefont {Nakahama}}, \bibinfo
  {author} {\bibfnamefont {D.}~\bibnamefont {Newbold}}, \bibinfo {author}
  {\bibfnamefont {K.}~\bibnamefont {Nordstrom}}, \bibinfo {author}
  {\bibfnamefont {P.}~\bibnamefont {Pani}}, \bibinfo {author} {\bibfnamefont
  {M.}~\bibnamefont {Papucci}}, \bibinfo {author} {\bibfnamefont
  {S.}~\bibnamefont {Pataraia}}, \bibinfo {author} {\bibfnamefont
  {B.}~\bibnamefont {Penning}}, \bibinfo {author} {\bibfnamefont
  {D.}~\bibnamefont {Pinna}}, \bibinfo {author} {\bibfnamefont
  {G.}~\bibnamefont {Polesello}}, \bibinfo {author} {\bibfnamefont
  {D.}~\bibnamefont {Racco}}, \bibinfo {author} {\bibfnamefont
  {E.}~\bibnamefont {Re}}, \bibinfo {author} {\bibfnamefont {A.~W.}\
  \bibnamefont {Riotto}}, \bibinfo {author} {\bibfnamefont {T.}~\bibnamefont
  {Rizzo}}, \bibinfo {author} {\bibfnamefont {D.}~\bibnamefont {Salek}},
  \bibinfo {author} {\bibfnamefont {S.}~\bibnamefont {Sarkar}}, \bibinfo
  {author} {\bibfnamefont {S.}~\bibnamefont {Schramm}}, \bibinfo {author}
  {\bibfnamefont {P.}~\bibnamefont {Skubic}}, \bibinfo {author} {\bibfnamefont
  {O.}~\bibnamefont {Slone}}, \bibinfo {author} {\bibfnamefont
  {J.}~\bibnamefont {Smirnov}}, \bibinfo {author} {\bibfnamefont
  {Y.}~\bibnamefont {Soreq}}, \bibinfo {author} {\bibfnamefont
  {T.}~\bibnamefont {Sumner}}, \bibinfo {author} {\bibfnamefont {T.~M.}\
  \bibnamefont {Tait}}, \bibinfo {author} {\bibfnamefont {M.}~\bibnamefont
  {Thomas}}, \bibinfo {author} {\bibfnamefont {I.}~\bibnamefont {Tomalin}},
  \bibinfo {author} {\bibfnamefont {C.}~\bibnamefont {Tunnell}}, \bibinfo
  {author} {\bibfnamefont {A.}~\bibnamefont {Vichi}}, \bibinfo {author}
  {\bibfnamefont {T.}~\bibnamefont {Volansky}}, \bibinfo {author}
  {\bibfnamefont {N.}~\bibnamefont {Weiner}}, \bibinfo {author} {\bibfnamefont
  {S.~M.}\ \bibnamefont {West}}, \bibinfo {author} {\bibfnamefont
  {M.}~\bibnamefont {Wielers}}, \bibinfo {author} {\bibfnamefont
  {S.}~\bibnamefont {Worm}}, \bibinfo {author} {\bibfnamefont {I.}~\bibnamefont
  {Yavin}}, \bibinfo {author} {\bibfnamefont {B.}~\bibnamefont {Zaldivar}},
  \bibinfo {author} {\bibfnamefont {N.}~\bibnamefont {Zhou}}, \ and\ \bibinfo
  {author} {\bibfnamefont {K.}~\bibnamefont {Zurek}},\ }\href {\doibase
  https://doi.org/10.1016/j.dark.2015.08.001} {\bibfield  {journal} {\bibinfo
  {journal} {Physics of the Dark Universe}\ }\textbf {\bibinfo {volume}
  {9-10}},\ \bibinfo {pages} {8} (\bibinfo {year} {2015})}\BibitemShut
  {NoStop}%
\bibitem [{\citenamefont {Arina}(2018)}]{Arina:2018zcq}%
  \BibitemOpen
  \bibfield  {author} {\bibinfo {author} {\bibfnamefont {C.}~\bibnamefont
  {Arina}},\ }\href {\doibase 10.3389/fspas.2018.00030} {\bibfield  {journal}
  {\bibinfo  {journal} {Front. Astron. Space Sci.}\ }\textbf {\bibinfo {volume}
  {5}},\ \bibinfo {pages} {30} (\bibinfo {year} {2018})},\ \Eprint
  {http://arxiv.org/abs/1805.04290} {arXiv:1805.04290 [hep-ph]} \BibitemShut
  {NoStop}%
\bibitem [{\citenamefont {Chang}\ \emph
  {et~al.}(2023{\natexlab{a}})\citenamefont {Chang}, \citenamefont {Scott},
  \citenamefont {Gonzalo}, \citenamefont {Kahlhoefer}, \citenamefont
  {Kvellestad},\ and\ \citenamefont {White}}]{Chang:2022jgo}%
  \BibitemOpen
  \bibfield  {author} {\bibinfo {author} {\bibfnamefont {C.}~\bibnamefont
  {Chang}}, \bibinfo {author} {\bibfnamefont {P.}~\bibnamefont {Scott}},
  \bibinfo {author} {\bibfnamefont {T.~E.}\ \bibnamefont {Gonzalo}}, \bibinfo
  {author} {\bibfnamefont {F.}~\bibnamefont {Kahlhoefer}}, \bibinfo {author}
  {\bibfnamefont {A.}~\bibnamefont {Kvellestad}}, \ and\ \bibinfo {author}
  {\bibfnamefont {M.}~\bibnamefont {White}},\ }\href {\doibase
  10.1140/epjc/s10052-023-11399-w} {\bibfield  {journal} {\bibinfo  {journal}
  {Eur. Phys. J. C}\ }\textbf {\bibinfo {volume} {83}},\ \bibinfo {pages} {249}
  (\bibinfo {year} {2023}{\natexlab{a}})},\ \Eprint
  {http://arxiv.org/abs/2209.13266} {arXiv:2209.13266 [hep-ph]} \BibitemShut
  {NoStop}%
\bibitem [{\citenamefont {Chang}\ \emph
  {et~al.}(2023{\natexlab{b}})\citenamefont {Chang}, \citenamefont {Scott},
  \citenamefont {Gonzalo}, \citenamefont {Kahlhoefer},\ and\ \citenamefont
  {White}}]{Chang:2023cki}%
  \BibitemOpen
  \bibfield  {author} {\bibinfo {author} {\bibfnamefont {C.}~\bibnamefont
  {Chang}}, \bibinfo {author} {\bibfnamefont {P.}~\bibnamefont {Scott}},
  \bibinfo {author} {\bibfnamefont {T.~E.}\ \bibnamefont {Gonzalo}}, \bibinfo
  {author} {\bibfnamefont {F.}~\bibnamefont {Kahlhoefer}}, \ and\ \bibinfo
  {author} {\bibfnamefont {M.}~\bibnamefont {White}},\ }\href {\doibase
  10.1140/epjc/s10052-023-11859-3} {\bibfield  {journal} {\bibinfo  {journal}
  {Eur. Phys. J. C}\ }\textbf {\bibinfo {volume} {83}},\ \bibinfo {pages} {692}
  (\bibinfo {year} {2023}{\natexlab{b}})},\ \bibinfo {note} {[Erratum:
  Eur.Phys.J.C 83, 768 (2023)]},\ \Eprint {http://arxiv.org/abs/2303.08351}
  {arXiv:2303.08351 [hep-ph]} \BibitemShut {NoStop}%
\bibitem [{\citenamefont {McDaniel}\ \emph {et~al.}(2024)\citenamefont
  {McDaniel}, \citenamefont {Ajello}, \citenamefont {Karwin}, \citenamefont
  {Di~Mauro}, \citenamefont {Drlica-Wagner},\ and\ \citenamefont
  {S\'anchez-Conde}}]{McDaniel:2023bju}%
  \BibitemOpen
  \bibfield  {author} {\bibinfo {author} {\bibfnamefont {A.}~\bibnamefont
  {McDaniel}}, \bibinfo {author} {\bibfnamefont {M.}~\bibnamefont {Ajello}},
  \bibinfo {author} {\bibfnamefont {C.~M.}\ \bibnamefont {Karwin}}, \bibinfo
  {author} {\bibfnamefont {M.}~\bibnamefont {Di~Mauro}}, \bibinfo {author}
  {\bibfnamefont {A.}~\bibnamefont {Drlica-Wagner}}, \ and\ \bibinfo {author}
  {\bibfnamefont {M.~A.}\ \bibnamefont {S\'anchez-Conde}},\ }\href {\doibase
  10.1103/PhysRevD.109.063024} {\bibfield  {journal} {\bibinfo  {journal}
  {Phys. Rev. D}\ }\textbf {\bibinfo {volume} {109}},\ \bibinfo {pages}
  {063024} (\bibinfo {year} {2024})},\ \Eprint
  {http://arxiv.org/abs/2311.04982} {arXiv:2311.04982 [astro-ph.HE]}
  \BibitemShut {NoStop}%
\bibitem [{\citenamefont {Aprile}\ \emph {et~al.}(2023)\citenamefont {Aprile}
  \emph {et~al.}}]{XENON:2023cxc}%
  \BibitemOpen
  \bibfield  {author} {\bibinfo {author} {\bibfnamefont {E.}~\bibnamefont
  {Aprile}} \emph {et~al.} (\bibinfo {collaboration} {XENON}),\ }\href
  {\doibase 10.1103/PhysRevLett.131.041003} {\bibfield  {journal} {\bibinfo
  {journal} {Phys. Rev. Lett.}\ }\textbf {\bibinfo {volume} {131}},\ \bibinfo
  {pages} {041003} (\bibinfo {year} {2023})},\ \Eprint
  {http://arxiv.org/abs/2303.14729} {arXiv:2303.14729 [hep-ex]} \BibitemShut
  {NoStop}%
\bibitem [{\citenamefont {Aprile}\ \emph {et~al.}(2025)\citenamefont {Aprile}
  \emph {et~al.}}]{XENON:2024znc}%
  \BibitemOpen
  \bibfield  {author} {\bibinfo {author} {\bibfnamefont {E.}~\bibnamefont
  {Aprile}} \emph {et~al.} (\bibinfo {collaboration} {XENON, (XENON
  Collaboration)**}),\ }\href {\doibase 10.1103/PhysRevLett.134.161004}
  {\bibfield  {journal} {\bibinfo  {journal} {Phys. Rev. Lett.}\ }\textbf
  {\bibinfo {volume} {134}},\ \bibinfo {pages} {161004} (\bibinfo {year}
  {2025})},\ \Eprint {http://arxiv.org/abs/2411.15289} {arXiv:2411.15289
  [hep-ex]} \BibitemShut {NoStop}%
\bibitem [{\citenamefont {Aalbers}\ \emph {et~al.}(2025)\citenamefont {Aalbers}
  \emph {et~al.}}]{LZ:2024zvo}%
  \BibitemOpen
  \bibfield  {author} {\bibinfo {author} {\bibfnamefont {J.}~\bibnamefont
  {Aalbers}} \emph {et~al.} (\bibinfo {collaboration} {LZ}),\ }\href {\doibase
  10.1103/4dyc-z8zf} {\bibfield  {journal} {\bibinfo  {journal} {Phys. Rev.
  Lett.}\ }\textbf {\bibinfo {volume} {135}},\ \bibinfo {pages} {011802}
  (\bibinfo {year} {2025})},\ \Eprint {http://arxiv.org/abs/2410.17036}
  {arXiv:2410.17036 [hep-ex]} \BibitemShut {NoStop}%
\bibitem [{\citenamefont {Arcadi}\ \emph {et~al.}(2018)\citenamefont {Arcadi},
  \citenamefont {Dutra}, \citenamefont {Ghosh}, \citenamefont {Lindner},
  \citenamefont {Mambrini}, \citenamefont {Pierre}, \citenamefont {Profumo},\
  and\ \citenamefont {Queiroz}}]{Arcadi:2017kky}%
  \BibitemOpen
  \bibfield  {author} {\bibinfo {author} {\bibfnamefont {G.}~\bibnamefont
  {Arcadi}}, \bibinfo {author} {\bibfnamefont {M.}~\bibnamefont {Dutra}},
  \bibinfo {author} {\bibfnamefont {P.}~\bibnamefont {Ghosh}}, \bibinfo
  {author} {\bibfnamefont {M.}~\bibnamefont {Lindner}}, \bibinfo {author}
  {\bibfnamefont {Y.}~\bibnamefont {Mambrini}}, \bibinfo {author}
  {\bibfnamefont {M.}~\bibnamefont {Pierre}}, \bibinfo {author} {\bibfnamefont
  {S.}~\bibnamefont {Profumo}}, \ and\ \bibinfo {author} {\bibfnamefont
  {F.~S.}\ \bibnamefont {Queiroz}},\ }\href {\doibase
  10.1140/epjc/s10052-018-5662-y} {\bibfield  {journal} {\bibinfo  {journal}
  {Eur. Phys. J. C}\ }\textbf {\bibinfo {volume} {78}},\ \bibinfo {pages} {203}
  (\bibinfo {year} {2018})},\ \Eprint {http://arxiv.org/abs/1703.07364}
  {arXiv:1703.07364 [hep-ph]} \BibitemShut {NoStop}%
\bibitem [{\citenamefont {Pospelov}\ \emph {et~al.}(2008)\citenamefont
  {Pospelov}, \citenamefont {Ritz},\ and\ \citenamefont
  {Voloshin}}]{Pospelov:2007mp}%
  \BibitemOpen
  \bibfield  {author} {\bibinfo {author} {\bibfnamefont {M.}~\bibnamefont
  {Pospelov}}, \bibinfo {author} {\bibfnamefont {A.}~\bibnamefont {Ritz}}, \
  and\ \bibinfo {author} {\bibfnamefont {M.~B.}\ \bibnamefont {Voloshin}},\
  }\href {\doibase 10.1016/j.physletb.2008.02.052} {\bibfield  {journal}
  {\bibinfo  {journal} {Phys. Lett. B}\ }\textbf {\bibinfo {volume} {662}},\
  \bibinfo {pages} {53} (\bibinfo {year} {2008})},\ \Eprint
  {http://arxiv.org/abs/0711.4866} {arXiv:0711.4866 [hep-ph]} \BibitemShut
  {NoStop}%
\bibitem [{\citenamefont {Belanger}\ \emph {et~al.}(2007)\citenamefont
  {Belanger}, \citenamefont {Boudjema}, \citenamefont {Pukhov},\ and\
  \citenamefont {Semenov}}]{Belanger:2006is}%
  \BibitemOpen
  \bibfield  {author} {\bibinfo {author} {\bibfnamefont {G.}~\bibnamefont
  {Belanger}}, \bibinfo {author} {\bibfnamefont {F.}~\bibnamefont {Boudjema}},
  \bibinfo {author} {\bibfnamefont {A.}~\bibnamefont {Pukhov}}, \ and\ \bibinfo
  {author} {\bibfnamefont {A.}~\bibnamefont {Semenov}},\ }\href {\doibase
  10.1016/j.cpc.2006.11.008} {\bibfield  {journal} {\bibinfo  {journal}
  {Comput. Phys. Commun.}\ }\textbf {\bibinfo {volume} {176}},\ \bibinfo
  {pages} {367} (\bibinfo {year} {2007})},\ \Eprint
  {http://arxiv.org/abs/hep-ph/0607059} {arXiv:hep-ph/0607059} \BibitemShut
  {NoStop}%
\bibitem [{\citenamefont {Belanger}\ \emph {et~al.}(2014)\citenamefont
  {Belanger}, \citenamefont {Boudjema}, \citenamefont {Pukhov},\ and\
  \citenamefont {Semenov}}]{Belanger:2013oya}%
  \BibitemOpen
  \bibfield  {author} {\bibinfo {author} {\bibfnamefont {G.}~\bibnamefont
  {Belanger}}, \bibinfo {author} {\bibfnamefont {F.}~\bibnamefont {Boudjema}},
  \bibinfo {author} {\bibfnamefont {A.}~\bibnamefont {Pukhov}}, \ and\ \bibinfo
  {author} {\bibfnamefont {A.}~\bibnamefont {Semenov}},\ }\href {\doibase
  10.1016/j.cpc.2013.10.016} {\bibfield  {journal} {\bibinfo  {journal}
  {Comput. Phys. Commun.}\ }\textbf {\bibinfo {volume} {185}},\ \bibinfo
  {pages} {960} (\bibinfo {year} {2014})},\ \Eprint
  {http://arxiv.org/abs/1305.0237} {arXiv:1305.0237 [hep-ph]} \BibitemShut
  {NoStop}%
\bibitem [{\citenamefont {B\'elanger}\ \emph {et~al.}(2018)\citenamefont
  {B\'elanger}, \citenamefont {Boudjema}, \citenamefont {Goudelis},
  \citenamefont {Pukhov},\ and\ \citenamefont {Zaldivar}}]{Belanger:2018ccd}%
  \BibitemOpen
  \bibfield  {author} {\bibinfo {author} {\bibfnamefont {G.}~\bibnamefont
  {B\'elanger}}, \bibinfo {author} {\bibfnamefont {F.}~\bibnamefont
  {Boudjema}}, \bibinfo {author} {\bibfnamefont {A.}~\bibnamefont {Goudelis}},
  \bibinfo {author} {\bibfnamefont {A.}~\bibnamefont {Pukhov}}, \ and\ \bibinfo
  {author} {\bibfnamefont {B.}~\bibnamefont {Zaldivar}},\ }\href {\doibase
  10.1016/j.cpc.2018.04.027} {\bibfield  {journal} {\bibinfo  {journal}
  {Comput. Phys. Commun.}\ }\textbf {\bibinfo {volume} {231}},\ \bibinfo
  {pages} {173} (\bibinfo {year} {2018})},\ \Eprint
  {http://arxiv.org/abs/1801.03509} {arXiv:1801.03509 [hep-ph]} \BibitemShut
  {NoStop}%
\bibitem [{\citenamefont {Alguero}\ \emph {et~al.}(2024)\citenamefont
  {Alguero}, \citenamefont {Belanger}, \citenamefont {Boudjema}, \citenamefont
  {Chakraborti}, \citenamefont {Goudelis}, \citenamefont {Kraml}, \citenamefont
  {Mjallal},\ and\ \citenamefont {Pukhov}}]{Alguero:2023zol}%
  \BibitemOpen
  \bibfield  {author} {\bibinfo {author} {\bibfnamefont {G.}~\bibnamefont
  {Alguero}}, \bibinfo {author} {\bibfnamefont {G.}~\bibnamefont {Belanger}},
  \bibinfo {author} {\bibfnamefont {F.}~\bibnamefont {Boudjema}}, \bibinfo
  {author} {\bibfnamefont {S.}~\bibnamefont {Chakraborti}}, \bibinfo {author}
  {\bibfnamefont {A.}~\bibnamefont {Goudelis}}, \bibinfo {author}
  {\bibfnamefont {S.}~\bibnamefont {Kraml}}, \bibinfo {author} {\bibfnamefont
  {A.}~\bibnamefont {Mjallal}}, \ and\ \bibinfo {author} {\bibfnamefont
  {A.}~\bibnamefont {Pukhov}},\ }\href {\doibase 10.1016/j.cpc.2024.109133}
  {\bibfield  {journal} {\bibinfo  {journal} {Comput. Phys. Commun.}\ }\textbf
  {\bibinfo {volume} {299}},\ \bibinfo {pages} {109133} (\bibinfo {year}
  {2024})},\ \Eprint {http://arxiv.org/abs/2312.14894} {arXiv:2312.14894
  [hep-ph]} \BibitemShut {NoStop}%
\bibitem [{\citenamefont {Aalbers}\ \emph {et~al.}(2022)\citenamefont {Aalbers}
  \emph {et~al.}}]{LZ:2022ufs}%
  \BibitemOpen
  \bibfield  {author} {\bibinfo {author} {\bibfnamefont {J.}~\bibnamefont
  {Aalbers}} \emph {et~al.} (\bibinfo {collaboration} {LZ}),\ }\href@noop {} {\
   (\bibinfo {year} {2022})},\ \Eprint {http://arxiv.org/abs/2207.03764}
  {arXiv:2207.03764 [hep-ex]} \BibitemShut {NoStop}%
\bibitem [{\citenamefont {Fitzpatrick}\ \emph {et~al.}(2013)\citenamefont
  {Fitzpatrick}, \citenamefont {Haxton}, \citenamefont {Katz}, \citenamefont
  {Lubbers},\ and\ \citenamefont {Xu}}]{Fitzpatrick:2012ix}%
  \BibitemOpen
  \bibfield  {author} {\bibinfo {author} {\bibfnamefont {A.~L.}\ \bibnamefont
  {Fitzpatrick}}, \bibinfo {author} {\bibfnamefont {W.}~\bibnamefont {Haxton}},
  \bibinfo {author} {\bibfnamefont {E.}~\bibnamefont {Katz}}, \bibinfo {author}
  {\bibfnamefont {N.}~\bibnamefont {Lubbers}}, \ and\ \bibinfo {author}
  {\bibfnamefont {Y.}~\bibnamefont {Xu}},\ }\href {\doibase
  10.1088/1475-7516/2013/02/004} {\bibfield  {journal} {\bibinfo  {journal}
  {JCAP}\ }\textbf {\bibinfo {volume} {02}},\ \bibinfo {pages} {004} (\bibinfo
  {year} {2013})},\ \Eprint {http://arxiv.org/abs/1203.3542} {arXiv:1203.3542
  [hep-ph]} \BibitemShut {NoStop}%
\bibitem [{\citenamefont {Arina}\ \emph {et~al.}(2015)\citenamefont {Arina},
  \citenamefont {Del~Nobile},\ and\ \citenamefont {Panci}}]{Arina:2014yna}%
  \BibitemOpen
  \bibfield  {author} {\bibinfo {author} {\bibfnamefont {C.}~\bibnamefont
  {Arina}}, \bibinfo {author} {\bibfnamefont {E.}~\bibnamefont {Del~Nobile}}, \
  and\ \bibinfo {author} {\bibfnamefont {P.}~\bibnamefont {Panci}},\ }\href
  {\doibase 10.1103/PhysRevLett.114.011301} {\bibfield  {journal} {\bibinfo
  {journal} {Phys. Rev. Lett.}\ }\textbf {\bibinfo {volume} {114}},\ \bibinfo
  {pages} {011301} (\bibinfo {year} {2015})},\ \Eprint
  {http://arxiv.org/abs/1406.5542} {arXiv:1406.5542 [hep-ph]} \BibitemShut
  {NoStop}%
\bibitem [{\citenamefont {Dolan}\ \emph {et~al.}(2015)\citenamefont {Dolan},
  \citenamefont {Kahlhoefer}, \citenamefont {McCabe},\ and\ \citenamefont
  {Schmidt-Hoberg}}]{Dolan:2014ska}%
  \BibitemOpen
  \bibfield  {author} {\bibinfo {author} {\bibfnamefont {M.~J.}\ \bibnamefont
  {Dolan}}, \bibinfo {author} {\bibfnamefont {F.}~\bibnamefont {Kahlhoefer}},
  \bibinfo {author} {\bibfnamefont {C.}~\bibnamefont {McCabe}}, \ and\ \bibinfo
  {author} {\bibfnamefont {K.}~\bibnamefont {Schmidt-Hoberg}},\ }\href
  {\doibase 10.1007/JHEP03(2015)171} {\bibfield  {journal} {\bibinfo  {journal}
  {JHEP}\ }\textbf {\bibinfo {volume} {03}},\ \bibinfo {pages} {171} (\bibinfo
  {year} {2015})},\ \bibinfo {note} {[Erratum: JHEP 07, 103 (2015)]},\ \Eprint
  {http://arxiv.org/abs/1412.5174} {arXiv:1412.5174 [hep-ph]} \BibitemShut
  {NoStop}%
\bibitem [{\citenamefont {Abe}\ \emph {et~al.}(2019)\citenamefont {Abe},
  \citenamefont {Fujiwara},\ and\ \citenamefont {Hisano}}]{Abe:2018emu}%
  \BibitemOpen
  \bibfield  {author} {\bibinfo {author} {\bibfnamefont {T.}~\bibnamefont
  {Abe}}, \bibinfo {author} {\bibfnamefont {M.}~\bibnamefont {Fujiwara}}, \
  and\ \bibinfo {author} {\bibfnamefont {J.}~\bibnamefont {Hisano}},\ }\href
  {\doibase 10.1007/JHEP02(2019)028} {\bibfield  {journal} {\bibinfo  {journal}
  {JHEP}\ }\textbf {\bibinfo {volume} {02}},\ \bibinfo {pages} {028} (\bibinfo
  {year} {2019})},\ \Eprint {http://arxiv.org/abs/1810.01039} {arXiv:1810.01039
  [hep-ph]} \BibitemShut {NoStop}%
\bibitem [{\citenamefont {Ertas}\ and\ \citenamefont
  {Kahlhoefer}(2019)}]{Ertas:2019dew}%
  \BibitemOpen
  \bibfield  {author} {\bibinfo {author} {\bibfnamefont {F.}~\bibnamefont
  {Ertas}}\ and\ \bibinfo {author} {\bibfnamefont {F.}~\bibnamefont
  {Kahlhoefer}},\ }\href {\doibase 10.1007/JHEP06(2019)052} {\bibfield
  {journal} {\bibinfo  {journal} {JHEP}\ }\textbf {\bibinfo {volume} {06}},\
  \bibinfo {pages} {052} (\bibinfo {year} {2019})},\ \Eprint
  {http://arxiv.org/abs/1902.11070} {arXiv:1902.11070 [hep-ph]} \BibitemShut
  {NoStop}%
\bibitem [{\citenamefont {Aghanim}\ \emph {et~al.}(2020)\citenamefont {Aghanim}
  \emph {et~al.}}]{Planck:2018vyg}%
  \BibitemOpen
  \bibfield  {author} {\bibinfo {author} {\bibfnamefont {N.}~\bibnamefont
  {Aghanim}} \emph {et~al.} (\bibinfo {collaboration} {Planck}),\ }\href
  {\doibase 10.1051/0004-6361/201833910} {\bibfield  {journal} {\bibinfo
  {journal} {Astron. Astrophys.}\ }\textbf {\bibinfo {volume} {641}},\ \bibinfo
  {pages} {A6} (\bibinfo {year} {2020})},\ \bibinfo {note} {[Erratum:
  Astron.Astrophys. 652, C4 (2021)]},\ \Eprint
  {http://arxiv.org/abs/1807.06209} {arXiv:1807.06209 [astro-ph.CO]}
  \BibitemShut {NoStop}%
\bibitem [{\citenamefont {Kolb}\ and\ \citenamefont
  {Turner}(1990)}]{Kolb:1990vq}%
  \BibitemOpen
  \bibfield  {author} {\bibinfo {author} {\bibfnamefont {E.~W.}\ \bibnamefont
  {Kolb}}\ and\ \bibinfo {author} {\bibfnamefont {M.~S.}\ \bibnamefont
  {Turner}},\ }\href {\doibase 10.1201/9780429492860} {\emph {\bibinfo {title}
  {{The Early Universe}}}},\ Vol.~\bibinfo {volume} {69}\ (\bibinfo
  {publisher} {CRC Press},\ \bibinfo {year} {1990})\BibitemShut {NoStop}%
\bibitem [{\citenamefont {Edsjo}\ and\ \citenamefont
  {Gondolo}(1997)}]{Edsjo:1997bg}%
  \BibitemOpen
  \bibfield  {author} {\bibinfo {author} {\bibfnamefont {J.}~\bibnamefont
  {Edsjo}}\ and\ \bibinfo {author} {\bibfnamefont {P.}~\bibnamefont
  {Gondolo}},\ }\href {\doibase 10.1103/PhysRevD.56.1879} {\bibfield  {journal}
  {\bibinfo  {journal} {Phys. Rev. D}\ }\textbf {\bibinfo {volume} {56}},\
  \bibinfo {pages} {1879} (\bibinfo {year} {1997})},\ \Eprint
  {http://arxiv.org/abs/hep-ph/9704361} {arXiv:hep-ph/9704361} \BibitemShut
  {NoStop}%
\bibitem [{\citenamefont {Binder}\ \emph {et~al.}(2017)\citenamefont {Binder},
  \citenamefont {Bringmann}, \citenamefont {Gustafsson},\ and\ \citenamefont
  {Hryczuk}}]{Binder:2017rgn}%
  \BibitemOpen
  \bibfield  {author} {\bibinfo {author} {\bibfnamefont {T.}~\bibnamefont
  {Binder}}, \bibinfo {author} {\bibfnamefont {T.}~\bibnamefont {Bringmann}},
  \bibinfo {author} {\bibfnamefont {M.}~\bibnamefont {Gustafsson}}, \ and\
  \bibinfo {author} {\bibfnamefont {A.}~\bibnamefont {Hryczuk}},\ }\href
  {\doibase 10.1103/PhysRevD.96.115010} {\bibfield  {journal} {\bibinfo
  {journal} {Phys. Rev. D}\ }\textbf {\bibinfo {volume} {96}},\ \bibinfo
  {pages} {115010} (\bibinfo {year} {2017})},\ \bibinfo {note} {[Erratum:
  Phys.Rev.D 101, 099901 (2020)]},\ \Eprint {http://arxiv.org/abs/1706.07433}
  {arXiv:1706.07433 [astro-ph.CO]} \BibitemShut {NoStop}%
\bibitem [{\citenamefont {Gondolo}\ and\ \citenamefont
  {Gelmini}(1991)}]{GONDOLO1991145}%
  \BibitemOpen
  \bibfield  {author} {\bibinfo {author} {\bibfnamefont {P.}~\bibnamefont
  {Gondolo}}\ and\ \bibinfo {author} {\bibfnamefont {G.}~\bibnamefont
  {Gelmini}},\ }\href {\doibase https://doi.org/10.1016/0550-3213(91)90438-4}
  {\bibfield  {journal} {\bibinfo  {journal} {Nuclear Physics B}\ }\textbf
  {\bibinfo {volume} {360}},\ \bibinfo {pages} {145} (\bibinfo {year}
  {1991})}\BibitemShut {NoStop}%
\bibitem [{\citenamefont {Binder}\ \emph {et~al.}(2021)\citenamefont {Binder},
  \citenamefont {Bringmann}, \citenamefont {Gustafsson},\ and\ \citenamefont
  {Hryczuk}}]{Binder:2021bmg}%
  \BibitemOpen
  \bibfield  {author} {\bibinfo {author} {\bibfnamefont {T.}~\bibnamefont
  {Binder}}, \bibinfo {author} {\bibfnamefont {T.}~\bibnamefont {Bringmann}},
  \bibinfo {author} {\bibfnamefont {M.}~\bibnamefont {Gustafsson}}, \ and\
  \bibinfo {author} {\bibfnamefont {A.}~\bibnamefont {Hryczuk}},\ }\href
  {\doibase 10.1140/epjc/s10052-021-09357-5} {\bibfield  {journal} {\bibinfo
  {journal} {Eur. Phys. J. C}\ }\textbf {\bibinfo {volume} {81}},\ \bibinfo
  {pages} {577} (\bibinfo {year} {2021})},\ \Eprint
  {http://arxiv.org/abs/2103.01944} {arXiv:2103.01944 [hep-ph]} \BibitemShut
  {NoStop}%
\bibitem [{\citenamefont {Barducci}\ \emph {et~al.}(2018)\citenamefont
  {Barducci}, \citenamefont {Belanger}, \citenamefont {Bernon}, \citenamefont
  {Boudjema}, \citenamefont {Da~Silva}, \citenamefont {Kraml}, \citenamefont
  {Laa},\ and\ \citenamefont {Pukhov}}]{Barducci:2016pcb}%
  \BibitemOpen
  \bibfield  {author} {\bibinfo {author} {\bibfnamefont {D.}~\bibnamefont
  {Barducci}}, \bibinfo {author} {\bibfnamefont {G.}~\bibnamefont {Belanger}},
  \bibinfo {author} {\bibfnamefont {J.}~\bibnamefont {Bernon}}, \bibinfo
  {author} {\bibfnamefont {F.}~\bibnamefont {Boudjema}}, \bibinfo {author}
  {\bibfnamefont {J.}~\bibnamefont {Da~Silva}}, \bibinfo {author}
  {\bibfnamefont {S.}~\bibnamefont {Kraml}}, \bibinfo {author} {\bibfnamefont
  {U.}~\bibnamefont {Laa}}, \ and\ \bibinfo {author} {\bibfnamefont
  {A.}~\bibnamefont {Pukhov}},\ }\href {\doibase 10.1016/j.cpc.2017.08.028}
  {\bibfield  {journal} {\bibinfo  {journal} {Comput. Phys. Commun.}\ }\textbf
  {\bibinfo {volume} {222}},\ \bibinfo {pages} {327} (\bibinfo {year}
  {2018})},\ \Eprint {http://arxiv.org/abs/1606.03834} {arXiv:1606.03834
  [hep-ph]} \BibitemShut {NoStop}%
\bibitem [{\citenamefont {Aalbers}\ \emph {et~al.}(2016)\citenamefont {Aalbers}
  \emph {et~al.}}]{DARWIN:2016hyl}%
  \BibitemOpen
  \bibfield  {author} {\bibinfo {author} {\bibfnamefont {J.}~\bibnamefont
  {Aalbers}} \emph {et~al.} (\bibinfo {collaboration} {DARWIN}),\ }\href
  {\doibase 10.1088/1475-7516/2016/11/017} {\bibfield  {journal} {\bibinfo
  {journal} {JCAP}\ }\textbf {\bibinfo {volume} {11}},\ \bibinfo {pages} {017}
  (\bibinfo {year} {2016})},\ \Eprint {http://arxiv.org/abs/1606.07001}
  {arXiv:1606.07001 [astro-ph.IM]} \BibitemShut {NoStop}%
\bibitem [{\citenamefont {Degrande}\ \emph {et~al.}(2012)\citenamefont
  {Degrande}, \citenamefont {Duhr}, \citenamefont {Fuks}, \citenamefont
  {Grellscheid}, \citenamefont {Mattelaer},\ and\ \citenamefont
  {Reiter}}]{Degrande:2011ua}%
  \BibitemOpen
  \bibfield  {author} {\bibinfo {author} {\bibfnamefont {C.}~\bibnamefont
  {Degrande}}, \bibinfo {author} {\bibfnamefont {C.}~\bibnamefont {Duhr}},
  \bibinfo {author} {\bibfnamefont {B.}~\bibnamefont {Fuks}}, \bibinfo {author}
  {\bibfnamefont {D.}~\bibnamefont {Grellscheid}}, \bibinfo {author}
  {\bibfnamefont {O.}~\bibnamefont {Mattelaer}}, \ and\ \bibinfo {author}
  {\bibfnamefont {T.}~\bibnamefont {Reiter}},\ }\href {\doibase
  10.1016/j.cpc.2012.01.022} {\bibfield  {journal} {\bibinfo  {journal}
  {Comput. Phys. Commun.}\ }\textbf {\bibinfo {volume} {183}},\ \bibinfo
  {pages} {1201} (\bibinfo {year} {2012})},\ \Eprint
  {http://arxiv.org/abs/1108.2040} {arXiv:1108.2040 [hep-ph]} \BibitemShut
  {NoStop}%
\bibitem [{\citenamefont {Belyaev}\ \emph {et~al.}(2013)\citenamefont
  {Belyaev}, \citenamefont {Christensen},\ and\ \citenamefont
  {Pukhov}}]{Belyaev:2012qa}%
  \BibitemOpen
  \bibfield  {author} {\bibinfo {author} {\bibfnamefont {A.}~\bibnamefont
  {Belyaev}}, \bibinfo {author} {\bibfnamefont {N.~D.}\ \bibnamefont
  {Christensen}}, \ and\ \bibinfo {author} {\bibfnamefont {A.}~\bibnamefont
  {Pukhov}},\ }\href {\doibase 10.1016/j.cpc.2013.01.014} {\bibfield  {journal}
  {\bibinfo  {journal} {Comput. Phys. Commun.}\ }\textbf {\bibinfo {volume}
  {184}},\ \bibinfo {pages} {1729} (\bibinfo {year} {2013})},\ \Eprint
  {http://arxiv.org/abs/1207.6082} {arXiv:1207.6082 [hep-ph]} \BibitemShut
  {NoStop}%
\bibitem [{\citenamefont {Alloul}\ \emph {et~al.}(2014)\citenamefont {Alloul},
  \citenamefont {Christensen}, \citenamefont {Degrande}, \citenamefont {Duhr},\
  and\ \citenamefont {Fuks}}]{Alloul:2013bka}%
  \BibitemOpen
  \bibfield  {author} {\bibinfo {author} {\bibfnamefont {A.}~\bibnamefont
  {Alloul}}, \bibinfo {author} {\bibfnamefont {N.~D.}\ \bibnamefont
  {Christensen}}, \bibinfo {author} {\bibfnamefont {C.}~\bibnamefont
  {Degrande}}, \bibinfo {author} {\bibfnamefont {C.}~\bibnamefont {Duhr}}, \
  and\ \bibinfo {author} {\bibfnamefont {B.}~\bibnamefont {Fuks}},\ }\href
  {\doibase 10.1016/j.cpc.2014.04.012} {\bibfield  {journal} {\bibinfo
  {journal} {Comput. Phys. Commun.}\ }\textbf {\bibinfo {volume} {185}},\
  \bibinfo {pages} {2250} (\bibinfo {year} {2014})},\ \Eprint
  {http://arxiv.org/abs/1310.1921} {arXiv:1310.1921 [hep-ph]} \BibitemShut
  {NoStop}%
\bibitem [{\citenamefont {Backovic}\ \emph {et~al.}(2014)\citenamefont
  {Backovic}, \citenamefont {Kong},\ and\ \citenamefont
  {McCaskey}}]{Backovic:2013dpa}%
  \BibitemOpen
  \bibfield  {author} {\bibinfo {author} {\bibfnamefont {M.}~\bibnamefont
  {Backovic}}, \bibinfo {author} {\bibfnamefont {K.}~\bibnamefont {Kong}}, \
  and\ \bibinfo {author} {\bibfnamefont {M.}~\bibnamefont {McCaskey}},\ }\href
  {\doibase 10.1016/j.dark.2014.04.001} {\bibfield  {journal} {\bibinfo
  {journal} {Physics of the Dark Universe}\ }\textbf {\bibinfo {volume}
  {5-6}},\ \bibinfo {pages} {18} (\bibinfo {year} {2014})},\ \Eprint
  {http://arxiv.org/abs/1308.4955} {arXiv:1308.4955 [hep-ph]} \BibitemShut
  {NoStop}%
\bibitem [{\citenamefont {Ambrogi}\ \emph {et~al.}(2019)\citenamefont
  {Ambrogi}, \citenamefont {Arina}, \citenamefont {Backovic}, \citenamefont
  {Heisig}, \citenamefont {Maltoni}, \citenamefont {Mantani}, \citenamefont
  {Mattelaer},\ and\ \citenamefont {Mohlabeng}}]{Ambrogi:2018jqj}%
  \BibitemOpen
  \bibfield  {author} {\bibinfo {author} {\bibfnamefont {F.}~\bibnamefont
  {Ambrogi}}, \bibinfo {author} {\bibfnamefont {C.}~\bibnamefont {Arina}},
  \bibinfo {author} {\bibfnamefont {M.}~\bibnamefont {Backovic}}, \bibinfo
  {author} {\bibfnamefont {J.}~\bibnamefont {Heisig}}, \bibinfo {author}
  {\bibfnamefont {F.}~\bibnamefont {Maltoni}}, \bibinfo {author} {\bibfnamefont
  {L.}~\bibnamefont {Mantani}}, \bibinfo {author} {\bibfnamefont
  {O.}~\bibnamefont {Mattelaer}}, \ and\ \bibinfo {author} {\bibfnamefont
  {G.}~\bibnamefont {Mohlabeng}},\ }\href {\doibase 10.1016/j.dark.2018.11.009}
  {\bibfield  {journal} {\bibinfo  {journal} {Phys. Dark Univ.}\ }\textbf
  {\bibinfo {volume} {24}},\ \bibinfo {pages} {100249} (\bibinfo {year}
  {2019})},\ \Eprint {http://arxiv.org/abs/1804.00044} {arXiv:1804.00044
  [hep-ph]} \BibitemShut {NoStop}%
\bibitem [{\citenamefont {Arina}\ \emph {et~al.}(2023)\citenamefont {Arina},
  \citenamefont {Heisig}, \citenamefont {Maltoni}, \citenamefont {Massaro},\
  and\ \citenamefont {Mattelaer}}]{Arina:2021gfn}%
  \BibitemOpen
  \bibfield  {author} {\bibinfo {author} {\bibfnamefont {C.}~\bibnamefont
  {Arina}}, \bibinfo {author} {\bibfnamefont {J.}~\bibnamefont {Heisig}},
  \bibinfo {author} {\bibfnamefont {F.}~\bibnamefont {Maltoni}}, \bibinfo
  {author} {\bibfnamefont {D.}~\bibnamefont {Massaro}}, \ and\ \bibinfo
  {author} {\bibfnamefont {O.}~\bibnamefont {Mattelaer}},\ }\href {\doibase
  10.1140/epjc/s10052-023-11377-2} {\bibfield  {journal} {\bibinfo  {journal}
  {Eur. Phys. J. C}\ }\textbf {\bibinfo {volume} {83}},\ \bibinfo {pages} {241}
  (\bibinfo {year} {2023})},\ \Eprint {http://arxiv.org/abs/2107.04598}
  {arXiv:2107.04598 [hep-ph]} \BibitemShut {NoStop}%
\bibitem [{\citenamefont {Chala}\ \emph {et~al.}(2015)\citenamefont {Chala},
  \citenamefont {Kahlhoefer}, \citenamefont {McCullough}, \citenamefont
  {Nardini},\ and\ \citenamefont {Schmidt-Hoberg}}]{chala2015}%
  \BibitemOpen
  \bibfield  {author} {\bibinfo {author} {\bibfnamefont {M.}~\bibnamefont
  {Chala}}, \bibinfo {author} {\bibfnamefont {F.}~\bibnamefont {Kahlhoefer}},
  \bibinfo {author} {\bibfnamefont {M.}~\bibnamefont {McCullough}}, \bibinfo
  {author} {\bibfnamefont {G.}~\bibnamefont {Nardini}}, \ and\ \bibinfo
  {author} {\bibfnamefont {K.}~\bibnamefont {Schmidt-Hoberg}},\ }\href
  {\doibase 10.1007/JHEP07(2015)089} {\bibfield  {journal} {\bibinfo  {journal}
  {Journal of High Energy Physics}\ }\textbf {\bibinfo {volume} {2015}},\
  \bibinfo {pages} {89} (\bibinfo {year} {2015})}\BibitemShut {NoStop}%
\bibitem [{\citenamefont {Shu}(2008)}]{PhysRevD.78.096004}%
  \BibitemOpen
  \bibfield  {author} {\bibinfo {author} {\bibfnamefont {J.}~\bibnamefont
  {Shu}},\ }\href {\doibase 10.1103/PhysRevD.78.096004} {\bibfield  {journal}
  {\bibinfo  {journal} {Phys. Rev. D}\ }\textbf {\bibinfo {volume} {78}},\
  \bibinfo {pages} {096004} (\bibinfo {year} {2008})}\BibitemShut {NoStop}%
\bibitem [{\citenamefont {Hosch}\ \emph {et~al.}(1997)\citenamefont {Hosch},
  \citenamefont {Whisnant},\ and\ \citenamefont {Young}}]{PhysRevD.55.3137}%
  \BibitemOpen
  \bibfield  {author} {\bibinfo {author} {\bibfnamefont {M.}~\bibnamefont
  {Hosch}}, \bibinfo {author} {\bibfnamefont {K.}~\bibnamefont {Whisnant}}, \
  and\ \bibinfo {author} {\bibfnamefont {B.-L.}\ \bibnamefont {Young}},\ }\href
  {\doibase 10.1103/PhysRevD.55.3137} {\bibfield  {journal} {\bibinfo
  {journal} {Phys. Rev. D}\ }\textbf {\bibinfo {volume} {55}},\ \bibinfo
  {pages} {3137} (\bibinfo {year} {1997})}\BibitemShut {NoStop}%
\bibitem [{\citenamefont {Babu}\ \emph {et~al.}(2012)\citenamefont {Babu},
  \citenamefont {Julio},\ and\ \citenamefont {Zhang}}]{BABU2012468}%
  \BibitemOpen
  \bibfield  {author} {\bibinfo {author} {\bibfnamefont {K.}~\bibnamefont
  {Babu}}, \bibinfo {author} {\bibfnamefont {J.}~\bibnamefont {Julio}}, \ and\
  \bibinfo {author} {\bibfnamefont {Y.}~\bibnamefont {Zhang}},\ }\href
  {\doibase https://doi.org/10.1016/j.nuclphysb.2012.01.018} {\bibfield
  {journal} {\bibinfo  {journal} {Nuclear Physics B}\ }\textbf {\bibinfo
  {volume} {858}},\ \bibinfo {pages} {468} (\bibinfo {year}
  {2012})}\BibitemShut {NoStop}%
\bibitem [{\citenamefont {Amole}\ \emph {et~al.}(2019)\citenamefont {Amole}
  \emph {et~al.}}]{PICO:2019vsc}%
  \BibitemOpen
  \bibfield  {author} {\bibinfo {author} {\bibfnamefont {C.}~\bibnamefont
  {Amole}} \emph {et~al.} (\bibinfo {collaboration} {PICO}),\ }\href {\doibase
  10.1103/PhysRevD.100.022001} {\bibfield  {journal} {\bibinfo  {journal}
  {Phys. Rev. D}\ }\textbf {\bibinfo {volume} {100}},\ \bibinfo {pages}
  {022001} (\bibinfo {year} {2019})},\ \Eprint
  {http://arxiv.org/abs/1902.04031} {arXiv:1902.04031 [astro-ph.CO]}
  \BibitemShut {NoStop}%
\bibitem [{\citenamefont {Blanchet}\ and\ \citenamefont
  {Lavalle}(2012)}]{Blanchet:2012vq}%
  \BibitemOpen
  \bibfield  {author} {\bibinfo {author} {\bibfnamefont {S.}~\bibnamefont
  {Blanchet}}\ and\ \bibinfo {author} {\bibfnamefont {J.}~\bibnamefont
  {Lavalle}},\ }\href {\doibase 10.1088/1475-7516/2012/11/021} {\bibfield
  {journal} {\bibinfo  {journal} {JCAP}\ }\textbf {\bibinfo {volume} {11}},\
  \bibinfo {pages} {021} (\bibinfo {year} {2012})},\ \Eprint
  {http://arxiv.org/abs/1207.2476} {arXiv:1207.2476 [astro-ph.HE]} \BibitemShut
  {NoStop}%
\bibitem [{\citenamefont {Koechler}\ and\ \citenamefont
  {Di~Mauro}(2025)}]{Koechler:2025ryv}%
  \BibitemOpen
  \bibfield  {author} {\bibinfo {author} {\bibfnamefont {J.}~\bibnamefont
  {Koechler}}\ and\ \bibinfo {author} {\bibfnamefont {M.}~\bibnamefont
  {Di~Mauro}},\ }\href@noop {} {\  (\bibinfo {year} {2025})},\ \Eprint
  {http://arxiv.org/abs/2508.02775} {arXiv:2508.02775 [hep-ph]} \BibitemShut
  {NoStop}%
\bibitem [{\citenamefont {Di~Mauro}\ and\ \citenamefont
  {Winkler}(2021)}]{DiMauro:2021qcf}%
  \BibitemOpen
  \bibfield  {author} {\bibinfo {author} {\bibfnamefont {M.}~\bibnamefont
  {Di~Mauro}}\ and\ \bibinfo {author} {\bibfnamefont {M.~W.}\ \bibnamefont
  {Winkler}},\ }\href {\doibase 10.1103/PhysRevD.103.123005} {\bibfield
  {journal} {\bibinfo  {journal} {Phys. Rev. D}\ }\textbf {\bibinfo {volume}
  {103}},\ \bibinfo {pages} {123005} (\bibinfo {year} {2021})},\ \Eprint
  {http://arxiv.org/abs/2101.11027} {arXiv:2101.11027 [astro-ph.HE]}
  \BibitemShut {NoStop}%
\bibitem [{\citenamefont {Arina}\ \emph {et~al.}(2024)\citenamefont {Arina},
  \citenamefont {Di~Mauro}, \citenamefont {Fornengo}, \citenamefont {Heisig},
  \citenamefont {Jueid},\ and\ \citenamefont {de~Austri}}]{Arina:2023eic}%
  \BibitemOpen
  \bibfield  {author} {\bibinfo {author} {\bibfnamefont {C.}~\bibnamefont
  {Arina}}, \bibinfo {author} {\bibfnamefont {M.}~\bibnamefont {Di~Mauro}},
  \bibinfo {author} {\bibfnamefont {N.}~\bibnamefont {Fornengo}}, \bibinfo
  {author} {\bibfnamefont {J.}~\bibnamefont {Heisig}}, \bibinfo {author}
  {\bibfnamefont {A.}~\bibnamefont {Jueid}}, \ and\ \bibinfo {author}
  {\bibfnamefont {R.~R.}\ \bibnamefont {de~Austri}},\ }\href {\doibase
  10.1088/1475-7516/2024/03/035} {\bibfield  {journal} {\bibinfo  {journal}
  {JCAP}\ }\textbf {\bibinfo {volume} {03}},\ \bibinfo {pages} {035} (\bibinfo
  {year} {2024})},\ \Eprint {http://arxiv.org/abs/2312.01153} {arXiv:2312.01153
  [astro-ph.HE]} \BibitemShut {NoStop}%
\bibitem [{\citenamefont {Di~Mauro}\ \emph {et~al.}(2025)\citenamefont
  {Di~Mauro}, \citenamefont {Fornengo}, \citenamefont {Jueid}, \citenamefont
  {de~Austri},\ and\ \citenamefont {Bellini}}]{DiMauro:2024kml}%
  \BibitemOpen
  \bibfield  {author} {\bibinfo {author} {\bibfnamefont {M.}~\bibnamefont
  {Di~Mauro}}, \bibinfo {author} {\bibfnamefont {N.}~\bibnamefont {Fornengo}},
  \bibinfo {author} {\bibfnamefont {A.}~\bibnamefont {Jueid}}, \bibinfo
  {author} {\bibfnamefont {R.~R.}\ \bibnamefont {de~Austri}}, \ and\ \bibinfo
  {author} {\bibfnamefont {F.}~\bibnamefont {Bellini}},\ }\href {\doibase
  10.1103/w6n5-vs4d} {\bibfield  {journal} {\bibinfo  {journal} {Phys. Rev.
  Lett.}\ }\textbf {\bibinfo {volume} {135}},\ \bibinfo {pages} {131002}
  (\bibinfo {year} {2025})},\ \Eprint {http://arxiv.org/abs/2411.04815}
  {arXiv:2411.04815 [astro-ph.HE]} \BibitemShut {NoStop}%
\bibitem [{\citenamefont {Di~Mauro}(2021{\natexlab{b}})}]{DiMauro:2021prd}%
  \BibitemOpen
  \bibfield  {author} {\bibinfo {author} {\bibfnamefont {M.}~\bibnamefont
  {Di~Mauro}},\ }\href {\doibase 10.1103/PhysRevD.103.063029} {\bibfield
  {journal} {\bibinfo  {journal} {Phys. Rev. D}\ }\textbf {\bibinfo {volume}
  {103}},\ \bibinfo {pages} {063029} (\bibinfo {year}
  {2021}{\natexlab{b}})}\BibitemShut {NoStop}%
\bibitem [{\citenamefont {Cholis}\ \emph
  {et~al.}(2022{\natexlab{b}})\citenamefont {Cholis}, \citenamefont {Zhong},
  \citenamefont {McDermott},\ and\ \citenamefont
  {Surdutovich}}]{Cholis:2022prd}%
  \BibitemOpen
  \bibfield  {author} {\bibinfo {author} {\bibfnamefont {I.}~\bibnamefont
  {Cholis}}, \bibinfo {author} {\bibfnamefont {Y.-M.}\ \bibnamefont {Zhong}},
  \bibinfo {author} {\bibfnamefont {S.~D.}\ \bibnamefont {McDermott}}, \ and\
  \bibinfo {author} {\bibfnamefont {J.~P.}\ \bibnamefont {Surdutovich}},\
  }\href {\doibase 10.1103/PhysRevD.105.103023} {\bibfield  {journal} {\bibinfo
   {journal} {Phys. Rev. D}\ }\textbf {\bibinfo {volume} {105}},\ \bibinfo
  {pages} {103023} (\bibinfo {year} {2022}{\natexlab{b}})}\BibitemShut
  {NoStop}%
\bibitem [{\citenamefont {Steigman}\ \emph {et~al.}(2012)\citenamefont
  {Steigman}, \citenamefont {Dasgupta},\ and\ \citenamefont
  {Beacom}}]{Steigman:2012nb}%
  \BibitemOpen
  \bibfield  {author} {\bibinfo {author} {\bibfnamefont {G.}~\bibnamefont
  {Steigman}}, \bibinfo {author} {\bibfnamefont {B.}~\bibnamefont {Dasgupta}},
  \ and\ \bibinfo {author} {\bibfnamefont {J.~F.}\ \bibnamefont {Beacom}},\
  }\href {\doibase 10.1103/PhysRevD.86.023506} {\bibfield  {journal} {\bibinfo
  {journal} {Phys. Rev. D}\ }\textbf {\bibinfo {volume} {86}},\ \bibinfo
  {pages} {023506} (\bibinfo {year} {2012})},\ \Eprint
  {http://arxiv.org/abs/1204.3622} {arXiv:1204.3622 [hep-ph]} \BibitemShut
  {NoStop}%
\bibitem [{\citenamefont {Bringmann}\ \emph {et~al.}(2021)\citenamefont
  {Bringmann}, \citenamefont {Depta}, \citenamefont {Hufnagel},\ and\
  \citenamefont {Schmidt-Hoberg}}]{BRINGMANN2021136341}%
  \BibitemOpen
  \bibfield  {author} {\bibinfo {author} {\bibfnamefont {T.}~\bibnamefont
  {Bringmann}}, \bibinfo {author} {\bibfnamefont {P.~F.}\ \bibnamefont
  {Depta}}, \bibinfo {author} {\bibfnamefont {M.}~\bibnamefont {Hufnagel}}, \
  and\ \bibinfo {author} {\bibfnamefont {K.}~\bibnamefont {Schmidt-Hoberg}},\
  }\href {\doibase https://doi.org/10.1016/j.physletb.2021.136341} {\bibfield
  {journal} {\bibinfo  {journal} {Physics Letters B}\ }\textbf {\bibinfo
  {volume} {817}},\ \bibinfo {pages} {136341} (\bibinfo {year}
  {2021})}\BibitemShut {NoStop}%
\bibitem [{\citenamefont {Macias}\ \emph {et~al.}(2018)\citenamefont {Macias},
  \citenamefont {Gordon}, \citenamefont {Crocker}, \citenamefont {Coleman},
  \citenamefont {Paterson}, \citenamefont {Horiuchi},\ and\ \citenamefont
  {Pohl}}]{Macias:2016nev}%
  \BibitemOpen
  \bibfield  {author} {\bibinfo {author} {\bibfnamefont {O.}~\bibnamefont
  {Macias}}, \bibinfo {author} {\bibfnamefont {C.}~\bibnamefont {Gordon}},
  \bibinfo {author} {\bibfnamefont {R.~M.}\ \bibnamefont {Crocker}}, \bibinfo
  {author} {\bibfnamefont {B.}~\bibnamefont {Coleman}}, \bibinfo {author}
  {\bibfnamefont {D.}~\bibnamefont {Paterson}}, \bibinfo {author}
  {\bibfnamefont {S.}~\bibnamefont {Horiuchi}}, \ and\ \bibinfo {author}
  {\bibfnamefont {M.}~\bibnamefont {Pohl}},\ }\href {\doibase
  10.1038/s41550-018-0414-3} {\bibfield  {journal} {\bibinfo  {journal} {Nat.
  Astron.}\ }\textbf {\bibinfo {volume} {2}},\ \bibinfo {pages} {387} (\bibinfo
  {year} {2018})},\ \Eprint {http://arxiv.org/abs/1611.06644} {arXiv:1611.06644
  [astro-ph.HE]} \BibitemShut {NoStop}%
\bibitem [{\citenamefont {Bartels}\ \emph {et~al.}(2018)\citenamefont
  {Bartels}, \citenamefont {Storm}, \citenamefont {Weniger},\ and\
  \citenamefont {Calore}}]{Bartels:2017vsx}%
  \BibitemOpen
  \bibfield  {author} {\bibinfo {author} {\bibfnamefont {R.}~\bibnamefont
  {Bartels}}, \bibinfo {author} {\bibfnamefont {E.}~\bibnamefont {Storm}},
  \bibinfo {author} {\bibfnamefont {C.}~\bibnamefont {Weniger}}, \ and\
  \bibinfo {author} {\bibfnamefont {F.}~\bibnamefont {Calore}},\ }\href
  {\doibase 10.1038/s41550-018-0531-z} {\bibfield  {journal} {\bibinfo
  {journal} {Nat. Astron.}\ }\textbf {\bibinfo {volume} {2}},\ \bibinfo {pages}
  {819} (\bibinfo {year} {2018})},\ \Eprint {http://arxiv.org/abs/1711.04778}
  {arXiv:1711.04778 [astro-ph.HE]} \BibitemShut {NoStop}%
\bibitem [{\citenamefont {Manconi}\ \emph {et~al.}(2024)\citenamefont
  {Manconi}, \citenamefont {Calore},\ and\ \citenamefont
  {Donato}}]{Manconi:2024tgh}%
  \BibitemOpen
  \bibfield  {author} {\bibinfo {author} {\bibfnamefont {S.}~\bibnamefont
  {Manconi}}, \bibinfo {author} {\bibfnamefont {F.}~\bibnamefont {Calore}}, \
  and\ \bibinfo {author} {\bibfnamefont {F.}~\bibnamefont {Donato}},\ }\href
  {\doibase 10.1103/PhysRevD.109.123042} {\bibfield  {journal} {\bibinfo
  {journal} {Phys. Rev. D}\ }\textbf {\bibinfo {volume} {109}},\ \bibinfo
  {pages} {123042} (\bibinfo {year} {2024})},\ \Eprint
  {http://arxiv.org/abs/2402.04733} {arXiv:2402.04733 [astro-ph.HE]}
  \BibitemShut {NoStop}%
\bibitem [{\citenamefont {Silveira}\ and\ \citenamefont
  {Zee}(1985)}]{SILVEIRA1985136}%
  \BibitemOpen
  \bibfield  {author} {\bibinfo {author} {\bibfnamefont {V.}~\bibnamefont
  {Silveira}}\ and\ \bibinfo {author} {\bibfnamefont {A.}~\bibnamefont {Zee}},\
  }\href {\doibase https://doi.org/10.1016/0370-2693(85)90624-0} {\bibfield
  {journal} {\bibinfo  {journal} {Physics Letters B}\ }\textbf {\bibinfo
  {volume} {161}},\ \bibinfo {pages} {136} (\bibinfo {year}
  {1985})}\BibitemShut {NoStop}%
\bibitem [{\citenamefont {Langacker}\ \emph {et~al.}(2008)\citenamefont
  {Langacker}, \citenamefont {Paz}, \citenamefont {Wang},\ and\ \citenamefont
  {Yavin}}]{Langacker:2007ac}%
  \BibitemOpen
  \bibfield  {author} {\bibinfo {author} {\bibfnamefont {P.}~\bibnamefont
  {Langacker}}, \bibinfo {author} {\bibfnamefont {G.}~\bibnamefont {Paz}},
  \bibinfo {author} {\bibfnamefont {L.-T.}\ \bibnamefont {Wang}}, \ and\
  \bibinfo {author} {\bibfnamefont {I.}~\bibnamefont {Yavin}},\ }\href
  {\doibase 10.1103/PhysRevLett.100.041802} {\bibfield  {journal} {\bibinfo
  {journal} {Phys. Rev. Lett.}\ }\textbf {\bibinfo {volume} {100}},\ \bibinfo
  {pages} {041802} (\bibinfo {year} {2008})},\ \Eprint
  {http://arxiv.org/abs/0710.1632} {arXiv:0710.1632 [hep-ph]} \BibitemShut
  {NoStop}%
\bibitem [{\citenamefont {Langacker}(2009)}]{Langacker:2008yv}%
  \BibitemOpen
  \bibfield  {author} {\bibinfo {author} {\bibfnamefont {P.}~\bibnamefont
  {Langacker}},\ }\href {\doibase 10.1103/RevModPhys.81.1199} {\bibfield
  {journal} {\bibinfo  {journal} {Rev. Mod. Phys.}\ }\textbf {\bibinfo {volume}
  {81}},\ \bibinfo {pages} {1199} (\bibinfo {year} {2009})},\ \Eprint
  {http://arxiv.org/abs/0801.1345} {arXiv:0801.1345 [hep-ph]} \BibitemShut
  {NoStop}%
\end{thebibliography}%

\end{document}